\documentclass[preprint]{aastex63}

\renewcommand{\deleted}[1]{}

\usepackage{color}
\usepackage{natbib}
\usepackage{amsmath}
\usepackage{nicefrac}
\tightenlines

\newcommand \Angstrom   {\,{\rm \AA}}

\newcommand \bE         {{\bf E}}

\newcommand \beq        {\begin{equation}}
\newcommand \beqa	{\begin{eqnarray}}

\newcommand \calP       {{\cal P}}

\newcommand \cm         {\,{\rm cm}}

\newcommand \eeq	{\end{equation}}
\newcommand \eeqa	{\end{eqnarray}}

\newcommand \erg	{\,{\rm erg}}

\newcommand \eV 	{\,{\rm eV}}
\newcommand \Fclip      {F_{\rm clip}}
\newcommand \fion       {f_{\rm ion}}
\newcommand \FTIR       {F_{\rm TIR}} 
\newcommand \G          {\,{\rm G}}  
\newcommand \gm         {\,{\rm g}}

\newcommand \gtsim	{\gtrsim}		 
\newcommand \Gyr        {\,{\rm Gyr}}
\newcommand \Ha 	{{\rm H}}

\newcommand \K  	{\,{\rm K}}

\newcommand \ltsim	{\lesssim}		 

\newcommand \Myr        {\,{\rm Myr}}

\newcommand \nH         {n_{\rm H}}

\newcommand \NH         {N_{\rm H}}

\newcommand \qpah       {q_{\rm PAH}}
\newcommand \s	        {\,{\rm s}}

\newcommand \aeff       {a_{\rm eff}}

\countdef\decade=200
\advance\decade by \year
\countdef\hours=201
\advance\hours by \time
\divide\hours by 60
\countdef\mins=202
\advance\mins by \hours
\multiply\mins by 60
\multiply\hours by 100
\countdef\miltime=203
\advance\miltime by \hours
\advance\miltime by \time
\advance\miltime by -\mins
\renewcommand\today{\number\decade.\number\month.\number\day.\number\miltime}
\begin{document}

\title{%
        \vspace*{-2.0em}
        {\normalsize\rm {\it The Astrophysical Journal}, {\bf 917:}3 (2021 Aug.\ 10)}\\ 
        \vspace*{1.0em}
        {\bf Excitation of Polycyclic Aromatic Hydrocarbon Emission: 
             Dependence on Size Distribution, Ionization,
             and Starlight Spectrum and Intensity
             }
	}

\author[0000-0002-0846-936X]{B.~T.~Draine}
\affiliation{Dept.\ of Astrophysical Sciences,
  Princeton University, Princeton, NJ 08544, USA}

\author[0000-0002-1119-642X]{Aigen~Li}
\affiliation{Dept.\ of Physics and Astronomy, University of Missouri,
Columbia, MO 62511, USA}

\author[0000-0001-7449-4638]{Brandon S. Hensley}
\affiliation{Dept.\ of Astrophysical Sciences,
  Princeton University, Princeton, NJ 08544, USA}
\affiliation{Spitzer Fellow}

\author[0000-0001-9162-2371]{L.~K.~Hunt}
\affiliation{INAF - Osservatorio Astrofisico di Arcetri,
  Largo E. Fermi 5, 50125 Firenze, Italy}

\author[0000-0002-4378-8534]{K.~Sandstrom}
\affiliation{Center for Astrophysics and Space Sciences,
  University of California, 9500 Gilman Drive, San Diego CA 92093, USA}

\author[0000-0003-1545-5078]{J.-D.~T.~Smith}
\affiliation{Dept.\ of Physics and Astronomy, University of Toledo,
  Toledo, OH 43606, USA}

\email{draine@astro.princeton.edu}

\begin{abstract}
Using physical models, we study the sensitivity of polycyclic aromatic
hydrocarbon (PAH) emission
spectra to the character of the illuminating starlight, to the PAH size
distribution, and to the PAH charge distribution.
The starlight models considered range from the emission from 
a $3\Myr$-old starburst, rich in far-ultraviolet (FUV) radiation, to the
FUV-poor spectrum of the very old population of the M31 bulge.  
A wide range of starlight intensities is considered.
The effects of reddening in dusty clouds are investigated for different
starlight spectra.  For a fixed PAH abundance parameter $\qpah$
(the fraction of the total dust mass in PAHs with $<10^3$ C atoms),
the fraction
of the IR power appearing in the PAH emission features can vary by a factor
of two as the starlight spectrum varies from FUV-poor (M31 bulge) to
FUV-rich (young starburst).
We show how $\qpah$ can be measured
from the strength of the 7.7$\micron$ emission.
The fractional power in the $17\micron$ feature can be suppressed by
high starlight intensities.
\end{abstract}

\let\svthefootnote\thefootnote
\let\thefootnote\relax\footnote{\textcopyright 2021.  All rights reserved.}
\let\thefootnote\svthefootnote

\section{Introduction
         \label{sec:intro}}
Strong infrared (IR) emission features at 3.3, 6.2, 7.7, 8.6, 11.2, 12.7, 
and 17.1$\micron$
are prominent in the spectra of normal star-forming galaxies
\citep[see the review by][]{Li_2020}, and these
emission features have been observed in the spectra of galaxies at
redshifts $z>4$ 
\citep{Riechers+Pope+Daddi+etal_2014,Armus+Charmandaris+Soifer_2020}.
\citet{Leger+Puget_1984} and \citet{Allamandola+Tielens+Barker_1985}
proposed that these features are radiated by the vibrational modes of
polycyclic aromatic hydrocarbon (PAH) molecules, and this
hypothesis is now generally accepted 
\citep{Tielens_2008,Li_2020}.

Except for regions with high-pressure hot plasma -- which are thought to
contribute negligibly to PAH emission from galaxies --
excitation of PAHs is dominated by  absorption of
starlight photons.
The photon absorption produces electronic excitation of the PAH, usually
followed by rapid ``internal conversion'' of the electronic excitation into
vibrational energy.
The vibrationally excited nanoparticle then cools
by infrared emission.  If ergodicity is assumed, a
realistic vibrational density of states 
(to relate the vibrational energy to temperature) and
assumed infrared band strengths then allow one to calculate the
time-averaged emission spectrum for PAHs heated by
starlight photons.

The emission spectrum of a galaxy, or of a region within a galaxy,
must depend not only on the abundance and
composition of the PAH population, but also on the spectrum of the
starlight responsible for exciting the PAHs.
For example, the PAH emission spectra in M31 
are seen to vary from the
central bulge to the star-forming rings 
\citep{Hemachandra+Barmby+Peeters+etal_2015}.
Most previous modeling of PAH excitation \citep[e.g.,][]{Li+Draine_2001b,
Draine+Li_2007} assumed a standard starlight spectrum estimated for
the diffuse starlight at the location of the Sun, with only a few
explorations of other illuminating spectra
\citep[e.g.,][]{Li+Draine_2002b,
                Galliano+Madden+Tielens+etal_2008,
                Draine_2011b,
                Mori+Sakon+Onaka+etal_2012,
                Draine+Aniano+Krause+etal_2014}.

The aim of the present paper is to calculate the PAH emission
for a range of starlight spectra and intensities
appropriate in different
environments,
to provide model results that may be useful in interpretation
of existing PAH emission spectra measured by ISO, Spitzer, and AKARI,
or by future facilities such as James Webb Space Telescope (JWST).
For the adopted physical model of PAHs, we seek to delineate the
sensitivity of PAH emission spectra to variations in the spectrum of
the starlight that is heating the PAHs.

In addition to using the local diffuse starlight spectrum 
(which continues to be a good proxy for the overall radiation in 
star-forming galaxies),
we also
consider extreme examples, ranging from the integrated light from a
very young starburst population, to the light from a very old evolved
stellar population.
We investigate the effects of reddening of the starlight by dust.
We also examine the sensitivity of the model emission spectra to 
possible changes
in the size distribution of the PAHs, as well as their degree of ionization.

The paper is organized as follows: Section \ref{sec:starlight spectra}
describes and characterizes the starlight spectra that are employed.
Section \ref{sec:PAH physics} describes the calculational approach,
with examples of temperature distributions shown in Section
\ref{sec:dPdT}, and time-averaged emission spectra for individual PAHs
presented in Section \ref{sec:single-PAH spectra}.  The adopted PAH
size distribution, and the emission spectra for such PAH mixtures, are
presented and discussed in Section \ref{sec:mixed-PAH spectra}.  The
effect of varying the PAH ionized fraction is examined in Section
\ref{sec:ionization}, and the sensitivity to the PAH size distribution
is investigated in Section \ref{sec:pah_size}.  
The dependence of the $F(11.2\micron)/F(7.7\micron)$, 
$F(6.2\micron)/F(7.7\micron)$, $F(3.3\micron)/F(7.7\micron)$, 
$F(3.3\micron)/F(11.2\micron)$, and
$F(17\micron)/F(11.2\micron)$ band ratios on the starlight spectrum,
intensity, PAH size distribution, and PAH ionization are evaluated
and discussed in Section \ref{sec:discussion}, and
summarized in Section \ref{sec:summary}.
\section{\label{sec:starlight spectra}
         Radiation Fields}

\subsection{Unreddened}

We consider starlight from various stellar populations.  We include
unreddened spectra from the single-age stellar population (``starburst'')
models of
\citet[][hereafter BC03]{Bruzual+Charlot_2003} 
for 
ages $t$ ranging from $3\Myr$ to $1\Gyr$.  
The BC03 models assume the stars to form with
a standard initial mass function
from gas with heavy-element mass fraction $Z=0.02$ 
(i.e., near-solar metallicity).
We also
include one 
very low-metallicity model with $Z=0.0004 \approx 0.02\,Z_\odot$
and $t=10\Myr$ to examine the effect of varying $Z$.  

We also
consider the ``BPASS'' single-age stellar population models
\citep{Eldridge+Stanway+Xiao+etal_2017,Stanway+Eldridge_2018},
which include the effects of binary stars.
We consider the same range of ages as for the BC03 models,
and include one BPASS low-metallicity example (for $t=10\Myr$, with
$Z=0.001 \approx 0.05\,Z_\odot$).

In addition to the BC03 and BPASS models, we consider 
the solar neighborhood spectrum as representative of the
typical interstellar radiation field in the diffuse 
interstellar medium (ISM) of 
a star-forming galaxy with more-or-less steady star formation for the
past $\sim$$10\Gyr$,  
with the starlight reddened by distributed interstellar dust.
We use the model of \citet[][hereafter MMP]{Mathis+Mezger+Panagia_1983}
for the starlight in the 
solar neighborhood, but with slightly modified parameters 
\citep[see discussion in][]{Draine_2011a}: 
the dilution factor for the 3000\,K
component is increased from $W=4\times10^{-13}$ to $7\times10^{-13}$,
and the dilution factor for the 4000\,K component is increased from
$W=1.0\times10^{-13}$ to $1.65\times10^{-13}$.
We refer to this as the modified MMP (mMMP) 
starlight spectrum,
with energy density per unit frequency $u_{{\rm mMMP},\nu}$, and
total starlight energy density
\beq
u_{\rm mMMP}=1.043\times10^{-12}\erg\cm^{-3}
~~~.
\eeq

We also consider starlight from a very
old stellar population, using the spectrum of stars in the bulge
population of M31 adopted by
\citet{Groves+Krause+Sandstrom+etal_2012}.  
The M31 bulge spectrum may be representative of the starlight heating dust in
an elliptical galaxy.

Because we are considering
the heating of dust grains in regions where the hydrogen is
predominantly H\,I or H$_2$, the starlight spectra in
all cases are cut off at the Lyman limit, $h\nu=I_{\rm H}=13.6\eV$.

The starlight intensity will be characterized by the heating
effect on the grains that dominate the far-infrared (FIR) emission.
We calculate the
rate of energy absorption by 
a specified ``standard grain.''
For the
standard grain we adopt a 1.6:1 oblate
``astrodust'' grain 
\citep{Draine+Hensley_2021a}, with porosity
$\calP=0.2$ and effective radius $a_{\rm eff}=0.1\micron$.
Let $C_{\rm abs}^{\rm (Ad)}(\nu)$ be the orientation-averaged
absorption cross section for this standard grain, 
$c$ the speed of light,
and $u_{\star,\nu}$ the starlight energy density per unit frequency.
The standard mMMP radiation field produces a heating rate
\beq
h_{\rm ref} \equiv 
\int d\nu \, u_{{\rm mMMP},\nu} \, c \,  C_{\rm abs}^{\rm (Ad)}(\nu)
=
1.958
\times10^{-12}\erg\s^{-1}
\eeq
for our standard astrodust grain 
($\aeff=0.1\micron$, $b/a=1.6$, $\calP=0.2$).

For each spectral shape $u_{\star,\nu}$
we define a dimensionless parameter
\beq
\gamma_{\star} \equiv
\frac{
[\int d\nu \, u_{\star,\nu} \, C_{\rm abs}^{\rm (Ad)}(\nu)]/
[\int d\nu \, u_{\star,\nu}]
}
{
[\int d\nu \, u_{{\rm mMMP},\nu} \, C_{\rm abs}^{\rm (Ad)}(\nu)]/
[\int d\nu \, u_{{\rm mMMP},\nu}]
}
~~~.
\eeq
$\gamma_\star$ is the spectrum-averaged absorption cross section
for the standard grain
relative to the spectrum-averaged absorption cross section for
the mMMP starlight spectrum.  
$\gamma_\star$ is a measure of how effective a given radiation
spectrum is (relative to the mMMP spectrum) 
for heating ``standard'' dust.
By definition, $\gamma_\star=1$ for the mMMP
spectrum;
$\gamma_\star>1$ for bluer starlight (more readily absorbed by dust), 
and $\gamma_\star < 1$ for redder
starlight (less effective for dust heating).

\begin{figure}
\begin{center}
\includegraphics[angle=0,width=8.9cm,
                 clip=true,trim=0.5cm 5.0cm 0.5cm 2.5cm]
{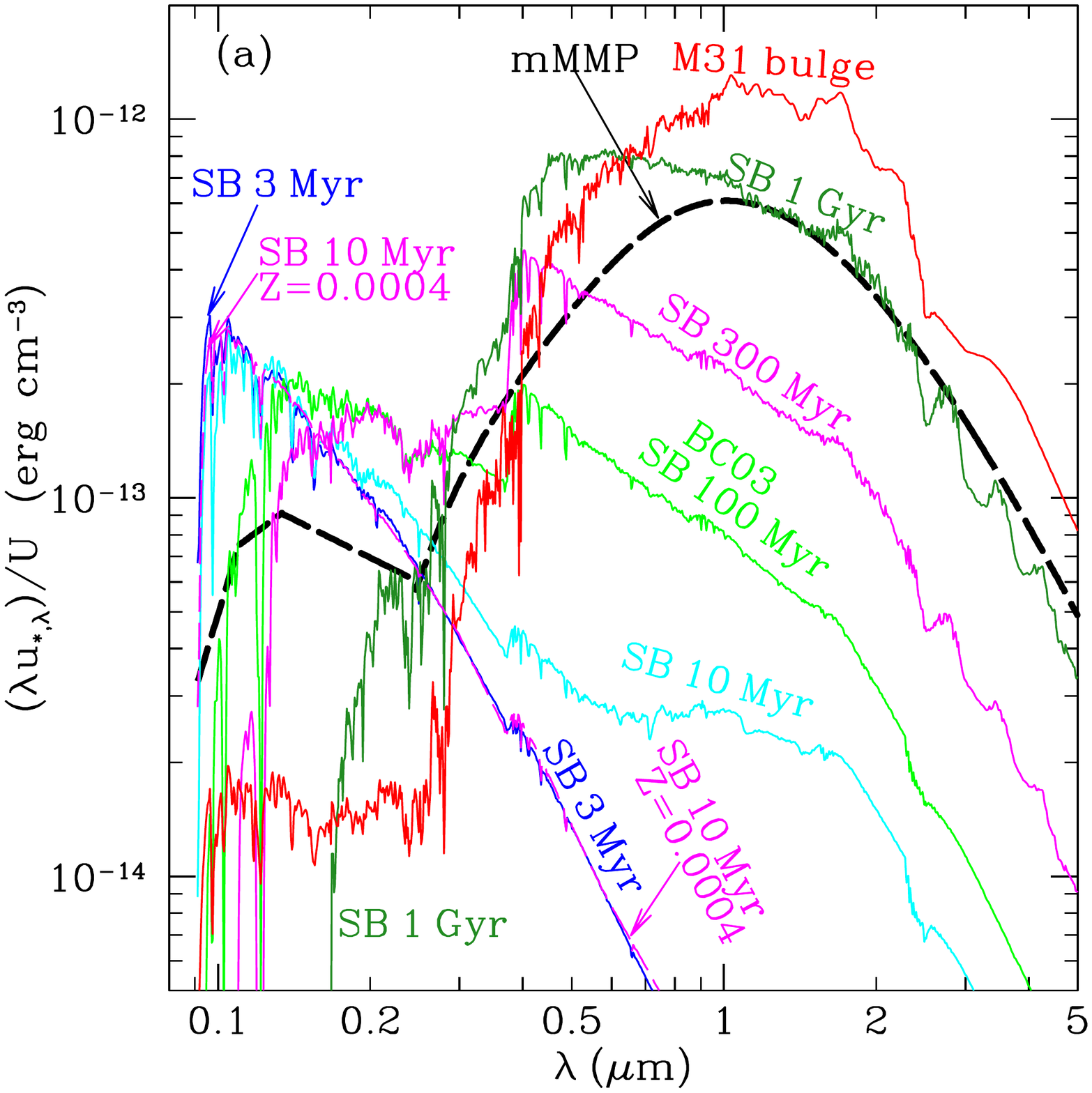}
\includegraphics[angle=0,width=8.9cm,
                 clip=true,trim=0.5cm 5.0cm 0.5cm 2.5cm]
{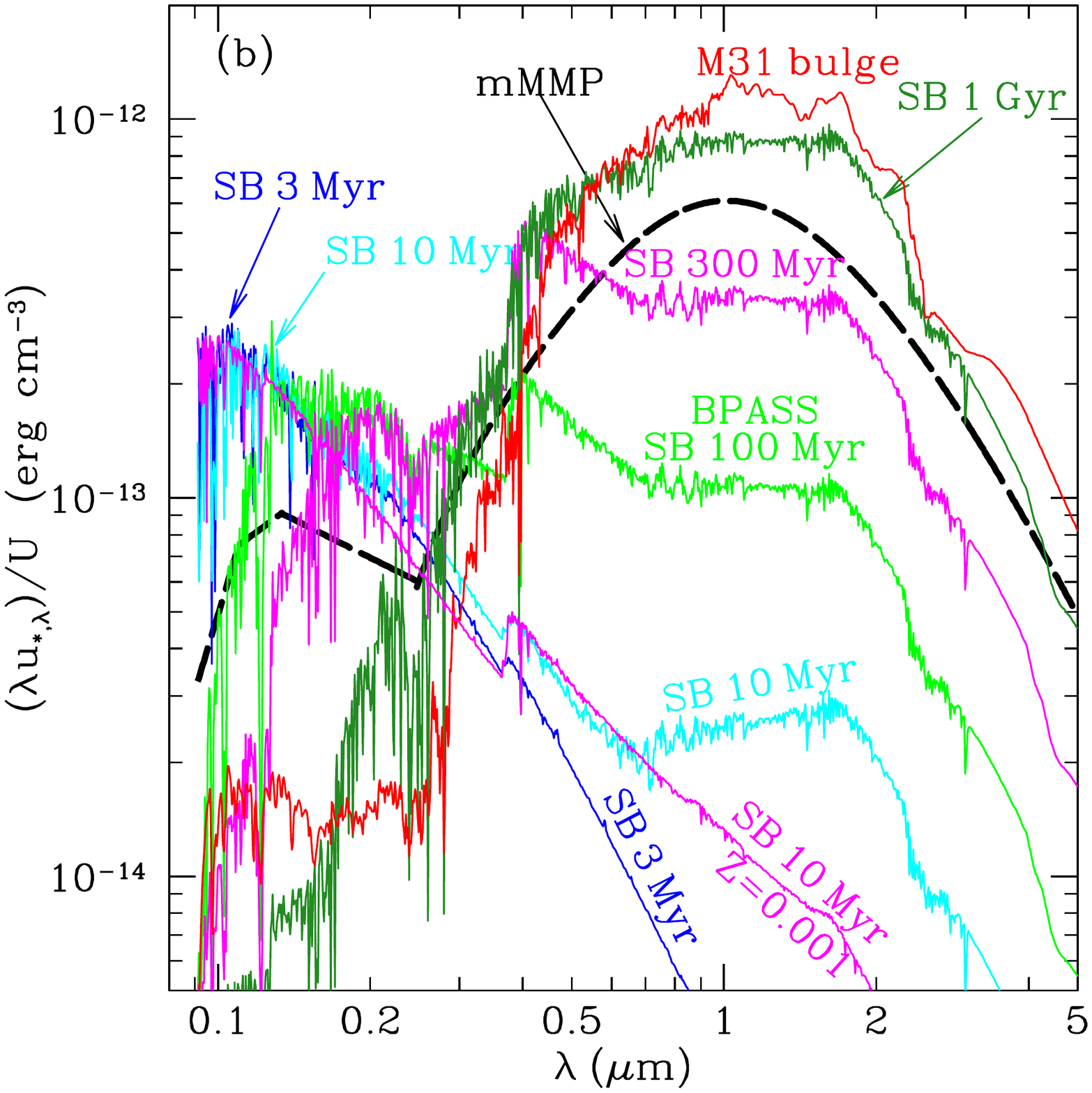}
\caption{\label{fig:isrf}\footnotesize
         (a)
         Spectra of starlight from various stellar populations, including
         single-age stellar populations \citep[][BC03]{Bruzual+Charlot_2003},
         with ages $t=3\Myr$, $10\Myr$, $100\Myr$, 
         $300\Myr$, and $1\Gyr$, and heavy-element mass fraction 
         $Z=0.02$.
         Also shown is the $t=10\Myr$, $Z=0.0004$ spectrum, 
         very similar to the
         $t=3\Myr$, $Z=0.02$ spectrum.
         The black dashed curve
         is the mMMP spectrum of starlight in the solar
         neighborhood (see text).
         and the red curve is
         the spectrum of starlight from the bulge population of M31
         \citep{Groves+Krause+Sandstrom+etal_2012}.
         For each spectrum, the intensity shown provides the same rate of
         heating $h_{\rm ref}$
         for a ``standard'' $a=0.1\micron$ grain (see text) as for
         the mMMP estimate for starlight in the solar neighborhood.
         (b) Same as (a), but with the BPASS single-age stellar
         populations 
         \citep{Eldridge+Stanway+Xiao+etal_2017,Stanway+Eldridge_2018}.
         }
\end{center}
\end{figure}
\begin{figure}
\begin{center}
\includegraphics[angle=0,width=8.0cm,
                 clip=true,trim=0.5cm 5.0cm 0.5cm 2.5cm]
{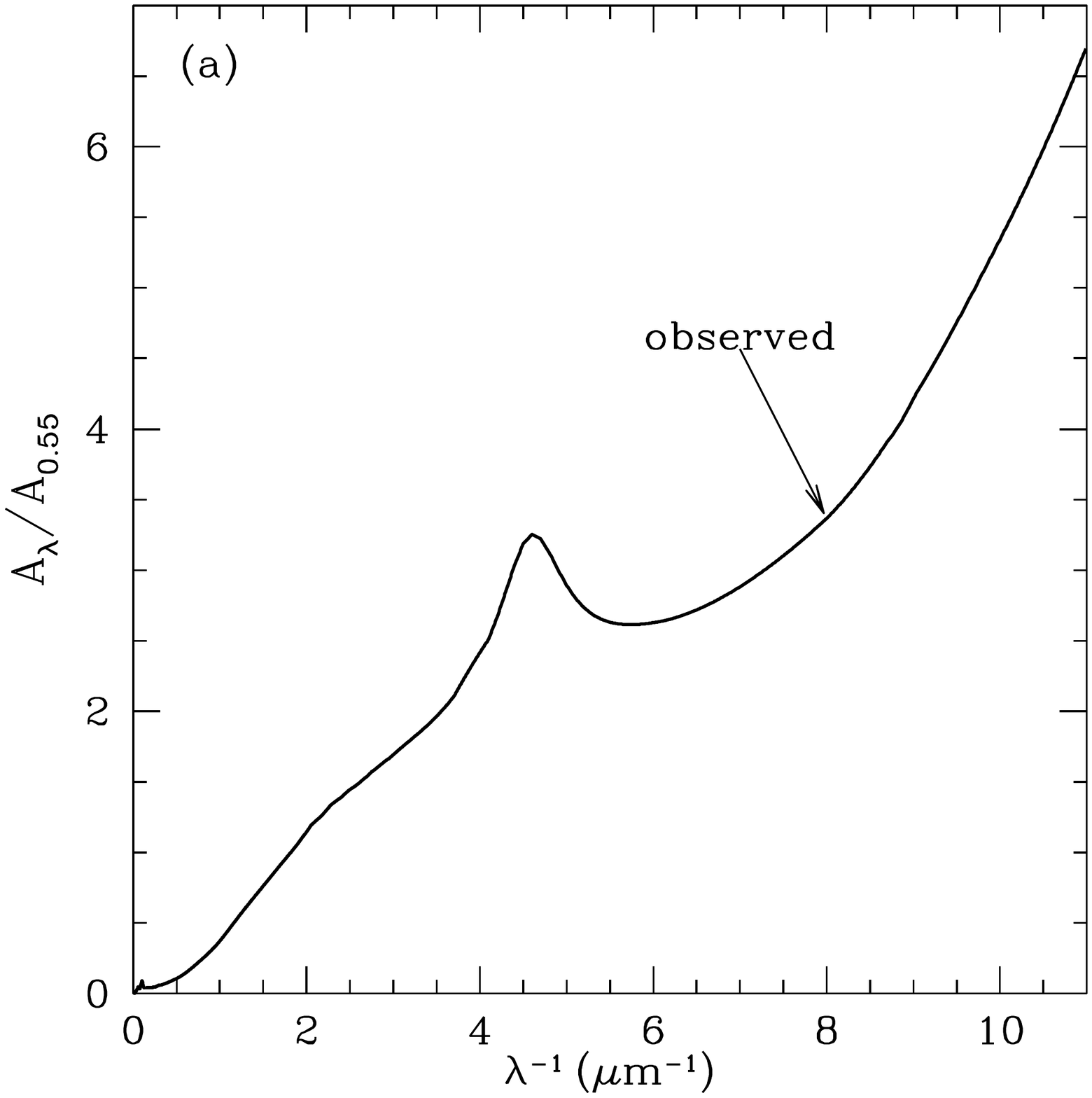}
\includegraphics[angle=0,width=8.0cm,
                 clip=true,trim=0.5cm 5.0cm 0.5cm 2.5cm]
{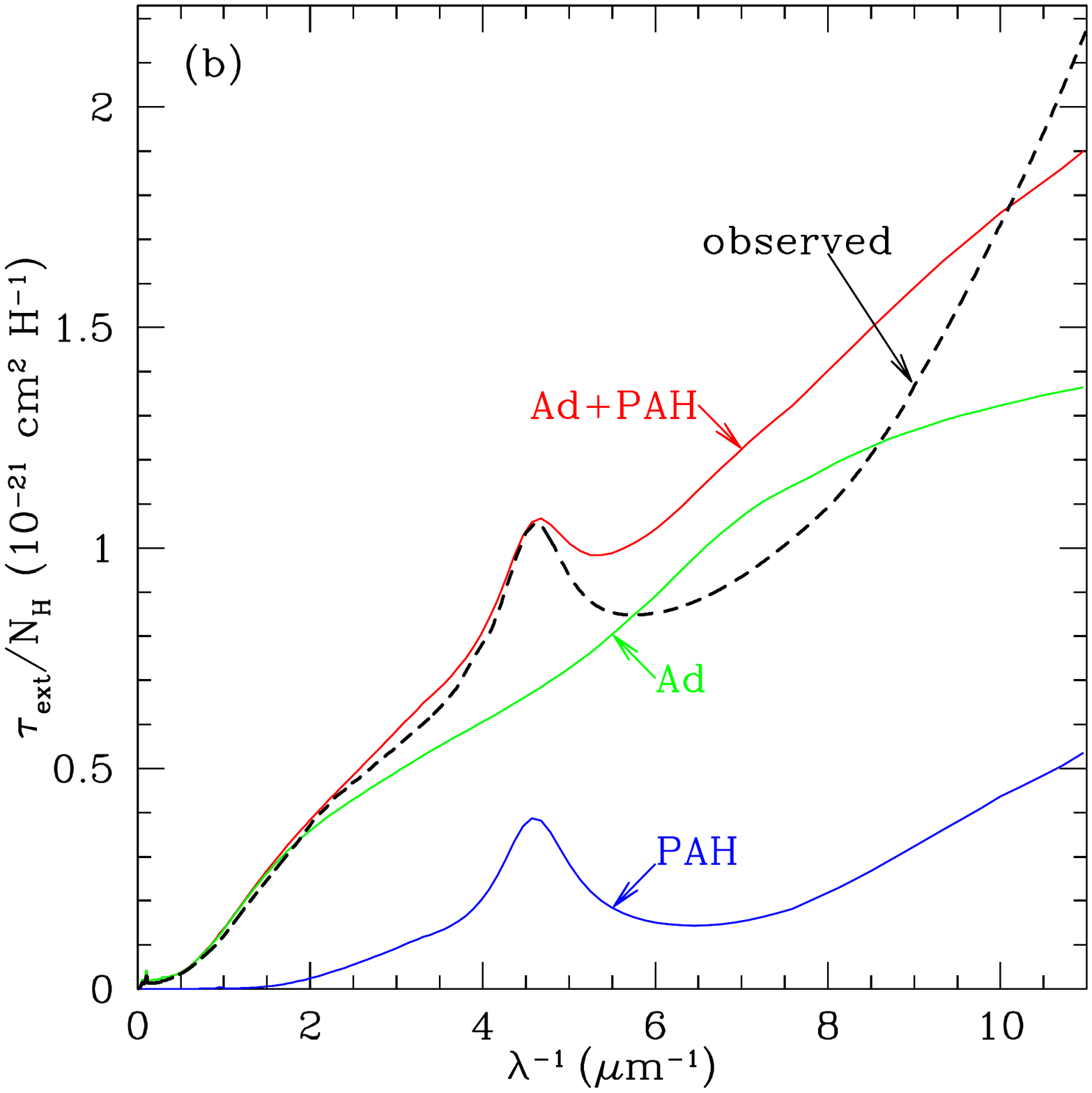}
\caption{\label{fig:ext}\footnotesize
         (a) 
         Dust extinction in the diffuse ISM used in Eq.\ (\ref{eq:unu vs x})
         \citep[from][]{Hensley+Draine_2021a}.
         (b)
         Extinction calculated for adopted size distributions
         of astrodust (Ad) and PAH 
         (Eqs.\ \ref{eq:dnda_Ad} and \ref{eq:dnda_pahs}).
         Broken curve is observed extinction (see panel (a)).
         }
\end{center}
\end{figure}

For each spectral shape, we define a reference energy density
\beq
u_{\rm ref,\star} \equiv \frac{u_{\rm mMMP}}{\gamma_\star}
\eeq
that produces the
standard amount of heating ($h=h_{\rm ref}$)
for our standard grain.
The intensity of radiation heating the dust in a region can be
estimated from the wavelength of the FIR emission peak, which
depends on the temperature (and therefore the heating rate) of the
$\sim$$0.1\micron$ grains that dominate the FIR emission.
We characterize the heating effect of a radiation field 
$u_{\star,\nu}$
by a
dimensionless intensity parameter
\beq \label{eq:U}
U \equiv 
\frac
{\int d\nu \, u_{\star,\nu} \, c \,  C_{\rm abs}^{\rm (Ad)}(\nu)}
{h_{\rm ref}}
=
\frac{u_\star}{u_{\rm ref,\star}}
=
\gamma_\star \frac{u_\star}{u_{\rm mMMP}}
\eeq 
where $u_\star$ is the energy density of the radiation.
With the definition in Equation (\ref{eq:U}), radiation fields with different
spectra but the same $U$ will heat the standard grain to the same temperature,
resulting in the same FIR emission spectrum.

The ``hardness'' of the starlight is indicated by 
the mean energy per absorbed photon for a grain X,
\beq
\langle h\nu\rangle_{\rm abs}^{\rm (X)} \equiv 
\frac{\int d\nu \, u_{\star,\nu} \, h\nu \, C_{\rm abs}^{\rm (X)}(\nu)}
{\int d\nu \, u_{\star,\nu} \, C_{\rm abs}^{\rm (X)}(\nu)}
~~~.
\eeq
Table \ref{tab:isrf} gives $\langle h\nu\rangle_{\rm abs}$ 
for our standard astrodust
grain, and also for a PAH cation.
As expected, the harder radiation fields (e.g, the $3\Myr$ old starburst)
have larger $\langle h\nu\rangle_{\rm abs}^{\rm (Ad)}$.
Because the opacity of the PAH
nanoparticles rises more rapidly in the UV, $\langle h\nu\rangle_{\rm abs}^{\rm (PAH)}$ 
can be significantly larger than $\langle h\nu\rangle_{\rm abs}^{\rm (Ad)}$
when the
radiation field is dominated by starlight from 
cool stars, as for the M31 bulge.

The starlight spectra, all scaled to give the same $U=1$ heating rate for
our standard astrodust grain, are shown in Figure \ref{fig:isrf}.
Because the $t=10\Myr$ $Z=0.0004$ 
and $t=3\Myr$ $Z=0.02$ starburst spectra are very similar for
$h\nu<13.6\eV$
(see Figure \ref{fig:isrf}a), results calculated below for the 
$t=3\Myr$ $Z=0.02$
spectrum may be taken to apply to the $t=10\Myr$ $Z=0.0004$ case.

\begin{table}
\begin{center}
\footnotesize
\caption{\label{tab:isrf} Selected Starlight Spectra}
\begin{tabular}{|cccccc|}
\hline
stellar population & ref
                   & $\gamma_\star$ 
                   & $u_{\rm ref,\star}$$^{\rm e}$
                   & $\langle h\nu\rangle_{\rm abs}^{\rm (Ad)}\,^{\rm f}$
                   & $\langle h\nu\rangle_{\rm abs}^{\rm (PAH)}\,^{\rm g}$\\
                   &&& $(10^{-13}\erg\cm^{-3})$ & $(\eV)$ & $(\eV)$ \\
\hline
$Z=0.02$, $t=3\Myr$   & a & 5.72  & $1.82$ & 6.73 & 8.64 \\
''                    & b & 5.45  & $1.92$ & 6.54 & 8.30 \\

$Z=0.0004$, $t=10\Myr$ & a & 5.73  & $1.82$ & 6.70 & 8.57 \\
$Z=0.001$, $t=10\Myr$  & b & 5.11  & $2.04$ & 6.35 & 8.42 \\

$Z=0.02$, $t=10\Myr$   & a & 4.64  & $2.25$ & 5.90 & 7.85 \\
''                     & b & 4.58  & $2.28$ & 5.84 & 7.80 \\

$Z=0.02$, $t=100\Myr$   & a & 2.94  & $3.54$ & 4.34 & 5.75 \\
''                      & b & 2.55  & $4.09$ & 3.94 & 5.69 \\

$Z=0.02$, $t=300\Myr$   & a & 1.70  & $6.13$ & 2.98 & 4.59 \\
''                      & b & 1.27  & $8.19$ & 2.38 & 4.29 \\

$Z=0.02$, $t=1\Gyr$   & a & 0.798  & $13.1$ & 1.63 & 3.04 \\
''                    & b & 0.675 & $15.5$ & 1.41 & 2.95 \\

mMMP ISRF              & c & 1    & $10.4$ & 1.93 & 4.54 \\
M31 bulge              & d & 0.580 & $18.0$ & 1.22 & 2.68 \\
\hline 
\multicolumn{6}{l}{a~BC03 \citep{Bruzual+Charlot_2003}}\\
\multicolumn{6}{l}{b~BPASS 
                   \citep{Eldridge+Stanway+Xiao+etal_2017,
                          Stanway+Eldridge_2018}}\\
\multicolumn{6}{l}{c~mMMP 
                   \citep{Mathis+Mezger+Panagia_1983,Draine_2011a}}\\
\multicolumn{6}{l}{d~\citet{Groves+Krause+Sandstrom+etal_2012}}\\
\multicolumn{6}{l}{e~Energy density corresponding to $U=1$}\\
\multicolumn{6}{l}{f~For $a=0.1\micron$ 1.6:1 oblate astrodust with porosity
$\calP=0.2$ \citep{Draine+Hensley_2021a}}\\
\multicolumn{6}{l}{g~For $N_{\rm C}=105$ PAH$^+$}
\end{tabular}
\end{center}
\end{table}


\subsection{Reddened Starlight}

In addition to studying the heating for the above-described starlight
spectra, we also consider the case of starlight incident on dust
clouds, with the radiation field within the cloud attenuated and
reddened by intervening dust.  As a representative case, we consider
dust clouds with extinction $A_V\approx 2\,{\rm mag}$.  For the
starlight radiation fields of interest, such a cloud is sufficiently
thick that, in the absence of scattering, the bulk of the incident starlight
energy will be absorbed by dust in the cloud and reradiated in the
infrared.  Grains near the cloud surface will be exposed to 
the unreddened spectrum;
grains deeper in the cloud will be heated by a weaker and redder
radiation field, and thus will be cooler.

We neglect scattering, and assume unidirectional radiation incident
normally on one cloud surface.
We take the extinction to have the wavelength dependence 
$A_\lambda/A_V$ adopted by
\citet{Hensley+Draine_2021a}
(based largely on studies by \citet{Schlafly+Meisner+Stutz+etal_2016}
and \citet{Fitzpatrick+Massa+Gordon+etal_2019})
for the diffuse interstellar medium
(see Figure \ref{fig:ext}a).  
The starlight energy density per unit
wavelength at a point within
the cloud is taken to be
\beq \label{eq:unu vs x}
u_\lambda(x) = u_\lambda(0)10^{-0.4 (A_\lambda/A_V) A_V x/L}
~~~,
\eeq
where $0\leq x\leq L$ is the distance from the slab surface,
$L$ is the thickness of the cloud, and $A_V$ is the
extinction at $V$ through the cloud.  
Eq.\ (\ref{eq:unu vs x})
treats scattering like absorption for estimating the attenuation of
the radiation field.  For normally-incident radiation, this will overestimate
the attenuation.  On the other hand, if some or all of the incoming radiation
is incident at appreciable angles relative to the normal, the
attenuation law of Eq.\ (\ref{eq:unu vs x}) will tend to underestimate
the attenuation within the cloud; with these two errors tending to partially compensate,
we use Eq.\ (\ref{eq:unu vs x}) 
to estimate the starlight intensity within slabs
with total extinction $A_V=2\,$mag. 

\section{\label{sec:PAH physics}
         PAH Physics}

We idealize the PAH population as consisting of either neutral or ionized PAHs,
with size-dependent H:C ratio as assumed by 
\citet[][hereafter DL07]{Draine+Li_2007}. 
We take the PAH nanoparticles to consist of hydrocarbon material with a
carbon mass density $\rho_{\rm C}=2.0\gm\cm^{-3}$, with 
\beq
N_{\rm C} = 418 \left(\frac{a}{10\Angstrom}\right)^3
\eeq
carbon atoms in a particle of nominal radius $a$.\footnote{%
   Our assumed PAH carbon mass density $\rho_{\rm C}=2.0\gm\cm^{-3}$ 
   is more appropriate
   for real hydrocarbon solids than the crystalline graphite
   value $2.24\gm\cm^{-3}$ used by DL07.
   E.g., $\rho\approx2.0\gm\cm^{-3}$ for evaporated amorphous carbon
   \citep{Fink+Muller-Heinzerling+Pfluger+etal_1983}.}
We adopt the 
PAH cross sections $C_{\rm abs}^{\rm (PAH)}(a,\lambda)$
from DL07, including both optical-UV
continuum and a set of resonance features, including a strong UV absorption feature at $2175\Angstrom$, and a number of infrared features corresponding
to PAH vibrational modes.
Because the observed $2-5\micron$ infrared emission from the ISM
appears to include a low-level continuum underlying the PAH features
\citep{Sellgren+Werner+Dinerstein_1983, Lu+Helou+Werner+Dinerstein+Dale_2003,Helou+Roussel+Appleton+etal_2004,Xie+Ho+Li+Shangguan_2018b}, 
we follow DL07 and take the absorption cross section to be
\beq \label{eq:Cabs}
C_{\rm abs}^{\rm PAH}(a,\lambda) = 
(1-\xi_{\rm gra})N_{\rm C}\Gamma(a,\lambda) + 
\xi_{\rm gra} C_{\rm abs}^{\rm (gra)}(a,\lambda)
~~~, 
\eeq
with
\beqa
\xi_{\rm gra} &~=~& 0.01 \hspace*{4.8cm} a < 50\Angstrom
\\ \label{eq:xi}
&=& 0.01 + 0.99\left[1-\left(\frac{50\Angstrom}{a}\right)^3\right]
~~~~~ a > 50\Angstrom
~~~.
\eeqa
where $\Gamma(a,\lambda)$ 
is the ``pure PAH'' absorption cross section per C atom, and
$C_{\rm abs}^{\rm (gra)}(a,\lambda)$ is the absorption cross section for
graphite spheres
(see Appendix \ref{app:Cabs for PAHs} for details).
Eq.\ (\ref{eq:Cabs}-\ref{eq:xi}) 
are entirely {\it ad hoc}, to provide a small amount of
``graphitic''
continuum opacity so that hot PAHs can provide the continuum emission that has
been observed.
DL07 used the optical properties of graphite estimated by \citet{Draine_2003a};
here we instead 
use the dielectric function
estimated by \citet{Draine_2016} for
polycrystalline graphite using 
Maxwell Garnett effective medium theory.\footnote{%
   The matrix is taken to have dielectric function $\epsilon=\epsilon_\perp$, 
   and the inclusions are taken
   to have $\epsilon=\epsilon_\parallel$, 
   where $\epsilon_\perp$ and $\epsilon_\parallel$ are the 
   eigenvalues of the dielectric tensor of crystalline graphite
   corresponding to
   $\bE\perp c$ and $\bE\parallel c$.}

We assume that the energy $h\nu$ of an absorbed photon is fully converted to
vibrational energy (i.e., ``heat'') in the PAH -- 
we neglect the energy lost in the
form of photoelectrons as well as possible 
fluorescent emission of optical photons.
To calculate the temperature fluctuations, we use the heat capacity
model from \citet{Draine+Li_2001}, based on realistic size-dependent
vibrational mode spectra for PAHs.

The interstellar PAH population likely includes a very large number of
distinct PAH-like particles, in multiple charge states, including
anions, neutrals, and cations.  Some may be partially dehydrogenated, while
others may be fully or even super-hydrogenated.

\citet{Maragkoudakis+Peeters+Ricca_2020} have modeled the
emission spectra from 308 specific PAHs, with sizes ranging from
$N_{\rm C}=22$ to $N_{\rm C}=216$, using band strengths 
from density functional theory calculations
in the NASA Ames PAH IR database 
\citep{Bauschlicher+Boersma+Ricca+etal_2010,
       Boersma+Bauschlicher+Ricca+etal_2014,
       Bauschlicher+Ricca+Boersma+Allamandola_2018}.
Their approach demonstrates trends with both size and ionization, and
also the considerable 
variation in spectra for different PAHs of similar size and ionization.

By adopting the DL07 model here, we idealize the PAHs as being characterized by
a single size parameter ($a$ or $N_{\rm C}$) and a binary charge
state parameter (neutral or ionized).  This is an extreme simplification, but
it allows us to investigate the effects of changes in the illuminating starlight
spectrum and intensity, changes in the overall PAH size distribution,
and changes in the ionized fraction.

\section{\label{sec:dPdT}
         Temperature Distribution Functions}

\begin{figure}
\begin{center}
\includegraphics[angle=0,width=8.5cm,
                 clip=true,trim=0.5cm 5.0cm 0.5cm 2.5cm]
{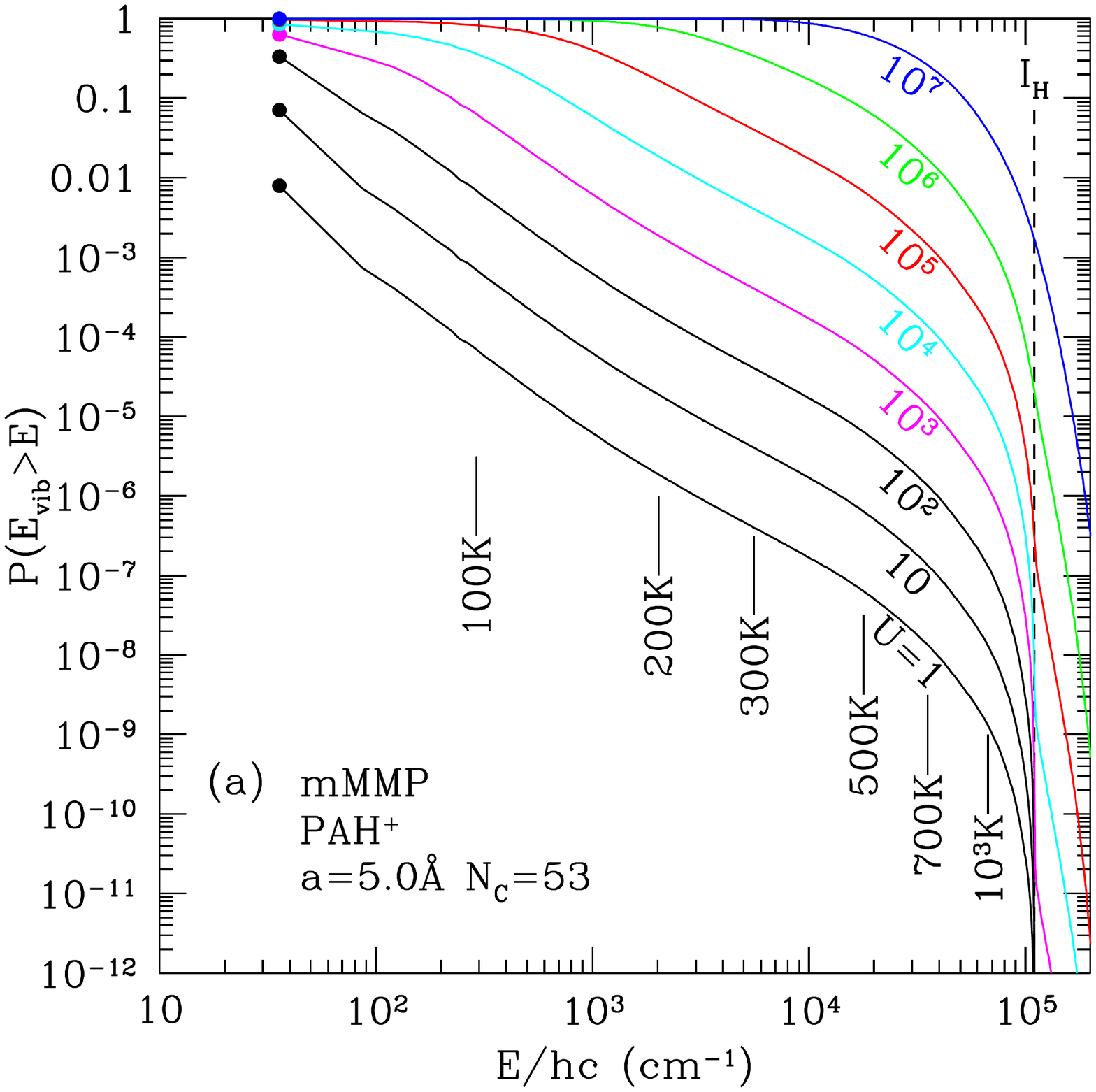}
\includegraphics[angle=0,width=8.5cm,
                 clip=true,trim=0.5cm 5.0cm 0.5cm 2.5cm]
{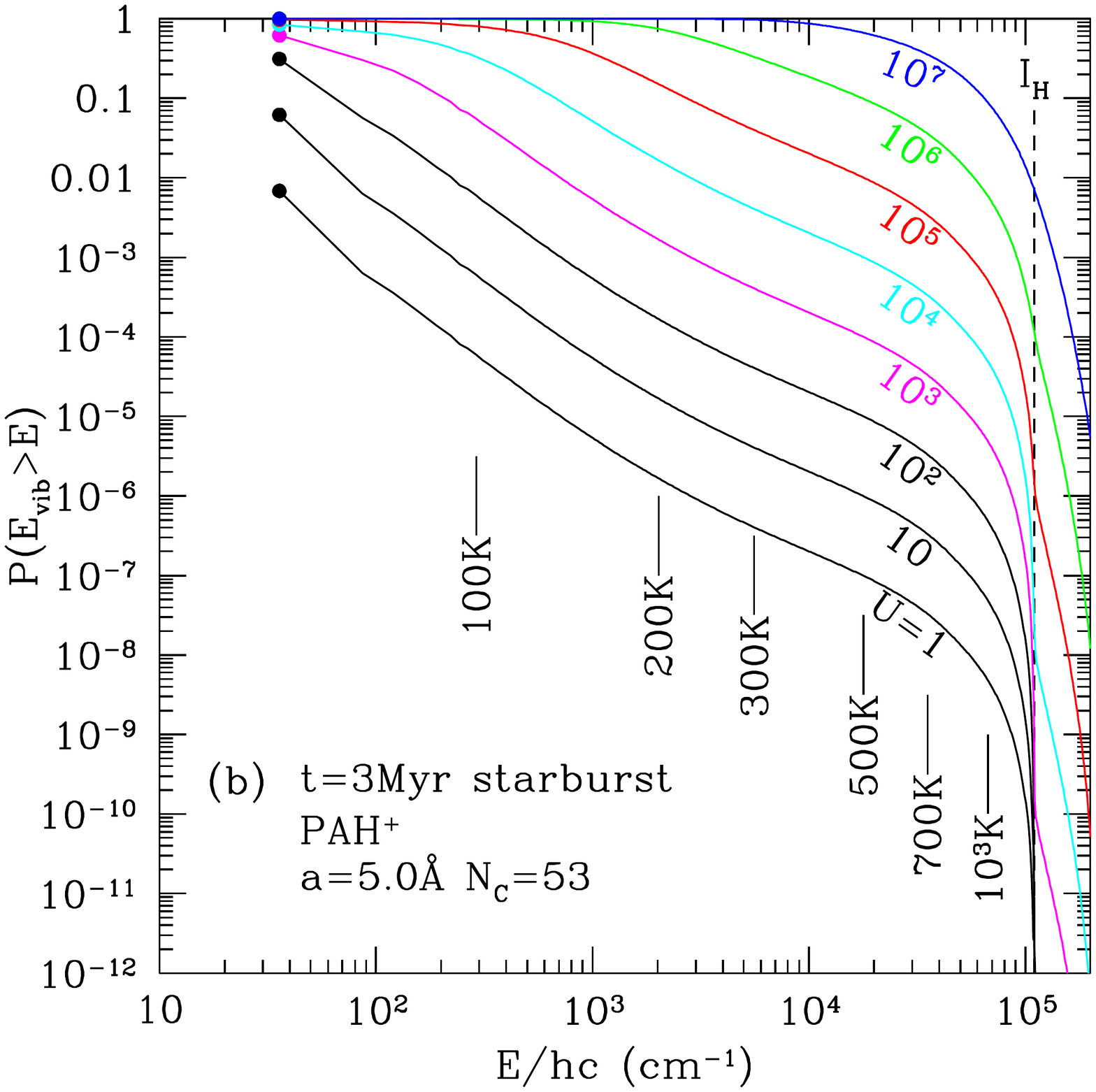}
\caption{\label{fig:pahi_dpdt}\footnotesize
         Probability $P(E_{\rm vib}>E)$ of having vibrational
         energy greater than $E$,
         for $a=5\Angstrom$ PAH$^+$ in (a) mMMP interstellar radiation
         field (ISRF),
         (b) unreddened spectrum from a 3\,Myr old starburst, cut off at
         the Lyman limit.  Results are labeled by radiation strength
         parameter $U$.
         Vibrational 
         energies corresponding to temperatures $T=100,200,300,500,700,$
         and $10^3\K$ are shown.
         For each case, 
         the dot indicates the location of the first energy bin above the
         vibrational ground state.  For $U=1$, a PAH$^+$ 
         nanoparticle with
         $N_{\rm C}=53$ C atoms
         spends $\sim$99\% of the time 
         in the vibrational ground
         state, with $P(E_{\rm vib}>0)\approx0.01$.
         The vertical dashed line shows the energy 
         $I_{\rm H}=13.6\eV$ of
         the highest-energy photons present in the illuminating starlight.
         }
\end{center}
\end{figure}
The compositions considered are PAH neutrals, PAH cations, and
``astrodust'' material.  
For each radiation field (spectrum and intensity), we find the
steady-state energy distribution function $(dP/dE)_{j,a}$
for grains of composition $j$ and size $a$.
This is accomplished
following the methods of \citet{Guhathakurta+Draine_1989}
and \citet{Draine+Li_2001}: choose an energy range 
$[E_{\rm min},E_{\rm max}]$ 
such that the grain is very unlikely to 
have $E<E_{\rm min}$ or $E>E_{\rm max}$,
divide this into $N=499$ energy bins (non-uniformly spaced), 
calculate the transition matrix $R_{\ell k}$
giving the probability per unit time
for a grain in a given bin $k$ to make a transition to a different bin 
$\ell$ as the result of either photon absorption or photon emission, and
then solve the system of equations to find 
the steady-state probability $P_k$
of finding the grain in energy bin $k$.
With each internal energy $E_k$ we associate a temperature 
$T(E_k)$,
defined to be the temperature such that for a thermal distribution
the expectation value of the energy
would be $E_k$.
\citet{Draine+Li_2001} showed that the instantaneous emission spectra for
vibrationally excited PAHs can be adequately approximated by thermal emission.

$E_{\rm min}$ and $E_{\rm max}$ are chosen adaptively according to the grain
size and radiation field; for small grains and weak radiation field, we take
$E_{\rm min}=0$ and $E_{\rm max}=13.6\eV$, because 
(for $U\ltsim 10^3$) the PAH molecule spends most of its
time at or close to the vibrational ground state, 
and single-photon heating by a starlight spectrum cut off at 
$13.6\eV$ will not raise the grain energy by more than $13.6\eV$.
However, for large grains, or intense radiation fields, 
the vibrational energy $E_{\rm vib}$ can
exceed
$13.6\eV$.
We adjust $E_{\rm max}$ so that $P(E>E_{\rm max})\ltsim 10^{-12}$.

Figure \ref{fig:pahi_dpdt} shows cumulative energy distribution functions for
PAH$^+$ with $N_{\rm C}=53$ in radiation fields with intensity parameters
ranging from $U=1$ to $U=10^7$, for the mMMP diffuse starlight spectrum
(Fig.\ \ref{fig:pahi_dpdt}a) and for the spectrum of a 3\,Myr old starburst
(Fig.\ \ref{fig:pahi_dpdt}b).  For $U>10^4$ we begin to see significant
populations 
$P > 10^{-8}$ at $E/hc>110,000\cm^{-1}$ ($E>13.6\eV$), 
resulting from absorption of $h\nu\approx13\eV$ photons by grains that
have not yet cooled back to the ground state
following a previous photon absorption.

\section{\label{sec:single-PAH spectra}
          Emission Spectra for Individual PAHs}

\subsection{Optically Thin Regions}

\begin{figure}
\begin{center}
\includegraphics[angle=0,width=8.5cm,
                 clip=true,trim=0.5cm 5.0cm 0.5cm 2.5cm]
{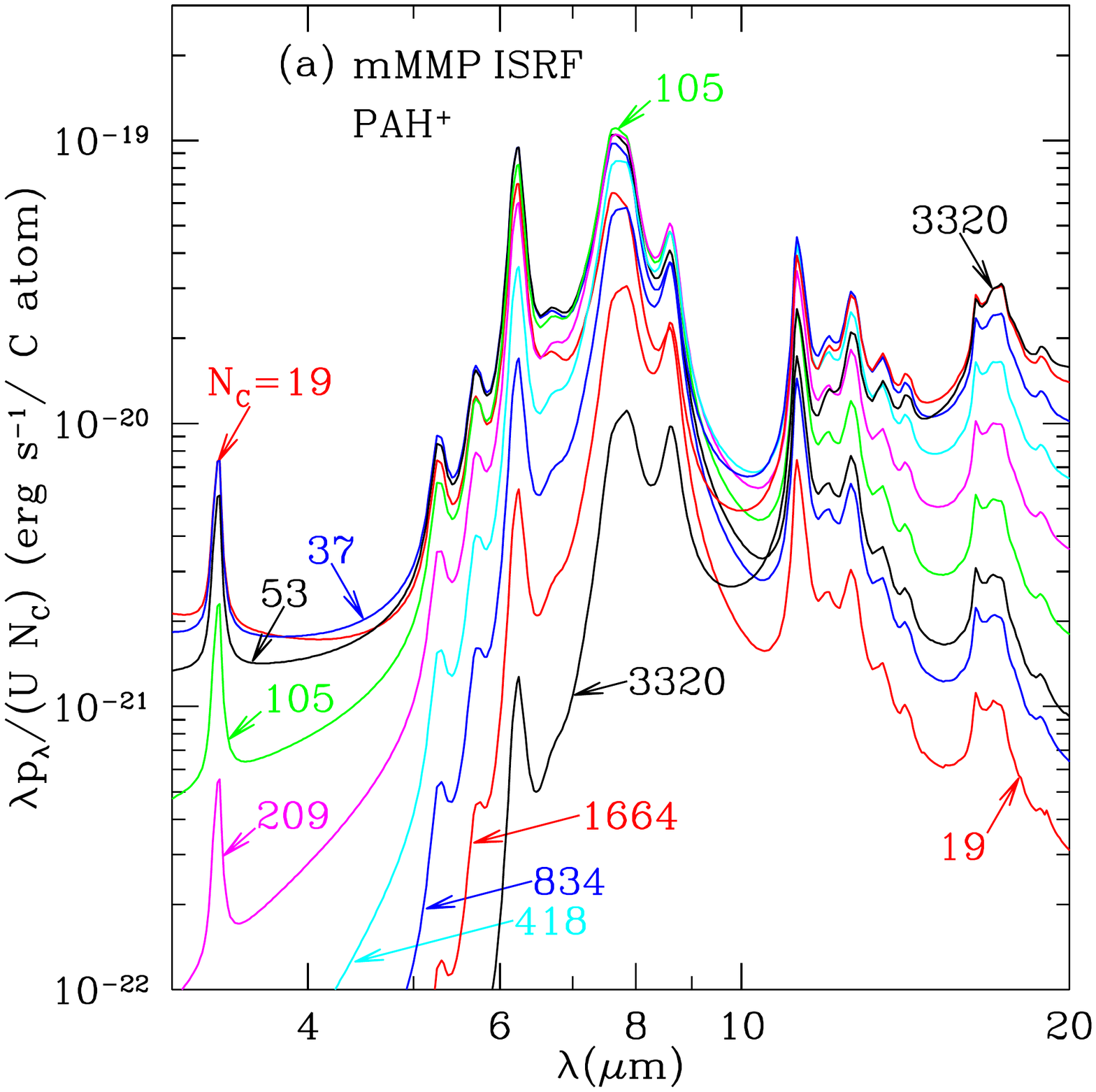}
\includegraphics[angle=0,width=8.5cm,
                 clip=true,trim=0.5cm 5.0cm 0.5cm 2.5cm]
{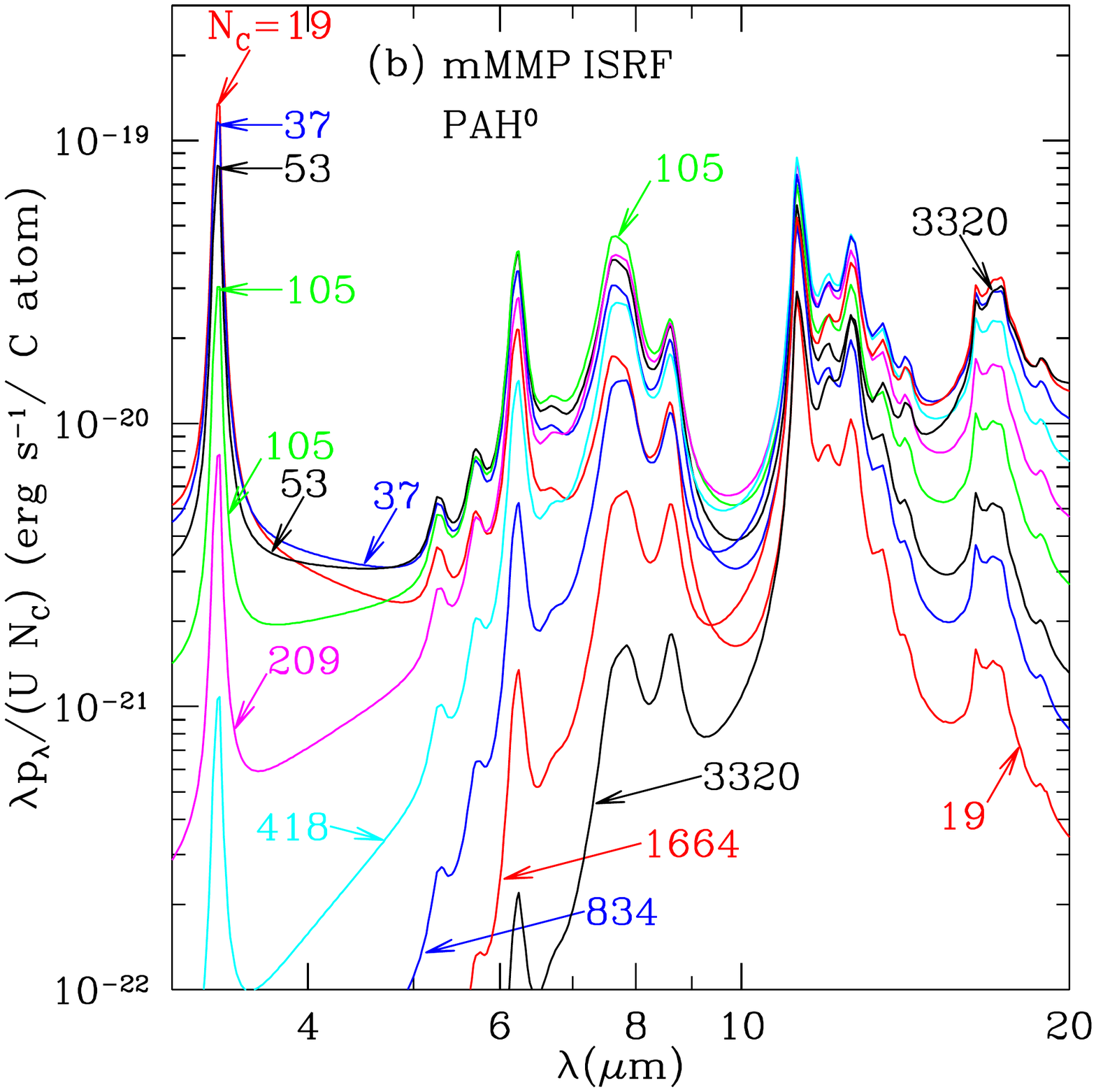}
\caption{\label{fig:pah_mmp}\footnotesize
         Emission per C atom for (a) 
         PAH ions and (b) neutrals of various sizes,
         when heated by the mMMP ISRF.
         The spectra shown are calculated 
         for $U=1$, but $p_\lambda/U$ 
         is nearly independent of $U$ for for $\lambda < 20\micron$ and
         $U\ltsim10^2$.
         }
\end{center}
\end{figure}

\begin{figure}
\begin{center}
\includegraphics[angle=0,width=8.5cm,
                 clip=true,trim=0.5cm 5.0cm 0.5cm 2.5cm]
{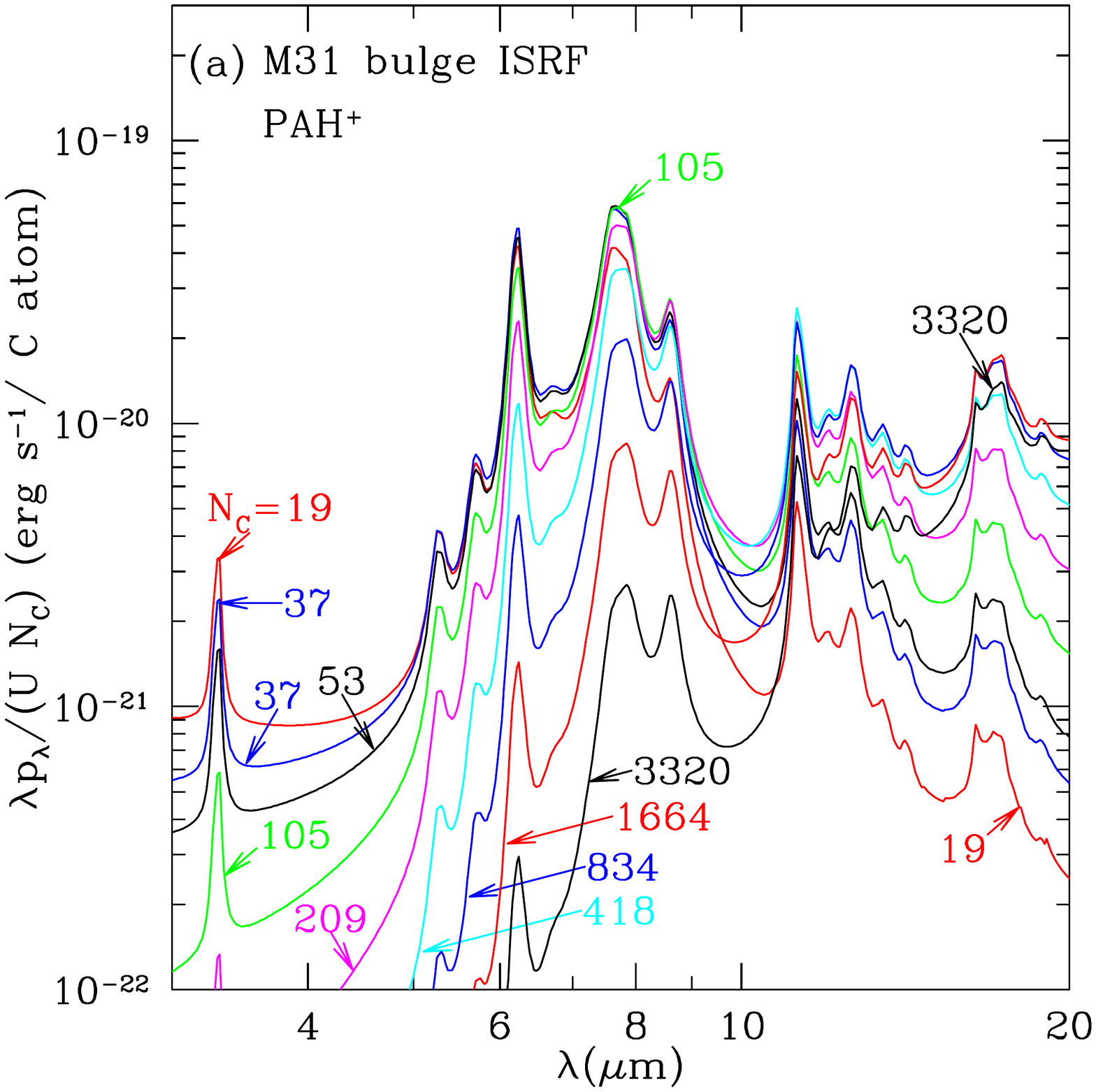}
\includegraphics[angle=0,width=8.5cm,
                 clip=true,trim=0.5cm 5.0cm 0.5cm 2.5cm]
{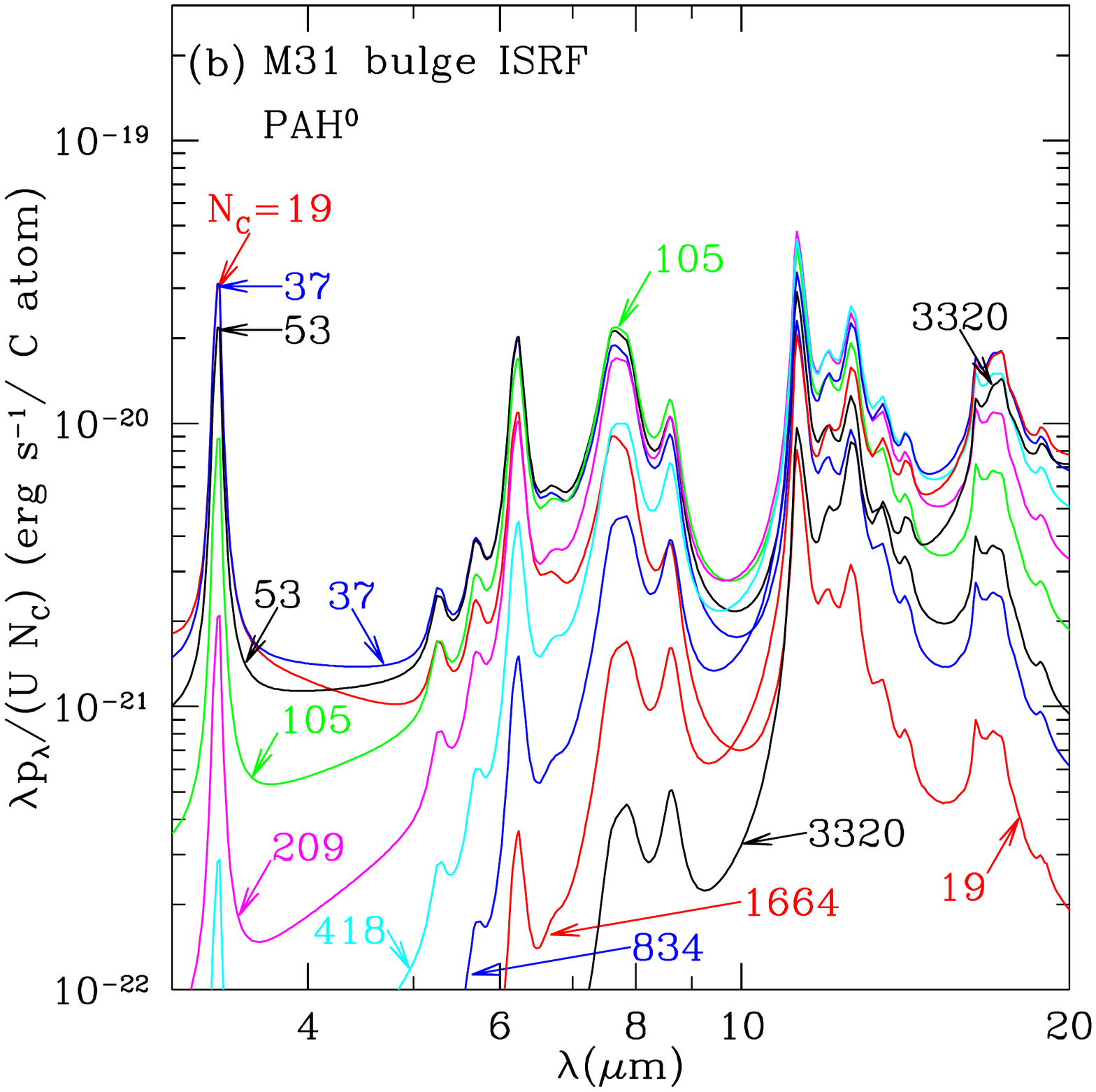}
\caption{\label{fig:pah_m31bulge}\footnotesize
         Same as Figure \ref{fig:pah_mmp}, but for heating
         by starlight from the M31 bulge population.
         }
\end{center}
\end{figure}

For optically thin dust, we find the temperature
distribution function $(dP/dT)_{j,a}$ and calculate the time-averaged
power per unit wavelength $p_\lambda$ radiated by one grain:
\beq
p_\lambda^{(j)}(a) = 4\pi C_{\rm abs}^{(j)}(a,\lambda) \int dT 
\left(\frac{dP}{dT}\right)_{j,a} B_\lambda(T)
~~~,
\eeq
where $B_\lambda(T)$ is the usual blackbody function.
Figure \ref{fig:pah_mmp} shows emission per C atom from ionized and neutral
PAHs of various sizes, when illuminated by the mMMP radiation field,
and Figure \ref{fig:pah_m31bulge} shows the emission per grain if heated by
starlight from the M31 bulge population.

Absorption of one photon can heat small PAHs to high enough
temperatures that they can efficiently radiate in the shortest wavelength
emission features 
(e.g., $3.3\micron$, $5.27\micron$, $5.7\micron$, $6.2\micron$) 
but larger PAHs radiate
primarily in longer-wavelength features
\citep{Schutte+Tielens+Allamandola_1993}.  This is evident in
Figures \ref{fig:pah_mmp} and \ref{fig:pah_m31bulge}.

The DL07 PAH opacity model for cations
includes features at $1.26\micron$ and
$1.905\micron$ as recommended by \citet{Mattioda+Allamandola+Hudgins_2005}.
According to our models, 
the smaller PAH cations do emit a modest fraction of their energy in
these features.  These emission features appear
unlikely to be observable on a galactic scale in the
presence of stellar continuum at these wavelengths, but
might be observable in reflection nebulae with JWST.

\begin{figure}
\begin{center}
\includegraphics[angle=0,width=8.5cm,
                 clip=true,trim=0.5cm 5.0cm 0.5cm 2.5cm]
{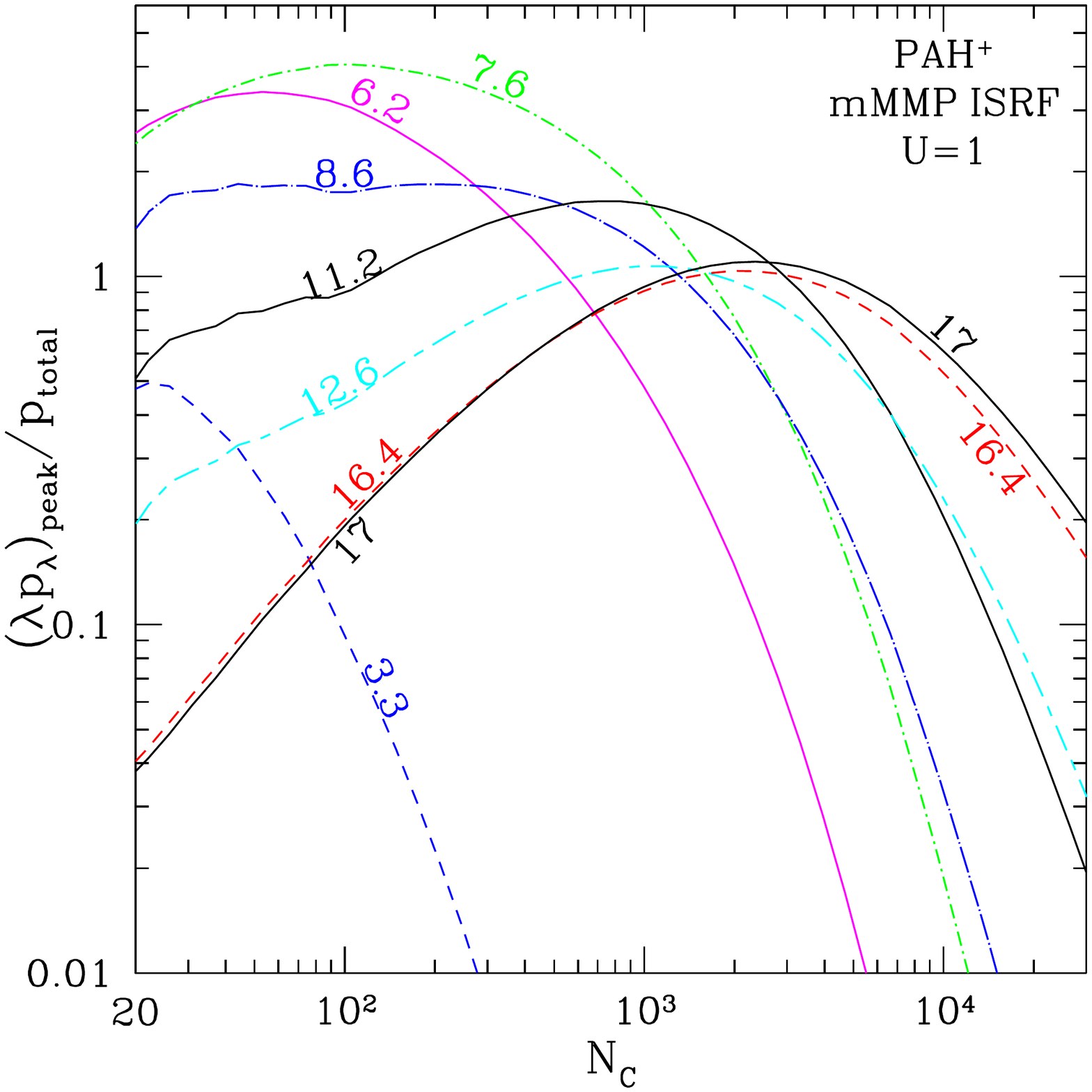}
\includegraphics[angle=0,width=8.5cm,
                 clip=true,trim=0.5cm 5.0cm 0.5cm 2.5cm]
{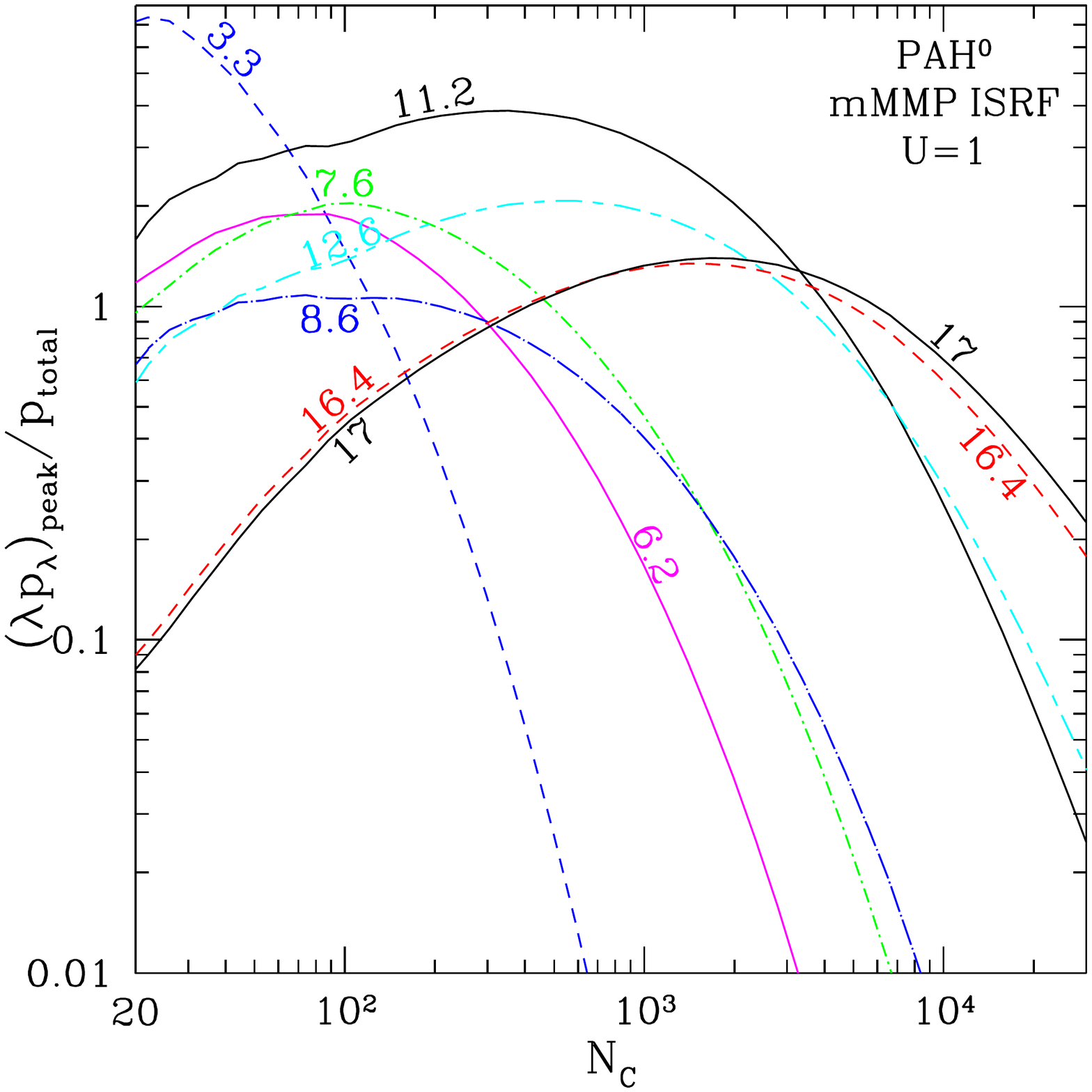}
\caption{\label{fig:conversion_efficiency}\footnotesize
         $\lambda p_\lambda$ evaluated at various PAH
         emission peaks $\lambda_{\rm peak}$,
         relative to the
         total radiated power $p_{\rm tot}=\int p_\lambda d\lambda$,
         as a function of the number $N_C$ of carbon atoms in the PAH,
         for PAH$^+$ cations (left) and PAH$^0$ neutrals (right)
         heated by the mMMP starlight spectrum with $U=1$.
         Curves are labelled by $\lambda_{\rm peak} (\micron)$.
         }
\end{center}
\end{figure}
To better compare the relative efficiency of PAHs of different sizes for
conversion of
absorbed starlight energy into PAH emission features, Figure
\ref{fig:conversion_efficiency} shows how the strengths of various
emission peaks depend on the PAH size by plotting
$(\lambda p_\lambda)_{\rm peak}/p_{\rm total}$ versus $N_{\rm C}$, where
$p_{\rm total}=\int d\lambda \, p_\lambda$ 
is the total time-averaged power radiated by the PAH.
Figure \ref{fig:conversion_efficiency} is similar
to Figure 6 of \citet{Draine+Li_2007}, differing only because of 
(1) differences
in treatment of the small ``continuum'' opacity 
[see Eq.\ (\ref{eq:Cabs}-\ref{eq:xi})],
(2) a small change in the band strengths adopted for features
at 14.19$\micron$ and the $17\micron$ complex 
(see Appendix \ref{app:Cabs for PAHs}), and (3)
replacement of the \citet{Mathis+Mezger+Panagia_1983}
spectrum used by DL07
by the mMMP spectrum used here.
We see that only the smallest PAH sizes ($N_{\rm C}\ltsim 100$)
radiate appreciably in the
$3.3\micron$ feature.

\citet{Schutte+Tielens+Allamandola_1993} found that PAHs with
$10^2$ -- $10^5$ C atoms could radiate in the IRAS 25$\micron$
photometric band.  Figure \ref{fig:conversion_efficiency} shows that
the $17\micron$ feature is efficiently
radiated by large PAHs, with $N_{\rm C}\gtsim 10^3$.
The strong $7.7\micron$ feature
is efficiently radiated by PAHs with
$N_{\rm C}\approx 10^2$.  This strong feature plays a large role in
observational studies of PAHs; because PAHs with $N_{\rm C}>10^3$ 
are relatively
inefficient at radiating in this feature, DL07 defined the PAH abundance
parameter $\qpah$ 
to be the ratio of the mass in PAHs with $N_{\rm C}<10^3$
to the total mass in dust.

\begin{figure}
\begin{center}
\includegraphics[angle=0,width=8.5cm,
                 clip=true,trim=0.5cm 5.0cm 0.5cm 2.5cm]
{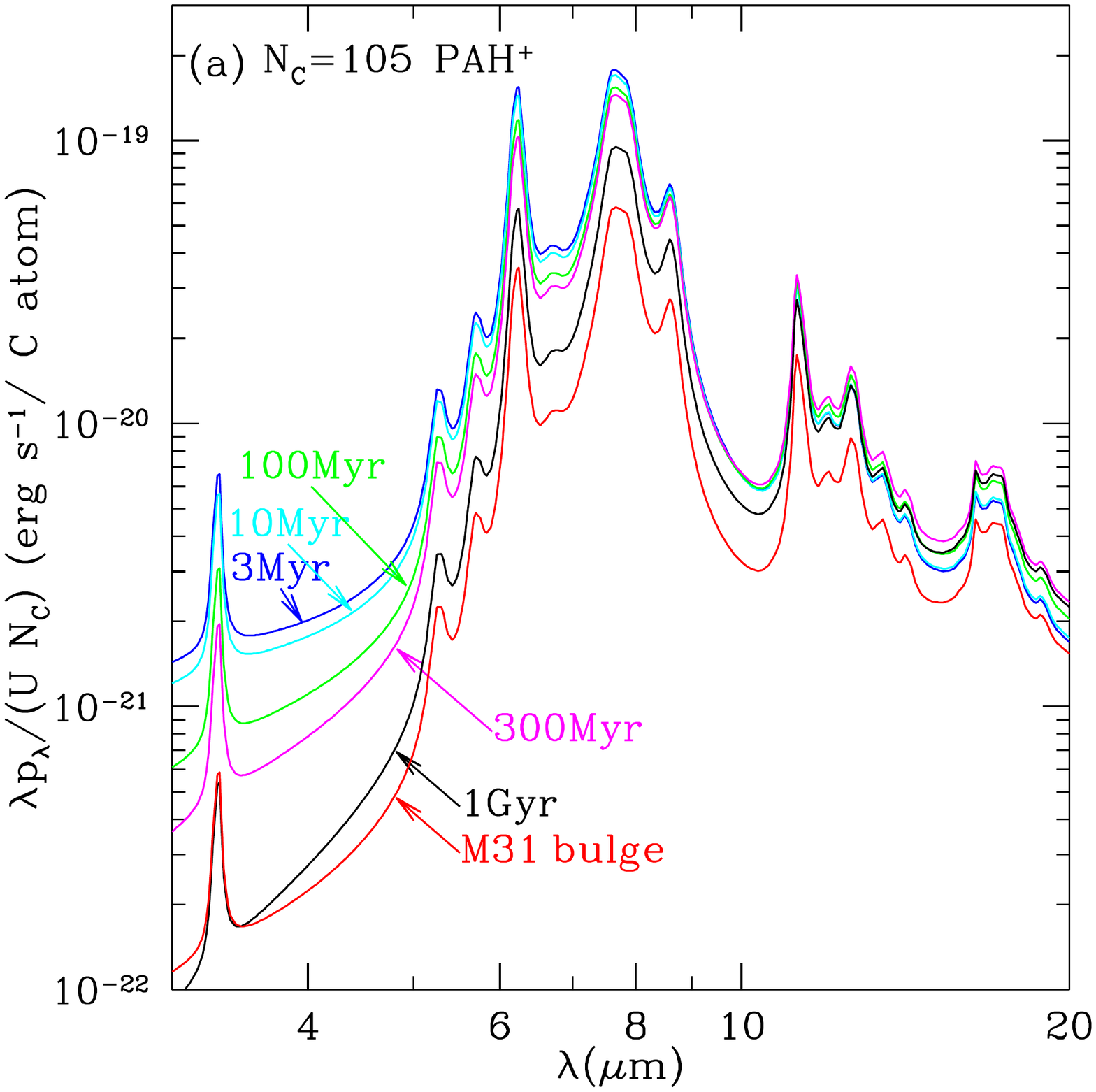}
\includegraphics[angle=0,width=8.5cm,
                 clip=true,trim=0.5cm 5.0cm 0.5cm 2.5cm]
{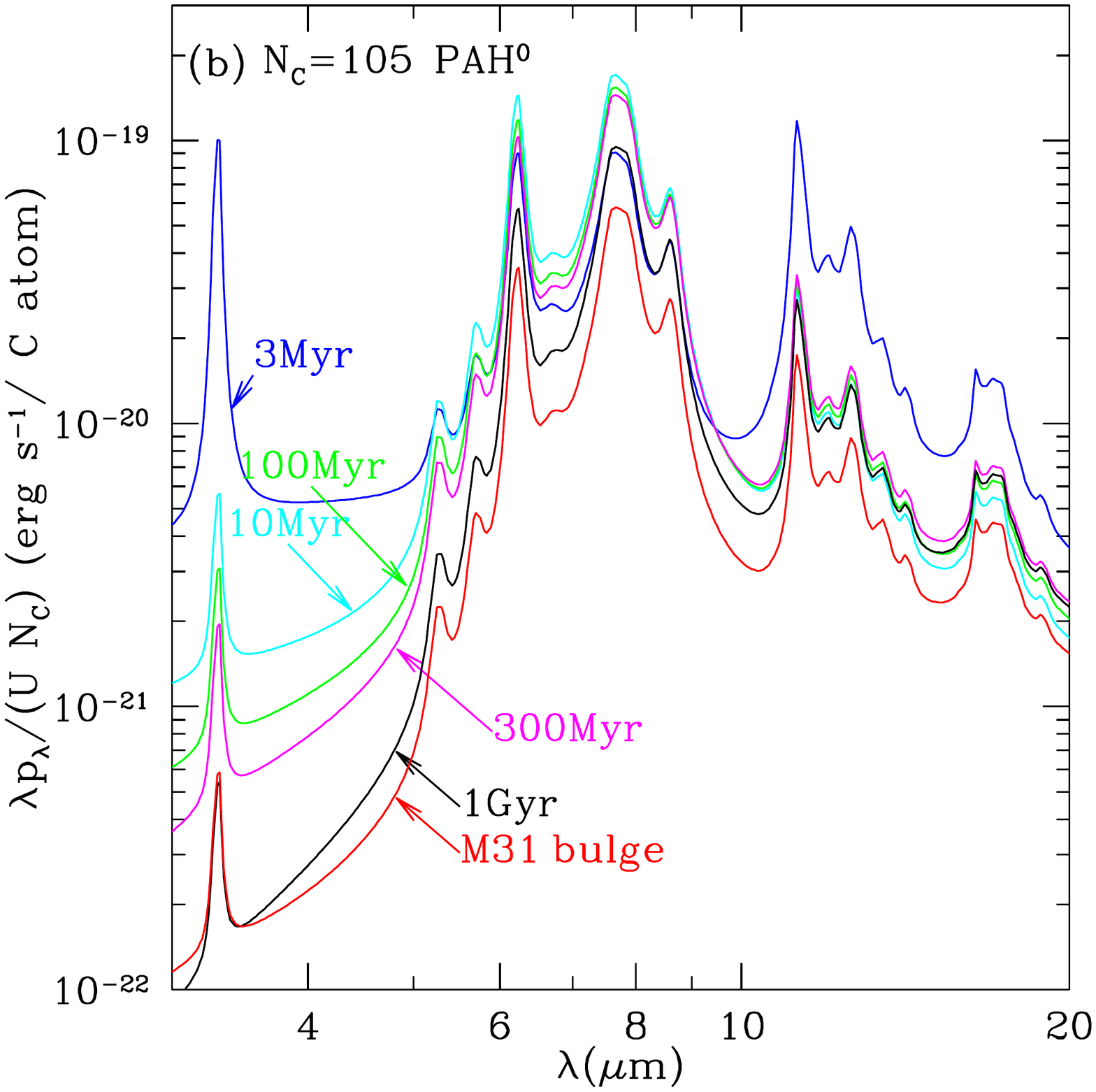}
\caption{\label{fig:pah,various rf}\footnotesize
         Emission per C atom for $N_{\rm C}=105$ PAH$^0$ and PAH$^+$
         when heated by 
         radiation with the
         spectrum of the M31 bulge, and single age $Z=0.02$ 
         stellar population models
         from BC03.
         }
\end{center}
\end{figure}

The time-averaged spectrum emitted by a PAH depends on the spectrum of the
starlight responsible for heating the PAH. 
Figure \ref{fig:pah,various rf} shows the time-averaged emission from
PAH neutrals and cations with $N_{\rm C}=105$ 
C atoms illuminated by different
radiation fields, all with heating parameter $U=1$.
For this particular PAH size, 
the ratio $I(6.2\micron)/I(7.7\micron)$ is approximately
the same for all of these radiation fields, but the ratio
$I(3.3\micron)/I(7.7\micron)$ varies dramatically between heating by
 a young starburst spectrum and heating by an old stellar population
(the M31 bulge spectrum).  Thus interpretation of PAH emission from
galaxies depends
on the spectrum of the starlight heating the dust.

\begin{figure}
\begin{center}
\includegraphics[angle=0,width=8.5cm,
                 clip=true,trim=0.5cm 5.0cm 0.5cm 2.5cm]
{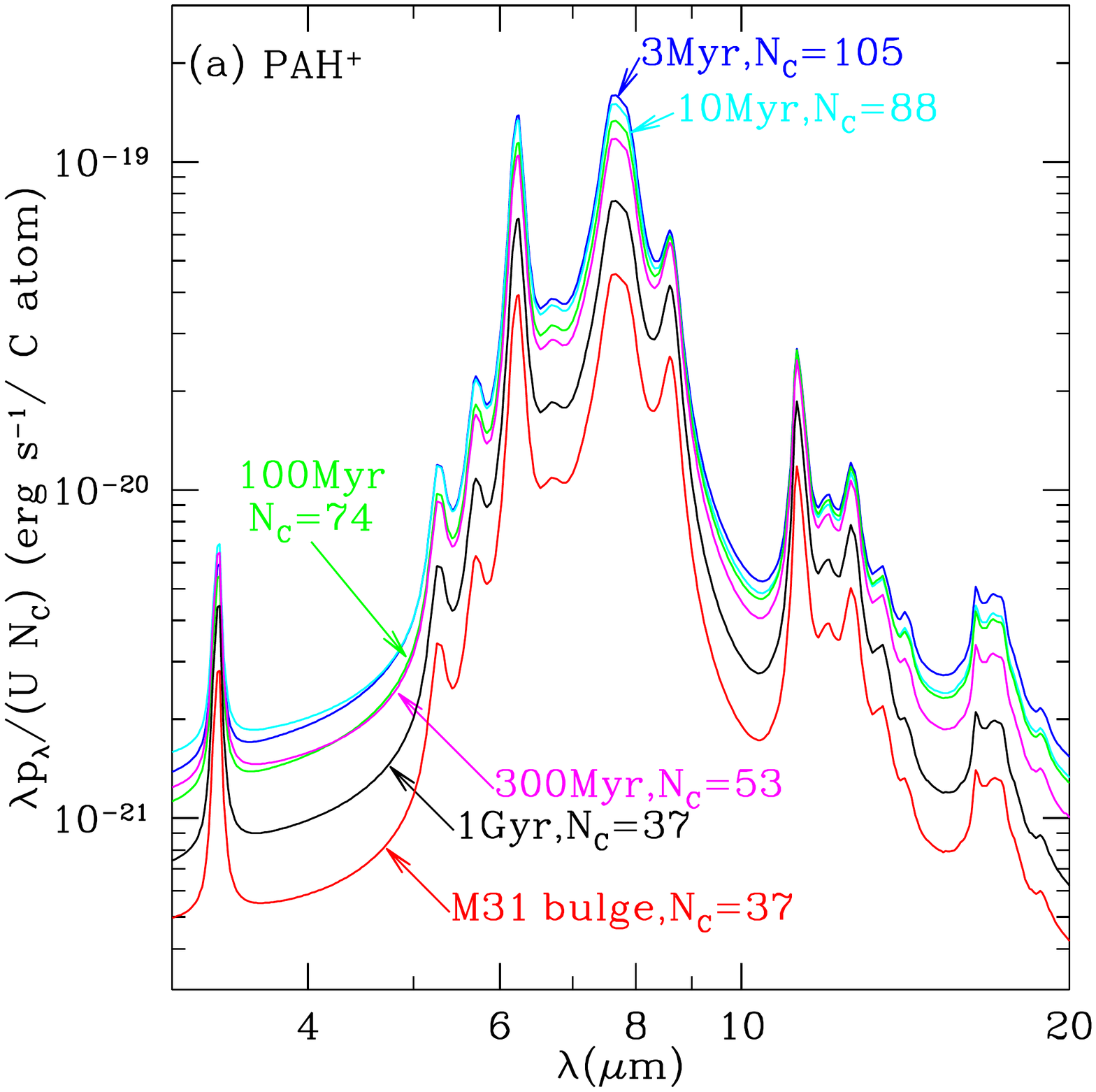}
\includegraphics[angle=0,width=8.5cm,
                 clip=true,trim=0.5cm 5.0cm 0.5cm 2.5cm]
{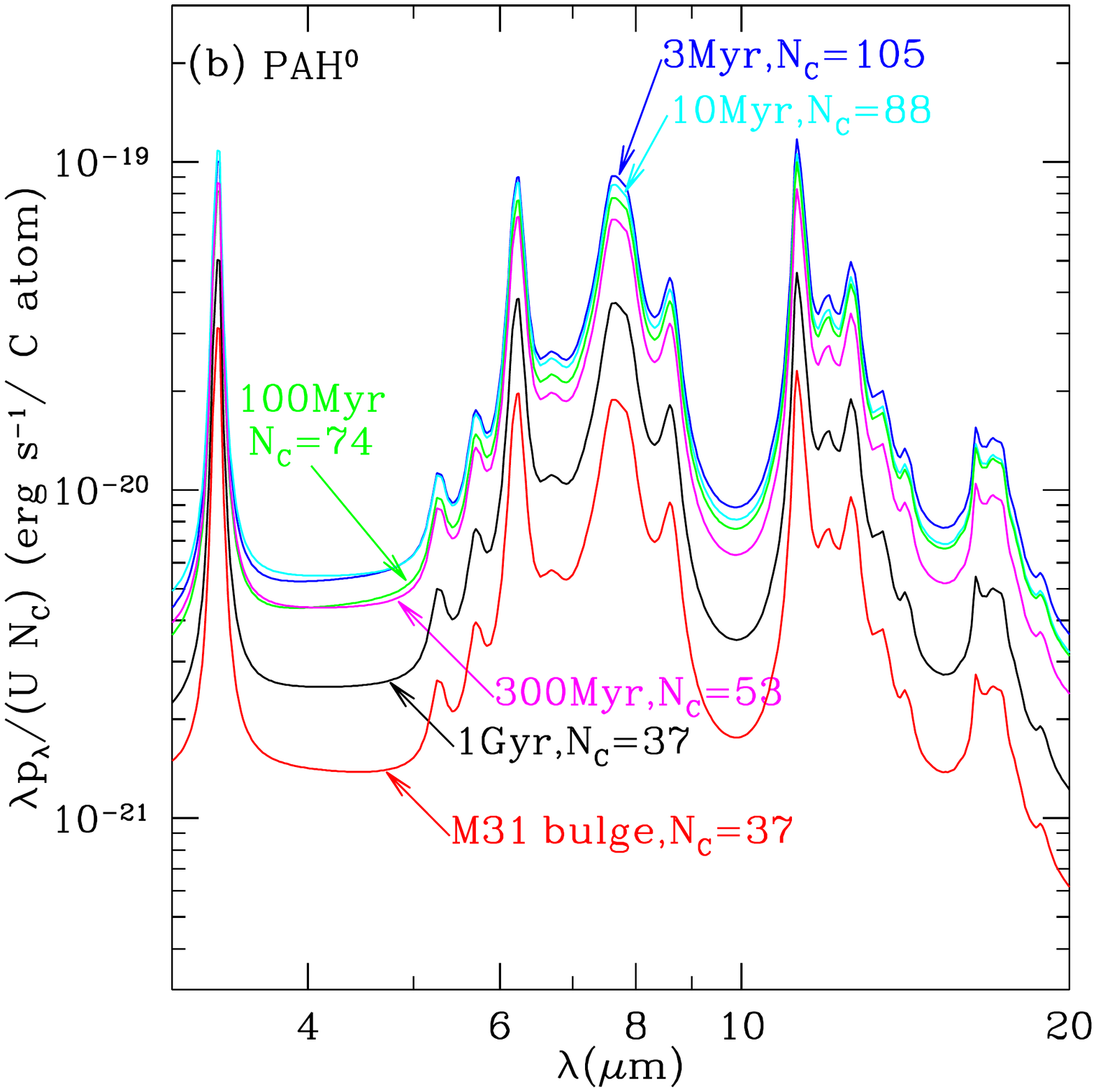}
\caption{\label{fig:pah,various rf, sizes tuned}\footnotesize
         Emission per C atom for PAH$^0$ and PAH$^+$
         when heated by various radiation fields, for sizes giving
         $\langle h\nu\rangle_{\rm abs}/N_{\rm C} \approx 0.08\eV$, 
         where $\langle h\nu\rangle_{\rm abs}$
         is the mean absorbed photon energy (see text).
         The single-age stellar population spectra are from BC03.
         Except for the reddest spectra (M31 bulge and 1 Gyr-old starburst)
         the PAH emission spectra are very similar provided
         $\langle h\nu\rangle_{\rm abs}/N_{\rm C} \approx const.$
         }
\end{center}
\end{figure}

The emission spectrum from a PAH depends on the distribution of
photon energies absorbed by the PAH, but to a considerable extent it 
depends on a single number: the mean absorbed photon energy 
$\langle h\nu\rangle_{\rm abs}$ per vibrational degree of freedom.
A PAH containing $N$ atoms has $3N-6$ vibrational degrees of freedom.
If the illuminating spectrum is varied, so that $\langle h\nu\rangle_{\rm abs}$
varies, we anticipate that the {\it shape} of the emission spectrum will be
approximately constant if 
$\langle h\nu\rangle_{\rm abs}/N_{\rm C} \approx const$.
\citet{Ricca+Bauschlicher+Boersma+etal_2012} illustrated this
with $I(3.3\micron)/I(11.2\micron)$ vs $N_{\rm C}$ for two different
initial energies.

In Figure \ref{fig:pah,various rf, sizes tuned} 
we vary $N_{\rm C}$ from 37 to 105 to keep
$\langle h\nu\rangle_{\rm abs}^{\rm PAH}/N_{\rm C}\approx 0.08\eV$ 
for the six different starlight spectra.
Figure \ref{fig:pah,various rf, sizes tuned} 
shows the emission from neutral and ionized PAHs. 
The shape of the emission spectrum 
[e.g., $I(3.3\micron)/I(11.2\micron)$]
is similar for all of the cases shown, as the radiation field heating the PAHs
is varied from
the very red M31 bulge spectrum 
($\langle h\nu\rangle_{\rm abs}^{\rm (PAH)}=2.7\eV$)
to the hard UV spectrum of a 3\,Myr old starburst
($\langle h\nu\rangle_{\rm abs}^{\rm (PAH)}=8.6\eV$).

\subsection{Clouds}

The spectrum of the starlight heating the dust will depend on the history
of star formation, but can also be affected by reddening if there is sufficient dust present so that the galaxy is not optically thin in the ultraviolet.
As an example, we consider dust clouds illuminated by stars external to
the cloud.

Because the starlight heating the dust is a function of depth $x$ into the
cloud [see Eq.\ (\ref{eq:unu vs x})], the temperature distribution function
$(dP/dT)_{j,a,x}$ is a function of $x$ in addition to depending on
composition $j$ and grain size $a$.
To obtain the emission spectrum, we must integrate over
the cloud volume.
As a representative example, we consider clouds that can be approximated
as slabs with total visual extinction $A_V=2\,{\rm mag}$ -- this is large
enough that most of the optical-UV energy incident on the cloud will be
absorbed by dust in the cloud.

We consider a slab with $A_V=2$\,mag 
illuminated by radiation incident normally on one surface,
with the energy density varying as in Eq.\ (\ref{eq:unu vs x}).
The emission by the dust is assumed to be optically-thin.
The average power per unit wavelength per grain of composition
$j$ and size $a$ is obtained by  integrating over the cloud volume:
\beq \label{eq:cloud-average p_nu}
\langle p_\lambda^{(j)}(a)\rangle
=  4\pi C_{\rm abs}^{(j)}(a,\lambda) \int_0^L \frac{dx}{L}
\int dT \left(\frac{dP}{dT}\right)_{j,a,x}
 B_\lambda(T)
~~~.
\eeq

\section{\label{sec:mixed-PAH spectra}
          Emission Spectra for Dust Mixtures}

\subsection{Size Distributions}

To calculate the total dust
emission we require the size distribution and optical
properties for
each dust component.  We assume the overall dust mass to be
dominated by grains composed of a mixture of amorphous silicate, other metal oxides,
and hydrocarbons -- the hypothetical material termed ``astrodust'' 
by \citet{Draine+Hensley_2021a}
and \citet{Hensley+Draine_2021b}.  
For astrodust we assume a porosity $\calP=0.20$,
and the astrodust mass/H estimated by \citet{Hensley+Draine_2021a} and
\citet{Draine+Hensley_2021a}.
The astrodust volume per H in the diffuse ISM in the solar neighborhood is
$V_{\rm Ad}=5.34\times10^{-27}\cm^3\Ha^{-1}$.

A simple power-law size distribution captures the overall balance
between small and large grains, but more complex size
distributions are required to
reproduce the observed interstellar extinction
\citep[e.g.,][]{Kim+Martin+Hendry_1994,
                Weingartner+Draine_2001a}
and infrared emission 
\citep[e.g.,][]{Draine+Anderson_1985,
       Desert+Boulanger+Puget_1990,
       Li+Draine_2001b,
       Siebenmorgen+Voshchinnikov+Bagnulo_2014}.  
Such size distributions are obtained
for the ``astrodust'' model in a following paper 
\citep{Hensley+Draine_2021b}.  However,
to model the infrared emission 
from the astrodust component, here 
we assume a simple power-law size distribution 
\beq  \label{eq:dnda_Ad}
\frac{1}{\nH}\frac{dn_{\rm Ad}}{da} = 
(4+p)\frac{(3/4\pi)V_{\rm Ad}}{a_{\rm max,Ad}^{4+p}-a_{\rm min,Ad}^{4+p}}
~a^{p} ~~~,~~~
a_{\rm min,Ad} < a < a_{\rm max,Ad} ~~~.  
\eeq 
We take $p=-3.3$,
$a_{\rm min,Ad}=4.5\Angstrom$, and 
$a_{\rm max,Ad}=0.4\micron$
to approximately reproduce the
average extinction law in the diffuse ISM.
For each astrodust grain size, 
we calculate the probability distribution $dP/dT$
for grain temperature.
For the larger grains, $dP/dT$ approaches a delta function.

\begin{table}
\begin{center}
\footnotesize
\caption{\label{tab:dnda}PAH Size Distribution Parameters$^{\rm a}$}
\begin{tabular}{|cccc|}
\hline
parameter & standard value & ``small PAHs'' & ``large PAHs''\\
\hline
$a_{01} (\Angstrom)$                        & $4.0$  & $3.0$  & $5.0$\\
$B_1 (10^{-7}\,\Ha^{-1})$                  & $6.134$ & $18.80$ & $2.893$ \\
$q_{\rm PAH}$                              & $0.0379$ & $0.0391$ & $0.0351$ \\ 
\hline
\multicolumn{4}{l}{$^a$\,\,All cases have: 
                         $a_{\rm min, PAH}=4.0\times 10^{-8}\cm$,\,\,
                         $a_{\rm max, PAH}=1.0\times 10^{-6}\cm$,\,\,$\sigma=0.40$,}\\
\multicolumn{4}{l}{\quad $V_{\rm PAH,1}=3.0\times10^{-28}\cm^3\,\Ha^{-1}$,\,\,
                         $V_{\rm PAH,2}=0.7\times10^{-28}\cm^3\,\Ha^{-1}$,}\\
\multicolumn{4}{l}{\quad $a_{02}=30\Angstrom$, $B_2=3.113\times10^{-10}\Ha^{-1}$}\\
\end{tabular}
\end{center}
\end{table}

\begin{figure}
\begin{center}
\includegraphics[angle=0,width=8.5cm,
                 clip=true,trim=0.5cm 5.0cm 0.5cm 2.5cm]
{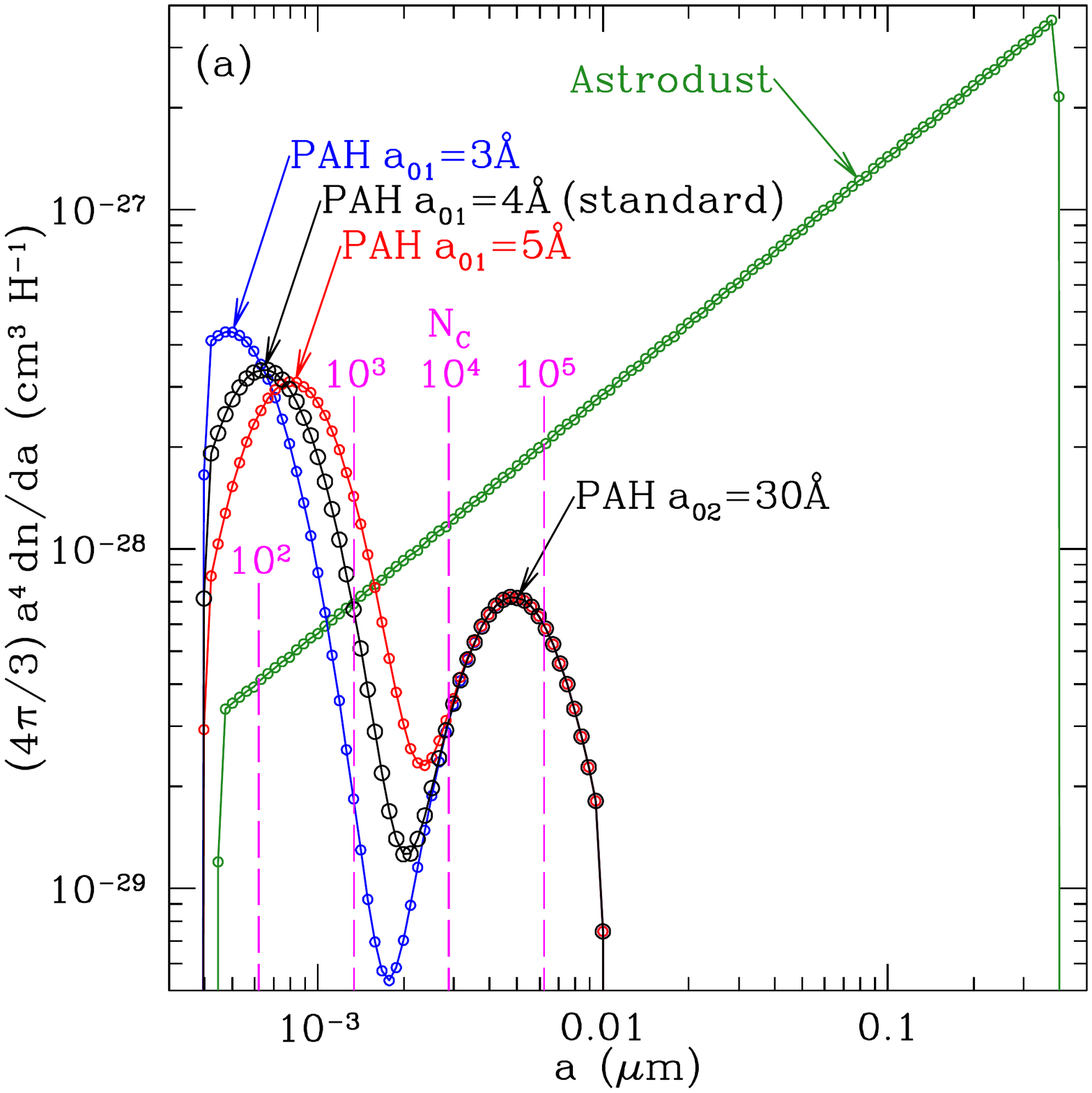}
\includegraphics[angle=0,width=8.5cm,
                 clip=true,trim=0.5cm 5.0cm 0.5cm 2.5cm]
{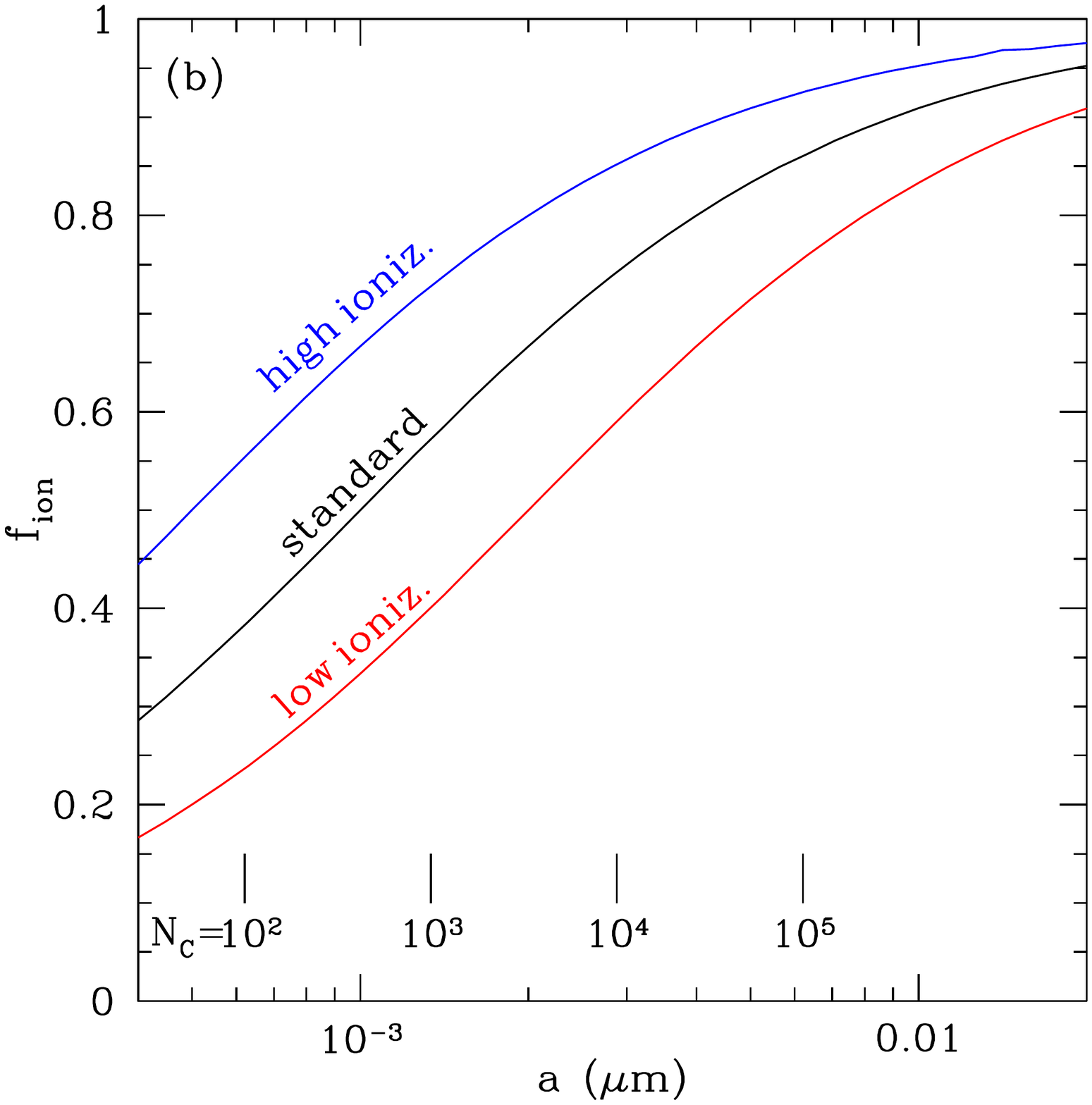}
\caption{\label{fig:dnda}\label{fig:fion}\footnotesize
         (a) Size distributions for astrodust and PAHs, for three different
         values of  parameter $a_{01}$ (see
         text).
         Sizes corresponding to $N_{\rm C}=10^2$, $10^3$, $10^4$,
         and $10^5$ are indicated.
         The dots show the discrete sizes used in our modeling.
         (b) Three examples of PAH ionization fractions (see text).
         }
\end{center}
\end{figure}

DL07 modeled the PAH population as the sum of
two log-normal size distributions: 
\beq \label{eq:dnda_pahs}
\frac{1}{\nH}
\frac{dn_{\rm PAH}}{da} = \sum_{j=1}^2\frac{B_j}{a}
\exp\left\{-\frac{[\ln(a/a_{0j})]^2}{2\sigma^2}\right\}
~~~~~~,~~~~~~ a_{\rm min,PAH} < a < a_{\rm max,PAH}   ~~~. 
\eeq
About 75\% of the PAH mass in the DL07 model was in a
log-normal population with $a_{01}=4.0\Angstrom$
and $\sigma=0.40$.
These are the PAH sizes that are needed to account for the
observed PAH emission features.
We retain this population in the present work, with
a volume $V_{{\rm PAH},1}=3.0\times10^{-28}\cm^3\Ha^{-1}$,
containing C/H $=30$\,ppm.

Following DL07, we add additional carbonaceous material in a second log-normal
size distribution.  For this component we take $a_{02}=30\Angstrom$, 
large enough to contribute minimal emission for
$\lambda < 20\micron$.
We take this component to contribute a volume
$V_{{\rm PAH},2}=0.7\times10^{-28}\cm^3\Ha^{-1}$, containing C/H = 7\,ppm.
The two log-normal components together reproduce the observed
strength of the $2175\Angstrom$ feature.

Below some critical size, transient heating events will lead to
``evaporation'' of atoms or groups of atoms, leading to
destruction of the nanoparticle.
\citet{Guhathakurta+Draine_1989} estimated that graphitic clusters 
need to have $N_{\rm C}\geq23$ to survive in the ISM.
We adopt a lower cutoff 
$a_{\rm min,PAH}=4.0\Angstrom$, corresponding to
$N_{\rm C,min}=27$ carbon atoms.

The size distributions are shown in Figure \ref{fig:dnda}a.
The parameters in Table \ref{tab:dnda} correspond to 
${\rm C/H} =$ 37$\,{\rm ppm}$ in the PAH population.  
This differs from
the value of 60 ppm adopted by DL07 because of recent findings that the
extinction per H in the general diffuse ISM is smaller than
previously thought.\footnote{%
    The value of $A_V/\NH=3.52\times10^{-22}{\rm mag}\cm^{-2}\Ha^{-1}$ 
    found by
    \citet{Lenz+Hensley+Dore_2017} is only 66\% of the
    \citet{Bohlin+Savage+Drake_1978} value that was long taken to be
    representative of the ISM in the solar neighborhood.}
For the standard parameters in Table \ref{tab:dnda}, 
the fraction of the total dust mass contributed by PAHs containing
$N_{\rm C}<10^3$ C atoms is $q_{\rm PAH}=0.038$.

Figure \ref{fig:ext}b shows 
the extinction curve corresponding to the size distributions of
Figure \ref{fig:dnda}a.
The detailed wavelength dependence of extinction is not accurately
reproduced, but the model extinction deviates
from the observed extinction by at most $\sim$20\% (near $\sim1400\Angstrom$),
and generally much less.

\subsection{PAH Ionization}

Interstellar PAHs will be present in a range of charge states, including
PAH$^-$ anions, neutral PAH$^0$ molecules, 
PAH$^+$ cations, and PAH$^{++}$ dications.
The present modeling collapses these to just PAH neutrals and PAH ions.

Because PAH neutrals and ions have quite different emission spectra
(compare Figs.\ \ref{fig:pah_mmp}a and \ref{fig:pah_mmp}b), we must specify the
PAH ionized fraction as well as the size distribution.  The PAH ionization will be determined by a statistical balance between photoionizations and
collisional charging by collisions with electrons and ions, and will therefore depend on electron density and electron kinetic temperature in addition to
the radiation intensity.
\cite{Li+Draine_2001b} estimated the ionized fraction $\fion(a)$, 
as a function of PAH radius $a$, 
for three sets of physical conditions for the diffuse
interstellar medium: ``cold neutral medium'' (CNM), ``warm neutral medium'' (WNM), and ``warm ionized medium'' (WNM).
A weighted average of these was used for the overall ISM, 
and this ionized fraction was adopted by DL07.

To explore the sensitivity of overall emission spectra to assumptions regarding
the ionization, we will consider three examples of $\fion(a)$, shown in
Figure \ref{fig:fion}b.
Our ``standard'' case is similar to $\fion(a)$ assumed by DL07.
The high and low $\fion(a)$ curves in Figure \ref{fig:fion}b
correspond to factor-of-two shifts in the PAH radius for which $\fion=0.5$.
Because the photoionization rate for a neutral PAH scales 
approximately as $Ua^3$, and the electron capture rate for a PAH cation
scales approximately as $a^2 n_e/\sqrt{T_e}$, 
a shift of a factor of two in
the size where $\fion=0.5$ corresponds approximately to a 
factor-of-two shift in the ratio $Ua\sqrt{T_e}/n_e$.  
It would not be surprising if
even larger variations in PAH ionization occurred.
The relative constancy of PAH spectra may therefore be an indication of
self-regulation in the ISM that limits variations in $U\sqrt{T_e}/n_e$.

\subsection{Diffuse ISM}

Let $4\pi j_\lambda$ be the power radiated per unit volume, 
per unit wavelength.
The emissivity per H nucleus is obtained by summing the radiated power
per grain over the grain size distribution:
\beq
\frac{4\pi j_\lambda}{\nH} = \sum_j \int da 
\left[ \frac{1}{\nH}\frac{dn_j}{da}\right] p_\lambda^{(j)}(a)
~~~.
\eeq

\begin{figure}
\begin{center}
\includegraphics[angle=0,width=8.5cm,
                 clip=true,trim=0.5cm 5.0cm 0.5cm 2.5cm]
{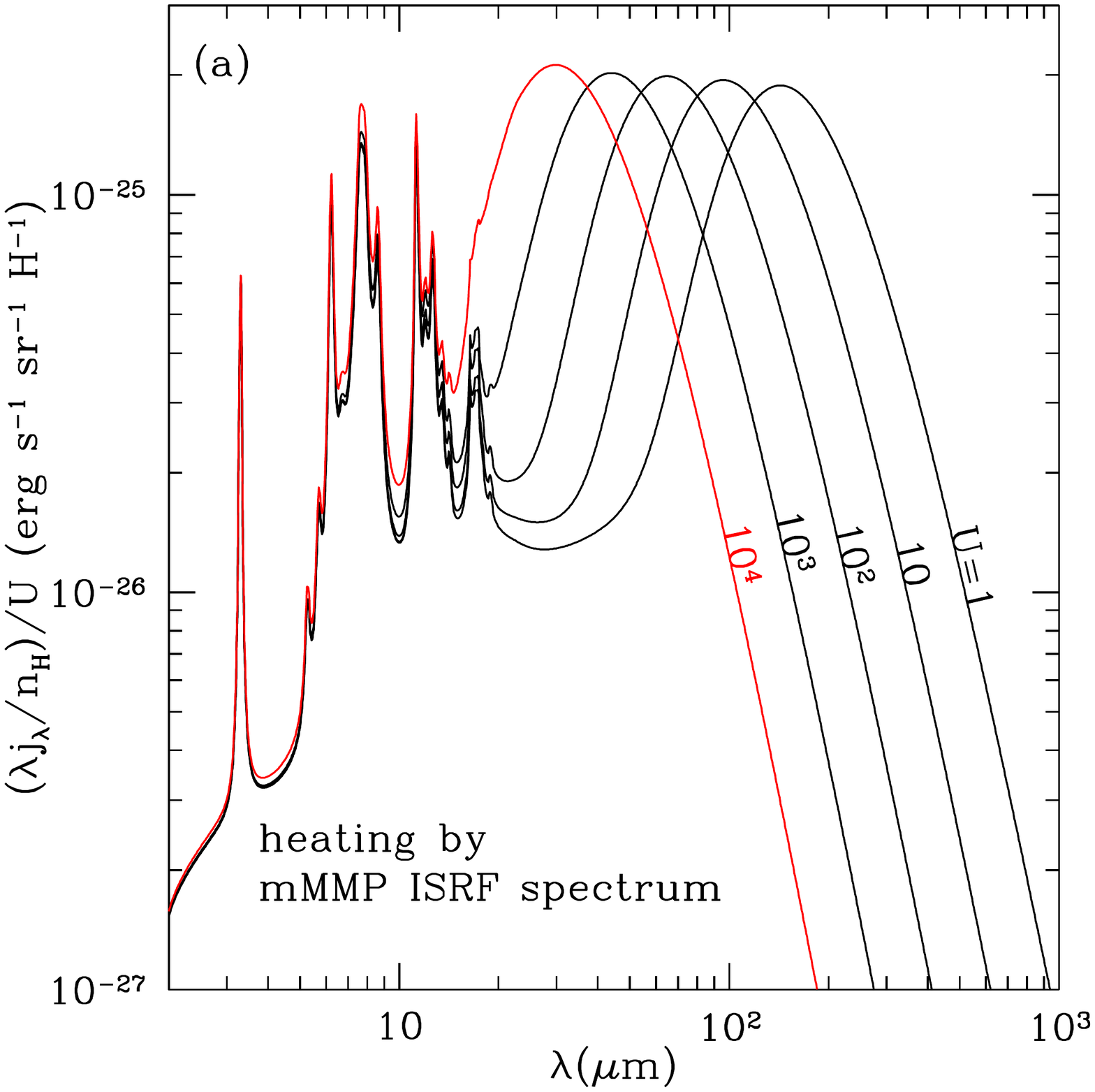}
\includegraphics[angle=0,width=8.5cm,
                 clip=true,trim=0.5cm 5.0cm 0.5cm 2.5cm]
{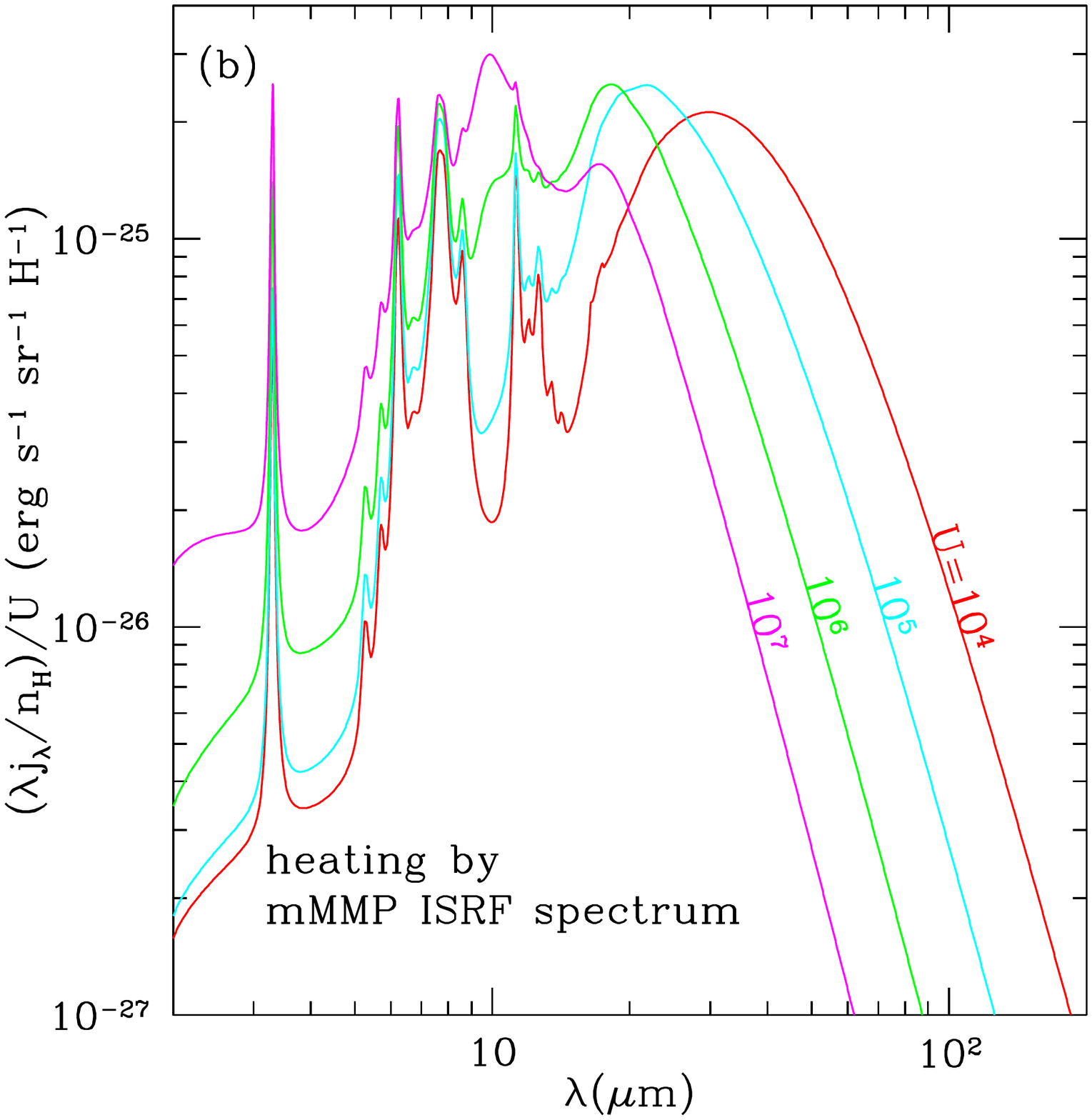}
\caption{\label{fig:dustmix}\footnotesize
         Emission spectra for standard dust mixture 
         heated by
         radiation with mMMP ISRF spectrum and varying $U$,
         for standard PAH size distribution 
         ($a_{01}=4\Angstrom$)
         and standard PAH$^+$ fraction $\fion$ (see text).
         (a) $U\leq10^4$; (b) $U\geq10^4$.
         }
\end{center}
\end{figure}

Figure \ref{fig:dustmix} shows emission spectra calculated for the
standard dust mixture heated by starlight with the mMMP spectrum, but
varying $U$.  
At far-infrared wavelengths, the opacity of astrodust varies approximately as
$\lambda^{-1.8}$ \citep{Hensley+Draine_2021a}.
Therefore,
as $U$ is increased, the temperature of the large grains increases
as $T \propto U^{1/5.8}$, with the wavelength where $\lambda p_\lambda$
peaks in the FIR varying as
\beq \label{eq:lambdapk}
\lambda_{\rm FIR\,peak} \approx 140\micron \times  U^{-1/5.8}
\hspace*{1.0cm}{\rm for~} U\ltsim 10^4
~~~.
\eeq
At the same time, for $U\ltsim 10^3$, the {\it shape} of the emission spectrum
at $\lambda \ltsim 15\micron$ remains nearly invariant, 
with the emission remaining dominated by cooling following single-photon heating.  
Only as $U$ exceeds $\sim$$10^3$ does
the heating become sufficient to modify the shape of the $\lambda < 15\micron$ emission, but the spectrum for $\lambda < 10\micron$ 
remains unaffected even for $U=10^4$.

For $U\gtsim10^5$, the silicate features begin to
appear in the emission spectrum
(see Figure \ref{fig:dustmix}b),
as the astrodust grains become warm enough ($T\gtsim 130\K$) to
radiate strongly in the $18\micron$ feature and, 
for $U\gtsim10^6$, in the
$9.7\micron$ feature as well.  
For $U\gtsim10^6$ 
the silicate $10\micron$ emission feature is comparable to or
stronger than
the strongest PAH emission features.

\begin{figure}
\begin{center}
\includegraphics[angle=0,width=8.5cm,
                 clip=true,trim=0.5cm 5.0cm 0.5cm 2.5cm]
{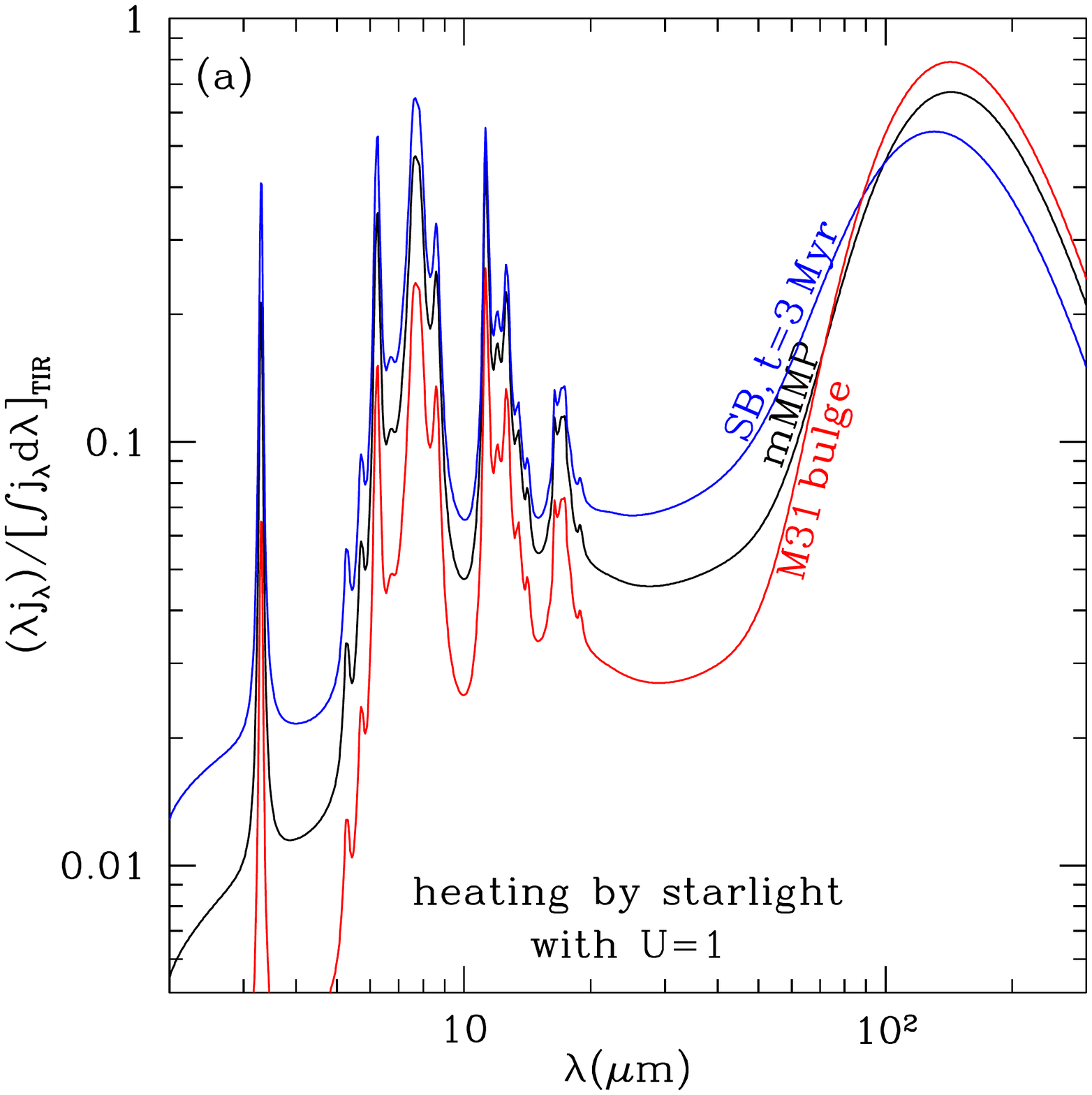}
\includegraphics[angle=0,width=8.5cm,
                 clip=true,trim=0.5cm 5.0cm 0.5cm 2.5cm]
{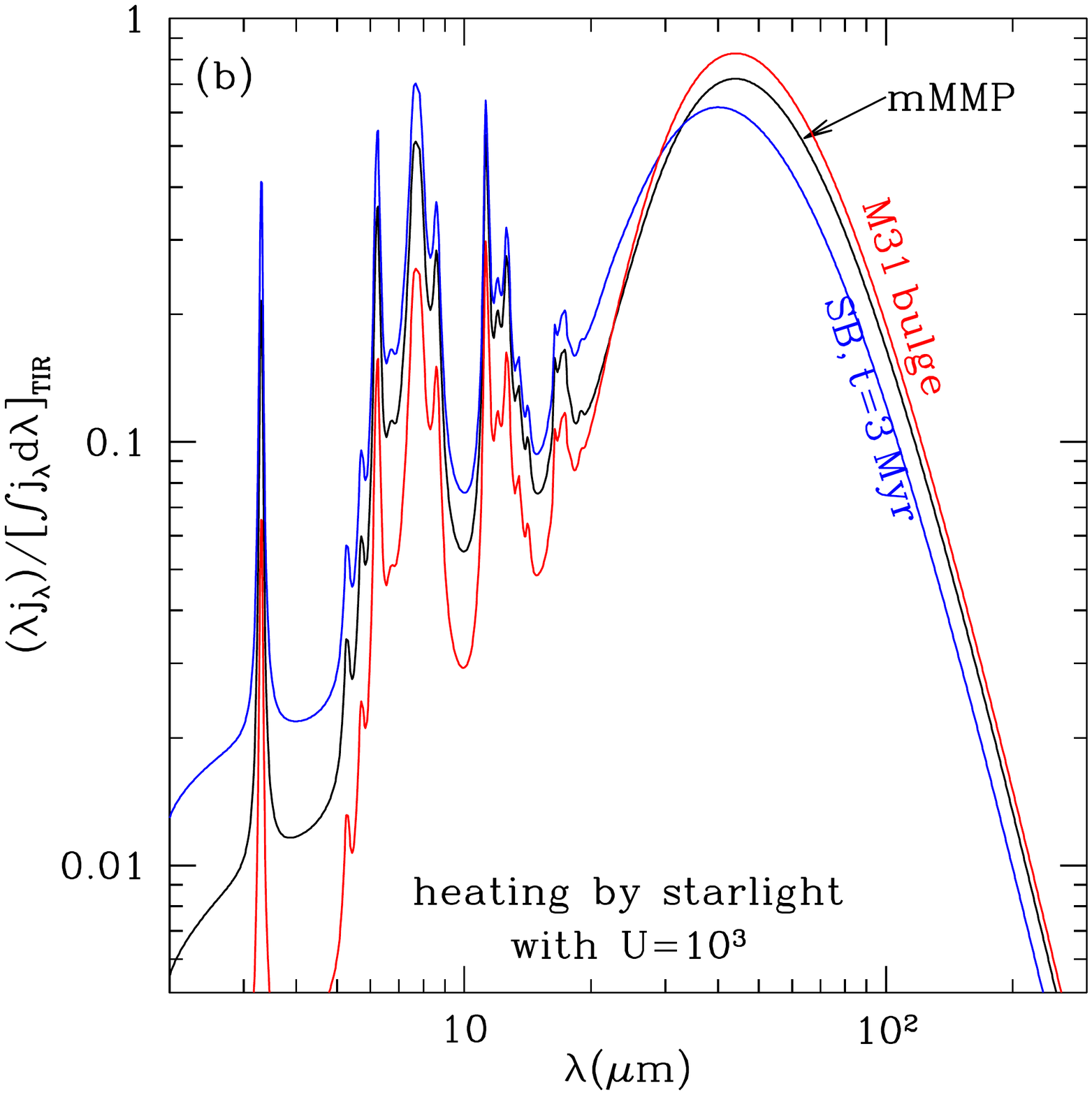}
\caption{\label{fig:varyisrf}\footnotesize
         Normalized emission spectra for standard dust mixture 
         heated by
         different starlight spectra with $U=1$ (left) and $U=10^3$ (right).
         }
\end{center}
\end{figure}

The normalized
emission 
$\lambda j_\lambda/[\int j_\lambda d\lambda]_{\rm TIR}=
\lambda L_\lambda/L_{\rm TIR}$, where $L_{\rm TIR}$ 
is the total infrared
power, is shown for the standard dust mixture in Figure \ref{fig:varyisrf}
for three different starlight spectra.
Results for
$U=1$ are shown in
Fig.\ \ref{fig:varyisrf}a,
and for $U=10^3$ in
Fig.\ \ref{fig:varyisrf}b.

The mMMP and
M31 bulge spectra both have $\lambda_{\rm FIR\,peak}\approx140\micron$ 
for $U=1$, but the 3\,Myr-old starburst spectrum
has $\lambda_{\rm FIR\,peak}\approx130\micron$, showing that
the single parameter $U$ does not completely capture the heating effects of
the radiation field on a dust mixture -- 
the shape of the spectrum also matters.
Recall that $U$ was defined in terms of the
rate of heating of $\aeff=0.1\micron$ astrodust grains.
For the standard dust mixture, the FIR emission from
$\aeff\approx 0.1\micron$ dust appears 
to be representative of the broad FIR peak
when the dust is heated by starlight spectra that are
broadly similar to the mMMP spectrum.  However, a very
young starburst has a starlight spectrum with far more energy in the
far-UV (see Figure \ref{fig:isrf}), which has the effect of 
increasing the fraction of the starlight absorption contributed by the
smaller grains.  
The smaller grains are somewhat warmer,
hence the peak of the SED is at 
$\sim$$130\micron$
for $U=1$ rather than $\sim$$140\micron$ as 
for the mMMP and M31 bulge
spectra.
Increasing $U$ to $U=10^3$, shifts $\lambda_{\rm FIR\,peak}$ 
to $\sim 42\micron$ for the mMMP spectrum.

The PAH emission features are sensitive 
to the starlight spectrum.
For our standard dust mixture,
Figure \ref{fig:varyisrf} shows that
the 3\,Myr old starburst spectrum produces 
normalized PAH features that are $\sim$20\% stronger than
for heating by the mMMP spectrum, 
while heating by the M31 bulge spectrum produces
normalized
PAH features that are a factor of $\sim$2 weaker than for the mMMP spectrum,
as previously found by \citet{Draine+Aniano+Krause+etal_2014}.
The PAH band ratios (e.g., $F(3.3\micron)/F(11.2\micron)$) are sensitive
to the spectrum of the illuminating starlight, and vary significantly as the
radiation field is varied from a 3\,Myr old starburst spectrum to the
very red M31 bulge spectrum.  The dependence of band ratios on the
starlight spectrum is discussed below in Section \ref{subsec:band_ratios}.
\begin{figure}
\begin{center}
\includegraphics[angle=0,width=8.5cm,
                 clip=true,trim=0.5cm 5.0cm 0.5cm 2.5cm]
{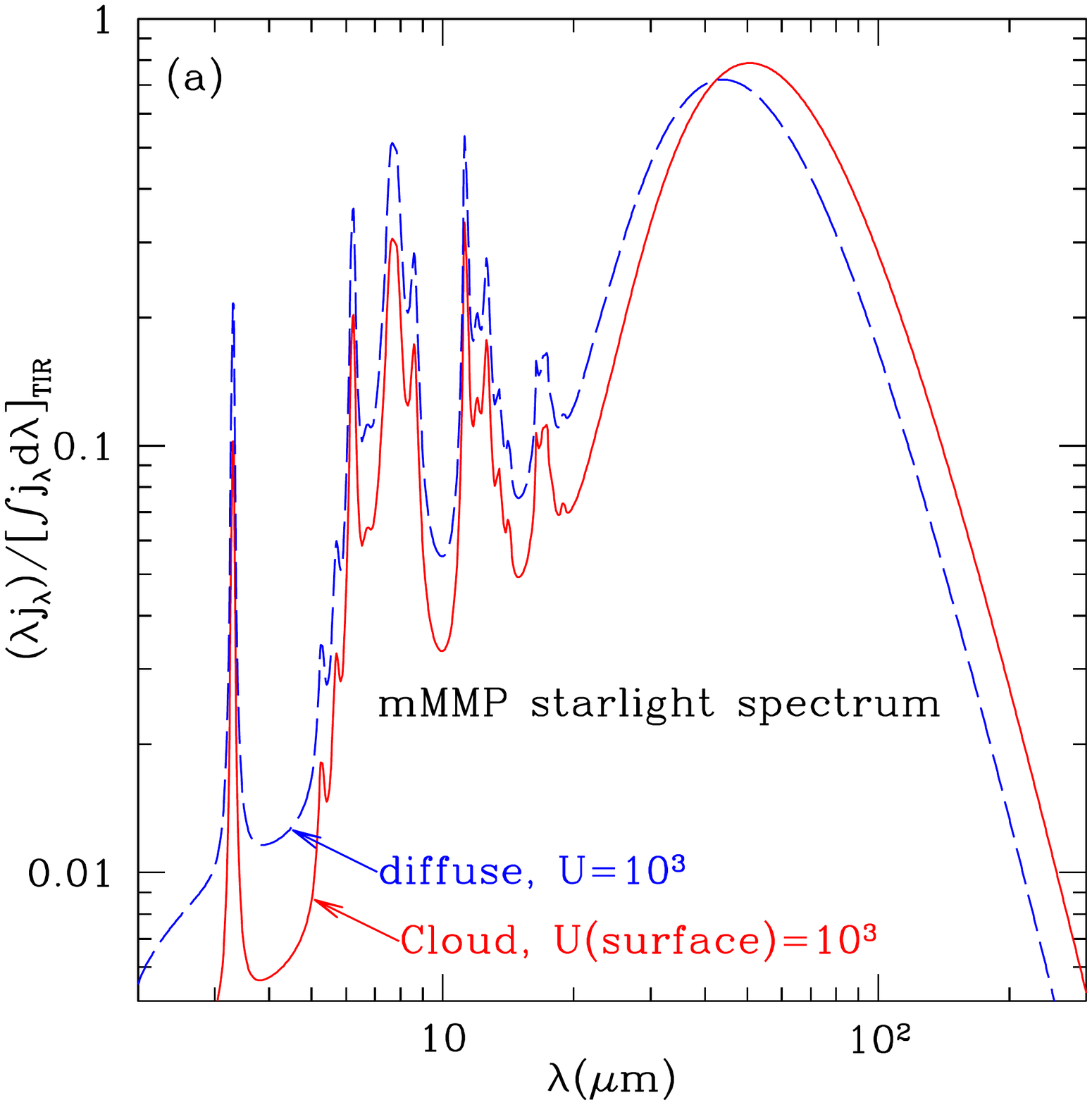}
\includegraphics[angle=0,width=8.5cm,
                 clip=true,trim=0.5cm 5.0cm 0.5cm 2.5cm]
{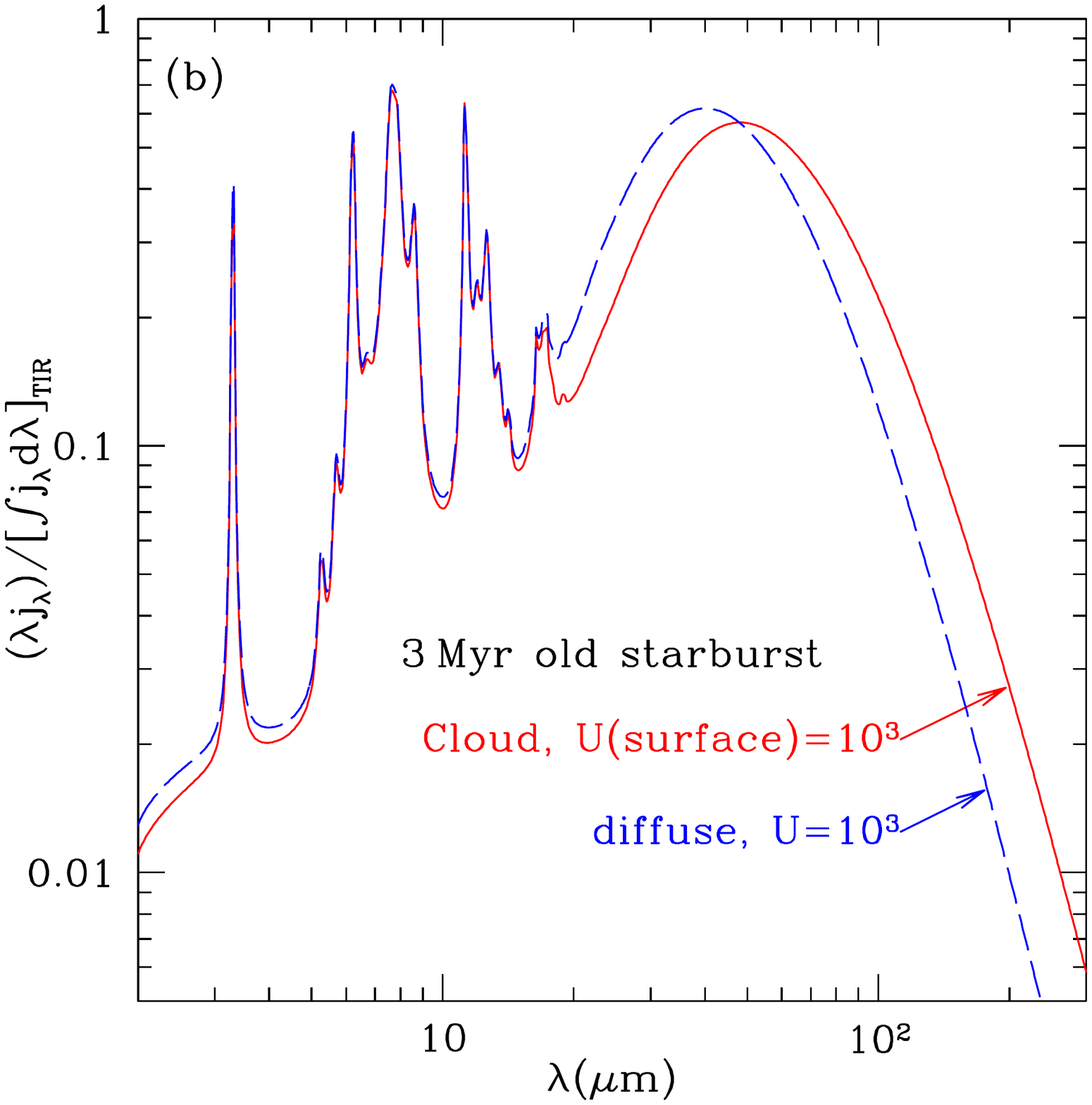}
\caption{\label{fig:diff_vs_slab}\footnotesize
    Emission (normalized to TIR) from diffuse dust exposed to starlight
    with $U=10^3$, and from a dust slab with $A_V=2\,{\rm mag}$.
    (a) Incident starlight with mMMP ISRF spectrum.
    The PAH features for the slab are weakened by a factor $\sim$$1.5$.
    (b) Incident starlight from 3 Myr old starburst.
    The normalized PAH features are 
    almost unaffected by reddening, because most
    of the stellar power is in the far-UV.
    }
\end{center}
\end{figure}

\subsection{Clouds}

Because the starlight heating the dust is a function of depth $x$ into the
cloud (see Equation (\ref{eq:unu vs x})), 
the temperature distribution function
$(dP/dT)_{j,a}$ is a function of $x$, and we must therefore average over
the cloud volume.
The infrared emission is assumed to be optically thin.
The cloud-averaged emissivity per H is
\beq
\Big\langle \frac{4\pi  j_\lambda}{\nH} \Big\rangle 
= \sum_j \int da \frac{1}{\nH}\frac{dn_j}{da} \langle p_\lambda^{(j)}(a)\rangle
~~~.
\eeq
with the cloud-averaged $\langle p_\lambda^{(j)}(a)\rangle$ given by 
Eq.\ (\ref{eq:cloud-average p_nu}).
Figure \ref{fig:diff_vs_slab} shows normalized emission spectra for 
diffuse dust
heated by starlight with $U=10^3$, and for a dust cloud with $A_V=2$ exposed
to the same radiation field at the cloud surface.
Figure \ref{fig:diff_vs_slab}a is for starlight with the mMMP spectrum,
appropriate for a mature star-forming galaxy.
As expected, the FIR peak shifts from 
$\lambda_{\rm FIR\,peak}\approx40\micron$ for the diffuse ISM starlight
to $\sim$$50\micron$ for the cloud,
because most of the starlight power absorbed in the clouds takes place 
in an attenuated (and reddened) radiation field, with dust temperatures
somewhat lower than at the cloud surface.  
The $3.3\micron$ PAH feature drops by about a factor of two, 
because excitation of the 3.3$\micron$ feature is dependent 
on the more energetic stellar photons, 
and these are attenuated relatively rapidly as one moves into the dust slab.

Figure \ref{fig:diff_vs_slab}b 
shows the same calculation, but now for a radiation
spectrum characteristic of a 3 Myr old starburst
and $U=10^3$.  Once again, 
attenuation within the cloud shifts
the peak of the
FIR emission longward by about a factor 1.2, but 
for this case the
normalized PAH features are nearly unchanged.  This is because {\it most}
of the starlight energy is in the UV, and capable of exciting the
PAH features via single-photon heating.  
Because the incident radiation field is dominated by FUV photons,
the PAH excitation falls off like the overall dust
heating rate, leaving the {\it ratio} of PAH emission to total emission
almost unaffected.
We will see below 
(in Section \ref{subsec:sensitivity_to_starlight_spectrum})
that the PAH band ratios remain nearly unchanged
by reddening when the unreddened starlight is dominated by UV.

\begin{figure}
\begin{center}
\includegraphics[angle=0,width=8.5cm,
                 clip=true,trim=0.5cm 5.0cm 0.5cm 2.5cm]
{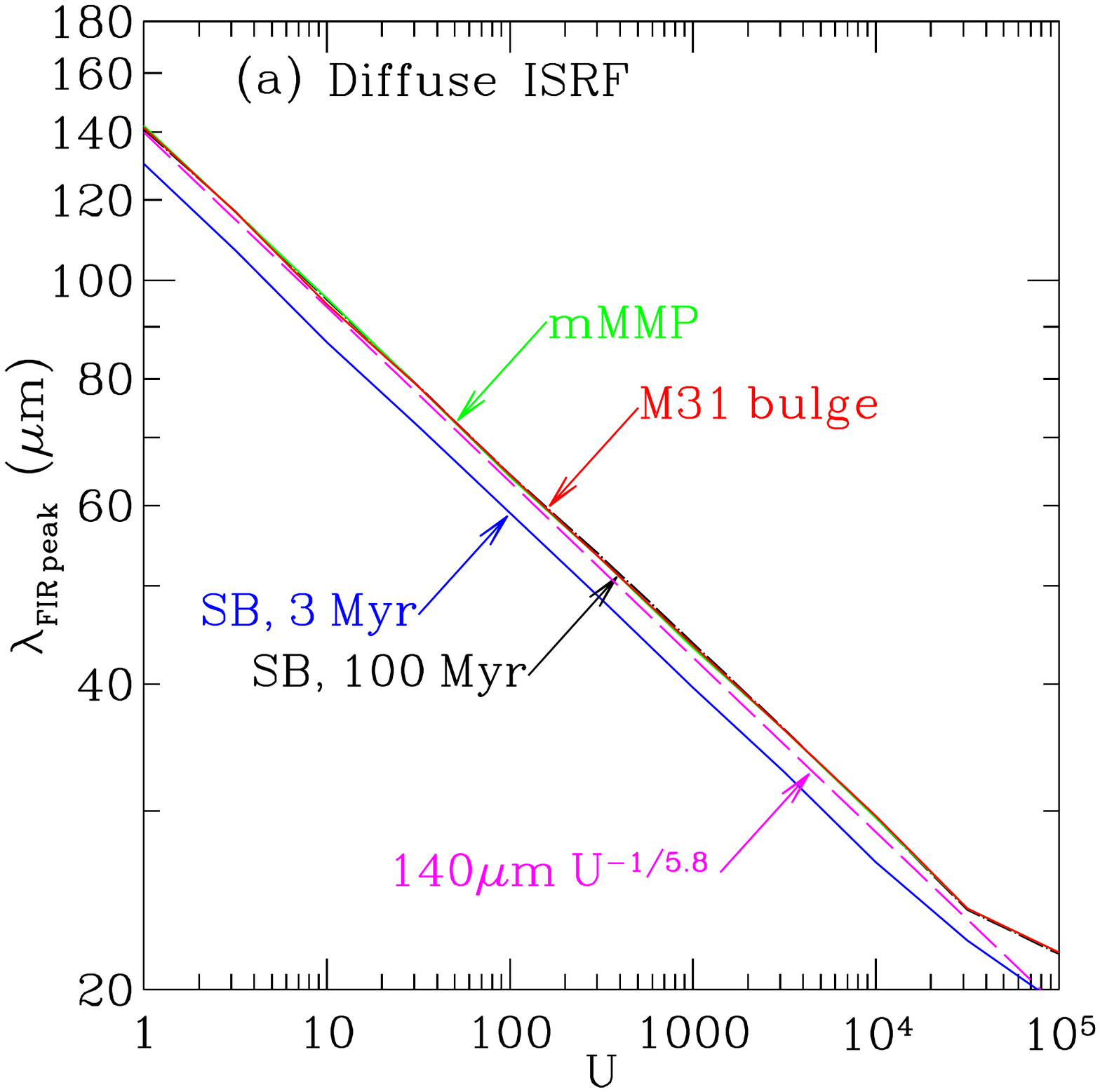}
\includegraphics[angle=0,width=8.5cm,
                 clip=true,trim=0.5cm 5.0cm 0.5cm 2.5cm]
{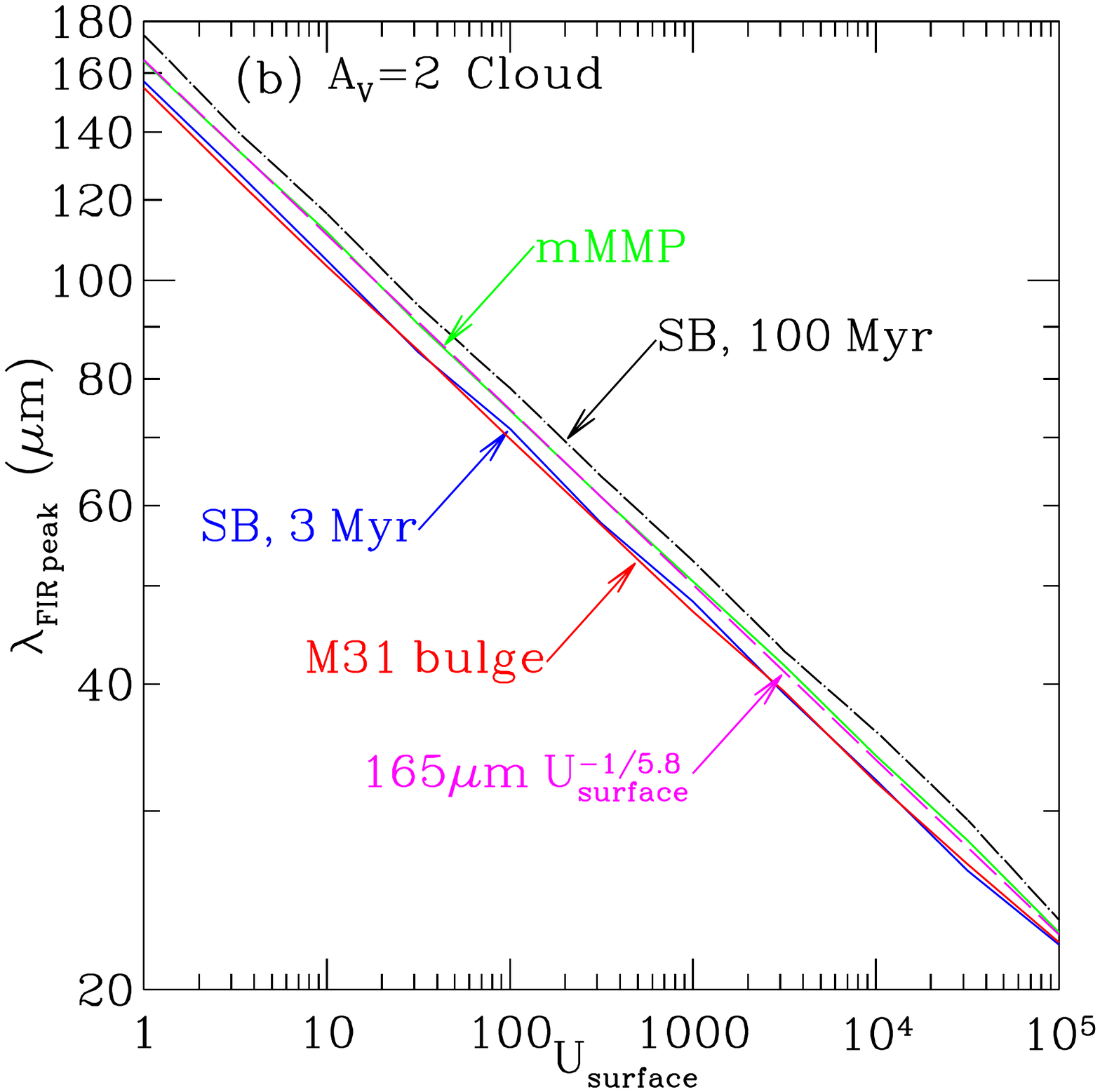}
\caption{\label{fig:lambdapk}\footnotesize
    Wavelength $\lambda_{\rm FIR peak}$ where $\lambda j_\lambda$ peaks
    in the far-infrared, 
    as a function of starlight heating parameter $U$, 
    for our standard model and various
    radiation fields.
    (a) Unattenuated rediation fields.
    (b) Emission summed over a cloud with $A_V=2$.
    In each panel, the magenta dashed line is a power-law fit.
    }
\end{center}
\end{figure}

Figure \ref{fig:lambdapk}a shows how $\lambda_{\rm FIR\,peak}$ depends on
the heating parameter $U$, and on the spectrum of the starlight.
The wavelength $\lambda_{\rm FIR\,peak}$ depends primarily on $U$ (see
Eq.\ \ref{eq:lambdapk}), but $\lambda_{\rm FIR\,peak}$ does depend
slightly on the spectrum of the radiation heating the dust.
Figure \ref{fig:lambdapk}b shows $\lambda_{\rm FIR\,peak}$
for emission from an $A_V=2$\,mag cloud, as a function of the
parameter $U$ for the radiation at the cloud surface. 
The emission from the cloud includes
radiation from dust grains heated by the attenuated radiation field
within the cloud, which shifts $\lambda_{\rm FIR\,peak}$ longward of
where it would be for only the emission from the grains at the cloud
surface.  Figure \ref{fig:lambdapk} shows that this shifts
$\lambda_{\rm FIR\,peak}$ longward by about a factor 1.2 for the mMMP spectrum.

\section{\label{sec:ionization}
         Sensitivity to Ionized Fraction}

The emission spectrum of a
vibrationally-excited PAH depends on whether the PAH is neutral or
ionized.  
Laboratory measurements and theoretical
calculations for small PAHs find that the band strengths for
some of the important spectral features depend strongly on the
ionization state
\citep{DeFrees+Miller+Talbi+etal_1993,Allamandola+Hudgins+Sandford_1999}.  
PAH$^+$ cations have smaller band
strengths for the C-H stretch at 3.3$\micron$
 than the corresponding neutral PAH.
Conversely, the C-C stretching modes at 6.2 and $7.7\micron$ 
have larger band strengths for PAH$^+$ cations than for
the corresponding neutral PAHs.  

Our present calculations are based on the PAH band strengths
adopted by DL07; for the 3.3$\micron$ C-H stretching feature, for
example, neutral PAHs are assumed to have a band strength 4.4 times
larger than for the corresponding PAH$^+$ cation, while the band strength
for the 7.7$\micron$ feature is taken to be nine times larger for cations
than for neutrals.  The values adopted by DL07 are consistent with
the range in results found in theoretical calculations
\citep[e.g.][]{Malloci+Joblin+Mulas_2007,
               Bauschlicher+Boersma+Ricca+etal_2010,
               Bauschlicher+Ricca+Boersma+Allamandola_2018}, and are
intended to represent general trends for the actual PAH-like nanoparticles
in the ISM.

As a result, the model emission spectrum depends on
the assumed fractional ionization of the PAHs, which will be dependent
on local conditions: the intensity of UV photons capable of
photoionizing the PAHs, and the density and temperature of the free
electrons that can be captured by the PAHs.  Because the PAH
ionization must depend on local conditions, one may expect regional
variations in PAH emission spectra arising from variations in the degree of
ionization of the PAHs.

\begin{figure}[t]
\begin{center}
\includegraphics[angle=0,width=8.5cm,
                 clip=true,trim=0.5cm 5.0cm 0.5cm 2.5cm]
{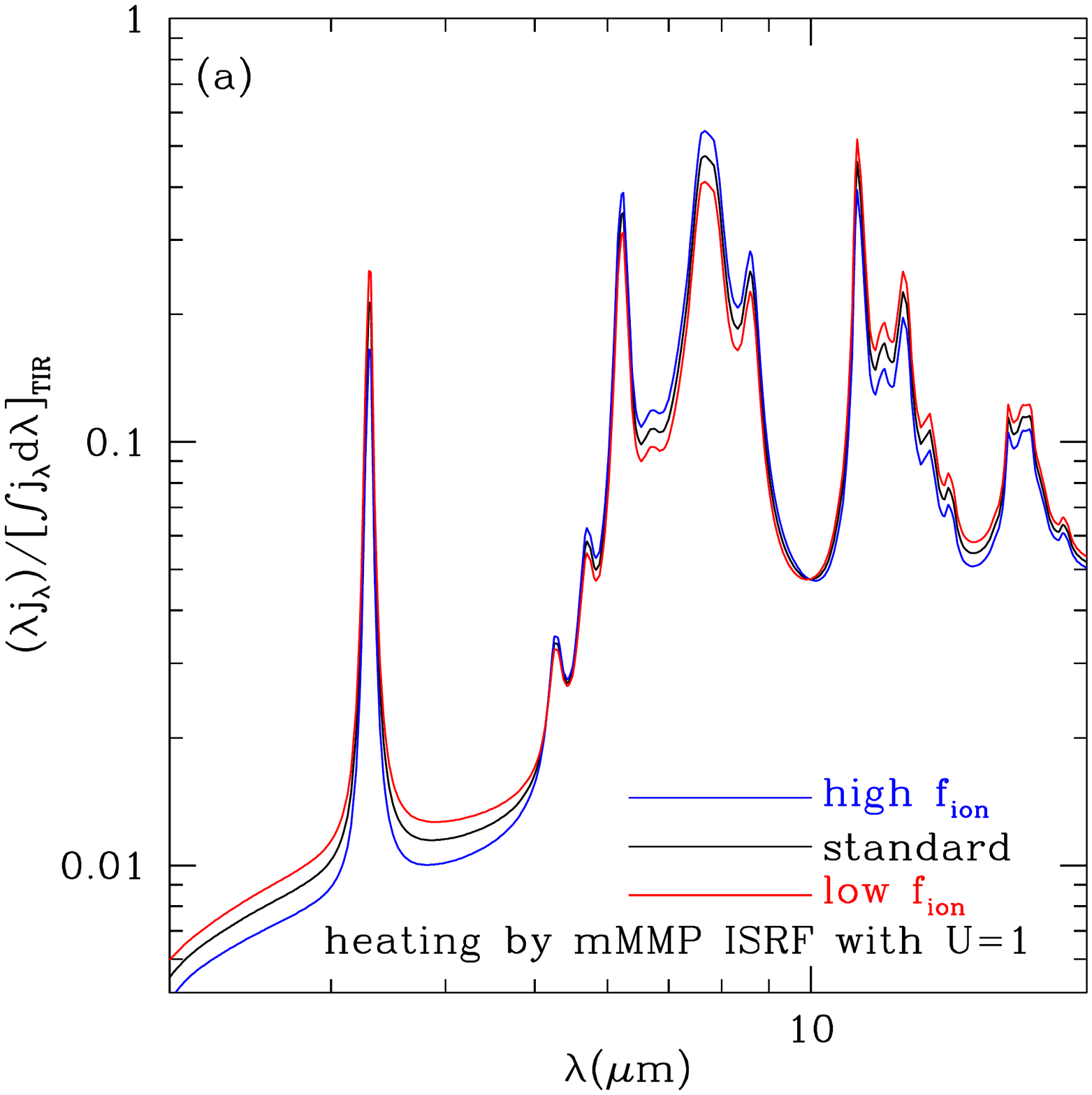}
\includegraphics[angle=0,width=8.5cm,
                 clip=true,trim=0.5cm 5.0cm 0.5cm 2.5cm]
{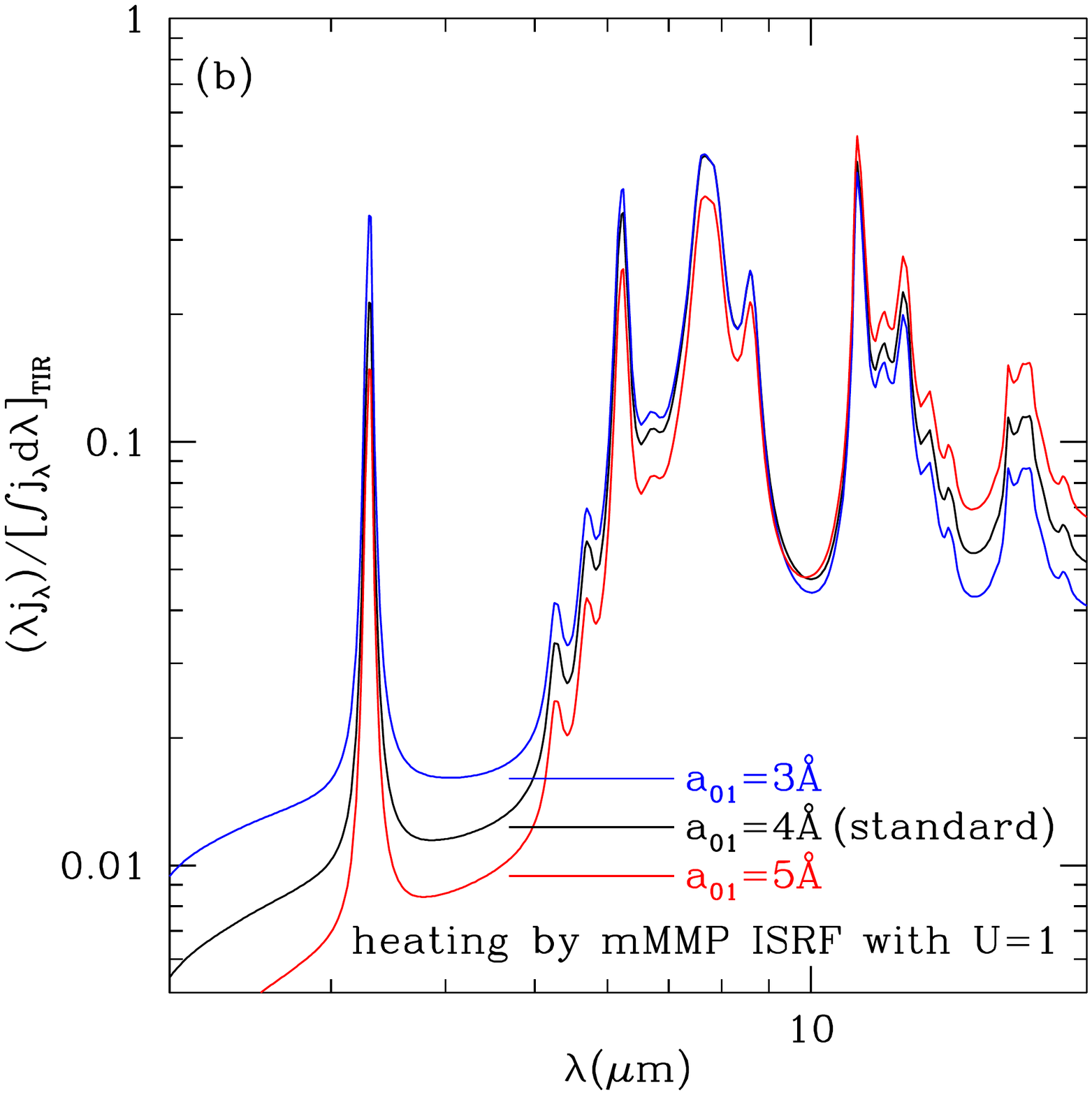}
\caption{\label{fig:varyfion}\label{fig:vary_pah_size}\footnotesize
         Emission spectra (normalized to total infrared emission)
         (a) for standard dust mixture heated by
         mMMP starlight spectrum, 
         for three 
         different PAH ionized fractions $\fion(a)$ (see text).
         (b) Three different PAH size distributions, parameterized by
         $a_{01}$ (see text).
         }
\end{center}
\end{figure}

Our standard dust model above included an assumption of how the PAH
ionization fraction depends on the PAH size (see
Fig.\ \ref{fig:fion}b).  Here we 
examine how varying $\fion(a)$ 
affects the relative strengths of PAH features
after integrating over the size distribution.
Fig.\ \ref{fig:varyfion}a shows model emission spectra for dust heated
by the mMMP radiation field, for the three different levels of PAH
ionization shown in Fig.\ \ref{fig:fion}b.  
As expected, increasing
(decreasing) the ionization increases (decreases) the
$7.7\micron$/$3.3\micron$ and $7.7\micron$/$11.2\micron$ band ratios.
This will be further explored in Section \ref{subsec:sensitivity_to_ionization}
below.

\section{\label{sec:pah_size}
         Sensitivity to PAH Size Distribution}

Single-photon heating raises smaller PAHs to higher peak temperatures than
larger PAHs.
Because emission 
in the shorter wavelength PAH bands can only be excited when the
nanoparticle is quite hot, the infrared emission spectrum
is sensitive to the PAH size distribution.

The PAH size
distribution is the result of competing processes including
fragmentation of large grains, chemisputtering by reaction with radicals,
physical sputtering in hot gas,
photodesorption of atoms or complexes, 
growth by accretion of individual atoms, 
agglomeration
with either other nanoparticles or larger grains, 
and perhaps other processes.  
Unfortunately, at this time
our limited theoretical understanding of these processes cannot provide
a credible
{\it a priori} expectation for the shape of the size distribution, 
other than that we expect a fairly sharp cutoff at the 
smallest sizes because below some
critical size particles cannot survive in the UV radiation field.

Changes in the PAH size distribution could be accompanied by
changes in the character of the PAHs.  PAH emission spectra have
been classified by \citet{Peeters+Hony+VanKerckhoven+etal_2002}
based on the 6--9$\micron$ spectrum.  Integrated spectra of galaxies generally
fall into class A, but individual regions (e.g., planetary nebulae,
Herbig Ae-Be stars, post-AGB stars) are often in classes B or
C.  An additional class D was introduced by
\citet{Matsuura+Bernard-Salas+LloydEvans+etal_2014} to accomodate the spectra
of certain post-AGB stars in the LMC.

\citet{Shannon+Boersma_2019} proposed that classes A and B may differ
in part because of the sizes of the PAHs emitting in the 6--9$\micron$
region, with class B PAHs being larger.  
\citet{Shannon+Boersma_2019} calculated the emission from PAHs excited by
monochromatic $5\eV$ photons.
Here we use a full spectrum of exciting radiation; we also examine the
complete 3--20$\micron$ spectra.

Our adopted ``standard'' size distribution 
[see Fig.\ \ref{fig:dnda}a and Eq.\ (\ref{eq:dnda_pahs})] 
is an empirical distribution, adopted by DL07
because it seemed to do a good job of reproducing observed 
$5$ -- $20\micron$ PAH
spectra of galaxies
\citep[e.g.][]{Smith+Draine+Dale+etal_2007}, 
if the PAHs are assumed to be heated by radiation with the
spectrum of the mMMP ISRF.  

To investigate the sensitivity of the
model spectra to the size distribution, we arbitrarily change the
parameter $a_{01}$ 
by $\pm25\%$, corresponding to
increasing the characteristic mass 
$\propto a_{01}^3$ by
$1.25^3=1.95$, or reducing it by a factor $0.75^3=0.42$.  In Figure
\ref{fig:vary_pah_size}b we see that 
even such small variations in PAH size result
in appreciable changes in PAH band ratios: as expected, reducing
$a_{01}$ 
results in an increase in the $3.3\micron$/$7.7\micron$ band
ratio, because the smallest PAHs are more readily excited to the
high temperatures required for appreciable emission at $3.3\micron$.
The effects on band ratios are discussed further in Section 
\ref{subsec:band_ratios}.

\section{\label{sec:discussion}
         Discussion}

The emission spectra $p_\lambda$ calculated here 
are available on-line\footnote{%
    Data are available for download from 
    \url{doi.org/10.7910/DVN/LPUHIQ}.
    }
for a grid of grain sizes and
starlight intensities, and for selected spectral shapes for the
starlight heating the dust.
We also make available emission spectra for complete dust models 
for three possible PAH size distributions 
(see Figure \ref{fig:dnda}a), and for
three possible PAH ionization fractions $\fion(a)$ 
(see Figure \ref{fig:dnda}b).
These emissivities can be used for modeling the emission 
from galaxies containing distributions of starlight intensities.

\subsection{\label{subsec:feature strengths}
            Feature Strengths}

Interpretation of the emission features in
observed infrared spectra 
requires quantitative determination of the
feature strengths.  
Feature ``extraction'' is complicated by presence of a ``continuum'', 
and because 
the features may be overlapping in wavelength.
In addition, the observed emission may be affected by extinction, 
particularly in the $9.7\micron$ silicate feature, 
that should be corrected for
\citep{Smith+Draine+Dale+etal_2007}.

Various approaches have been taken.
Some studies employ spline fits to the
presumed continuum so that it can be subtracted to reveal the emission features
\citep[e.g.,][]{Peeters+Hony+VanKerckhoven+etal_2002,
                Brandl+Bernard-Salas+Spoon+etal_2006}.
Another approach is to fit the
observed spectra, including the continuum, 
using a physically-motivated set of fitting functions 
with a tractable number of free parameters;
PAHFIT \citep{Smith+Draine+Dale+etal_2007} is one such spectral-fitting code.
Different feature extraction techniques may differ appreciably in
estimation of the power in the different PAH bands.
Paper II
%
(J.-D.T. Smith et al., in preparation)
applies an expanded PAHFIT decomposition
to our model spectra.

\begin{figure}
\begin{center}
\includegraphics[angle=0,width=8.5cm,
                 clip=true,trim= 0.5cm 5.0cm 0.5cm 2.5cm]
{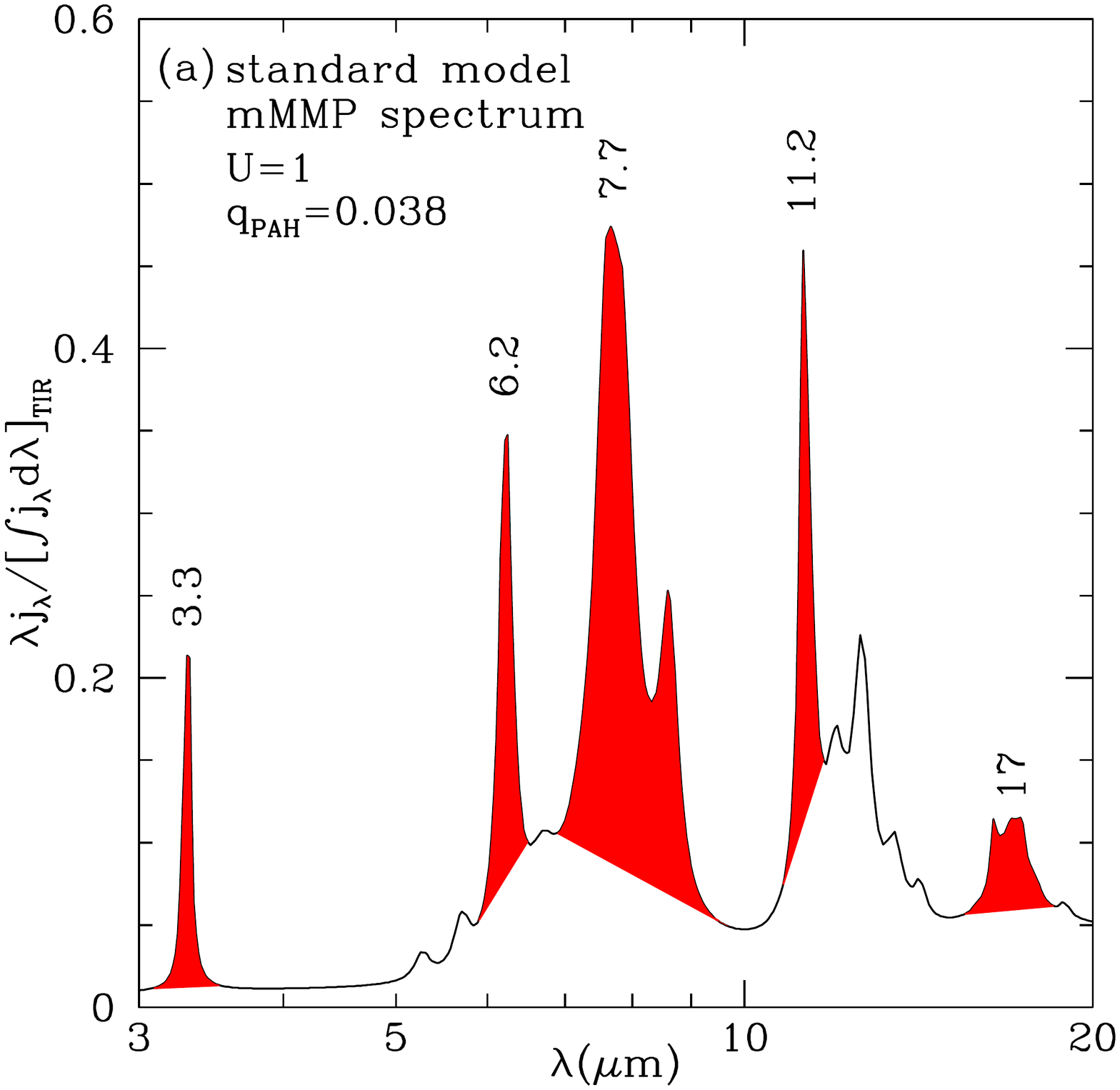}
\includegraphics[angle=0,width=8.5cm,
                 clip=true,trim= 0.5cm 5.0cm 0.5cm 2.5cm]
{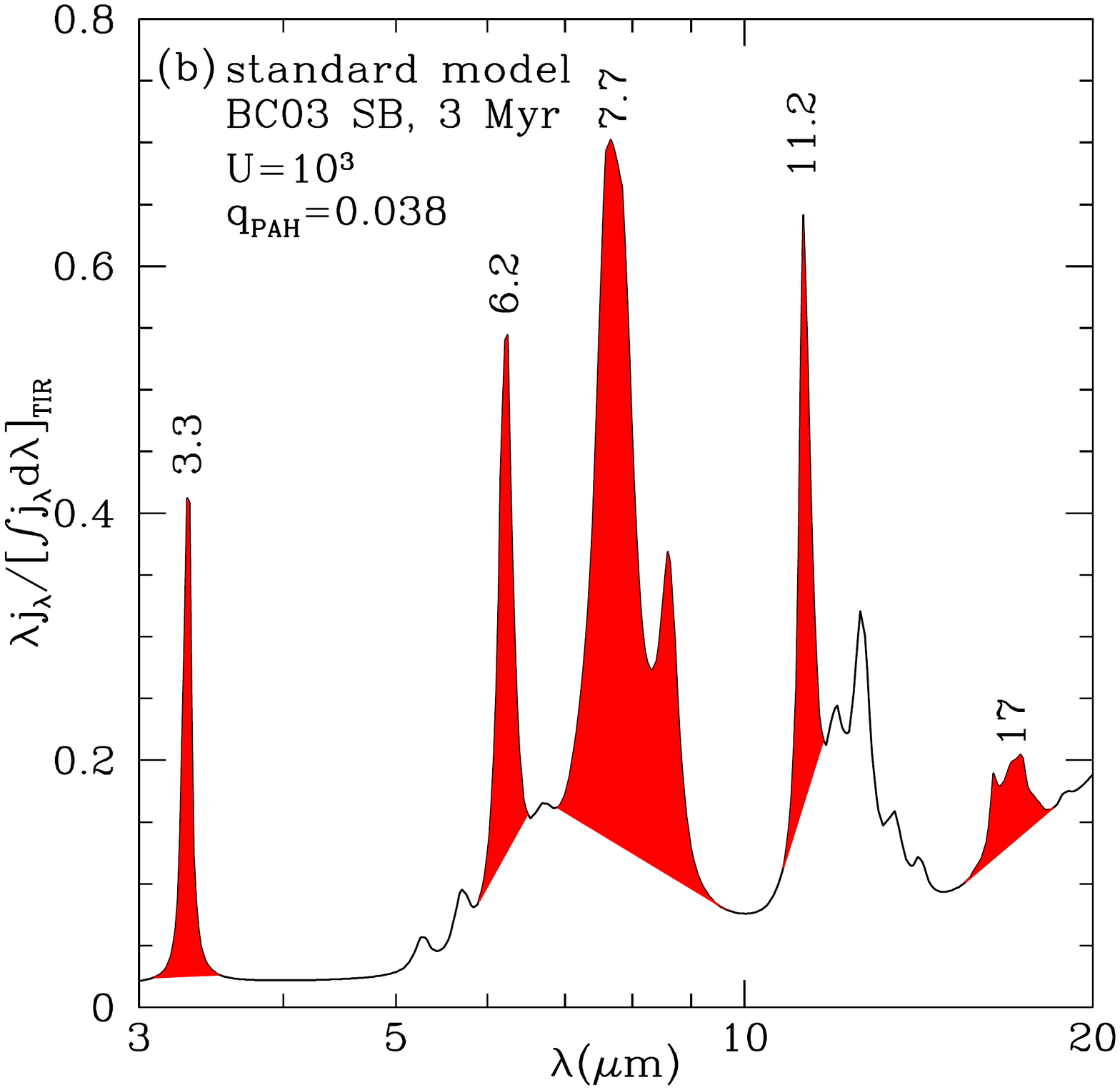}
\caption{\label{fig:PAH_features}\footnotesize
         Normalized emission spectra for our standard model.
         (a) For the mMMP starlight
         spectrum with $U=1$.
         The red line indicates the baselines used for extraction of
         the power is various emission features.
         (b) For heating by the $h\nu<13.6\eV$ radiation from a
         3\,Myr-old starburst, with $U=10^3$.
         }
\end{center}
\end{figure}

Component-fitting procedures such as PAHFIT provide the best
measure of the power $F$ radiated in a spectral
feature, but require sophisticated
fitting of a multicomponent model to the measured spectra.
Since our modeled emission spectra do not include contributions from direct
starlight, line emission, differential attenuation of the emission by
dust (with the 10$\micron$ silicate feature), and emission from other hot dust
(e.g., from AGN), a simpler approach can illustrate the main trends.
Here we adopt a very simple method, to obtain 
the ``clipped'' flux $\Fclip$ in a feature: 
we specify points $\lambda_1$ and $\lambda_2$
on either side of the feature where the feature strength
will be taken to be zero, 
define a ``clip-line'' $\lambda F_\lambda^{\rm (c.l.)}$
between these two wavelengths
to be 
a linear function of $\log\lambda$ connecting $\lambda F_\lambda$
at the clip points
(see Figure \ref{fig:PAH_features}), and define
\beq
\Fclip({\rm band}) \equiv 
\int_{\lambda_1}^{\lambda_2} 
\left( F_\lambda - F_\lambda^{\rm (c.l.)}\right) d\lambda
~~~.
\eeq
Our adopted ``clip points'' $\lambda_1,\lambda_2$ 
for each feature or ``band''
are given in Table \ref{tab:features}.
We treat the broad $7.7\micron$ complex 
(with overlapping subfeatures at $7.417$, $7.598$ and $7.85\micron$) 
{\it and} the $8.6\micron$
feature as a single ``$7.7\micron$'' feature
extending from $6.9$ to $9.7\micron$.  
Similarly, we aggregate emission features at
$15.9$, $16.4$, $17.04$, and $17.375\micron$ into a single
``$17\micron$'' feature
extending from $15.5$ to $18.5\micron$.
Values of $\Fclip({\rm band})/\FTIR$ for our
standard model heated by the mMMP ISRF with $U=1$
are given for five features in Table \ref{tab:features}.

\begin{table}
\begin{center}
\footnotesize
\caption{\label{tab:features}Selected PAH Emission Components}
\begin{tabular}{|cccc|}
\hline
& \multicolumn{2}{c}{\underline{\quad Clip Points \quad}}&\\
Feature      & $\lambda_1 (\micron)$ & $\lambda_2 (\micron)$ & 
$\Fclip/\FTIR$$^a$ \\
\hline
$3.3\micron$ & $3.09$ & $3.52$ & $0.0044$ \\
$6.2\micron$ & $5.90$  & $6.50$  & $0.0130$ \\
$7.7\micron$ & $6.90$  & $9.70$  & $0.0394$ \\
$11.2\micron$& $10.80$ & $11.70$ & $0.0103$ \\
$17\micron$  & $15.50$ & $18.50$ & $0.0061$ \\
\hline
\multicolumn{4}{l}{$^a$ Standard model ($\qpah=0.0379$),
 mMMP starlight, $U=1$.}
\end{tabular}
\end{center}
\end{table}
Figure \ref{fig:PAH_features}a shows our model spectrum for dust heated by
the mMMP starlight spectrum, with our standard PAH size distribution,
and our standard PAH ionization fraction $\fion$;
Figure \ref{fig:PAH_features}b shows the same dust model but heated by
starlight with the spectrum of a 3 Myr-old starburst and 
heating parameter $U=10^3$.
We examine the strengths of the 5 features listed in Table \ref{tab:features}.
Figure \ref{fig:PAH_features} shows these 5 features with 
the adopted baselines.
Because the real feature profiles may have
broad wings (note the substantial power below the red
``features'' in Figure \ref{fig:PAH_features}),
the present approach will significantly 
underestimate the actual power in the features.
The present approach does, however, provide a simple
systematic way to quantify
feature strengths, thereby allowing us to discuss variations of those 
feature strengths, and to compare to observed spectra.

\begin{figure}
\begin{center}
\includegraphics[angle=0,width=10.0cm,
                 clip=true,trim= 0.5cm 5.0cm 0.5cm 2.5cm]
{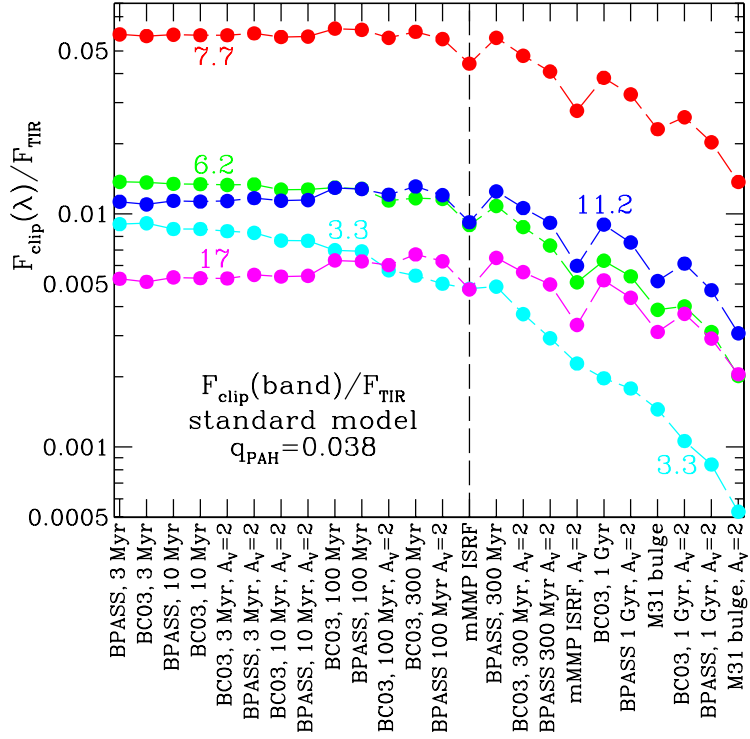} 
\caption{\label{fig:bandstrengths_vs_starlight}\footnotesize
         Normalized ``clipped'' band strengths 
         for the $3.3\micron$, $6.2\micron$,
         and $11.2\micron$ features, and for the $7.7\micron$ and
         $17\micron$ complexes (see Figure \ref{fig:PAH_features}).
         Models are calculated for unattenuated starlight spectra with
         $U=1$, and for $A_V=2$ clouds with $U_{\rm surface}=1$.
         Cases are shown in order of decreasing $\Fclip(3.3)/\FTIR$.
         }
\end{center}
\end{figure}

\begin{figure}
\begin{center}
\includegraphics[angle=0,width=8.5cm,
                 clip=true,trim= 0.5cm 5.0cm 0.5cm 2.5cm]
{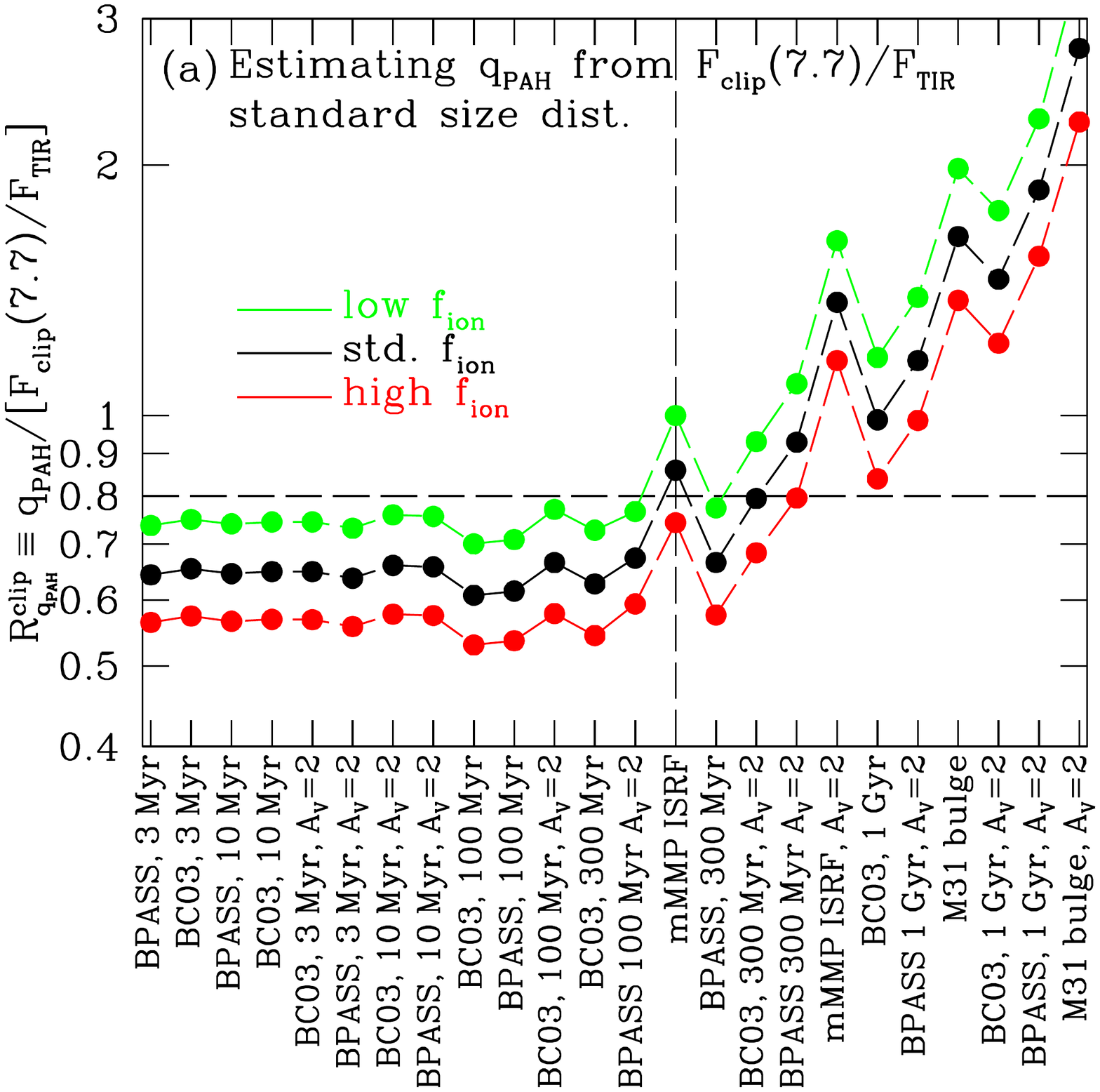} 
\includegraphics[angle=0,width=8.5cm,
                 clip=true,trim= 0.5cm 5.0cm 0.5cm 2.5cm]
{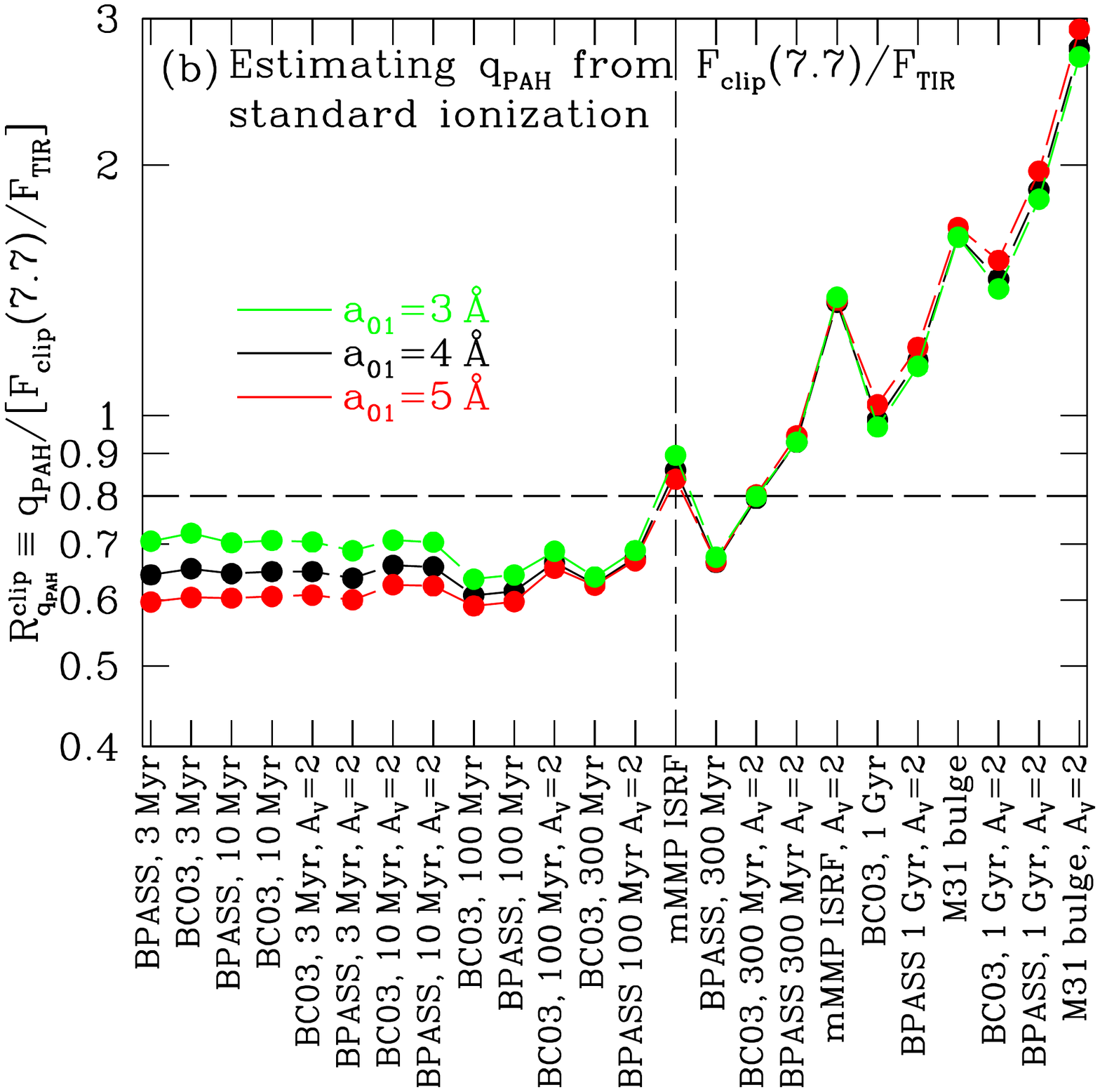} 
\caption{\label{fig:qpah_estimator}\footnotesize
         $R_{\qpah}^{\rm clip}\equiv\qpah/[\Fclip(7.7)/\FTIR]$, 
         where $\Fclip(7.7)$ is extracted as shown
         in Figure \ref{fig:PAH_features}.
         (a) Standard size distribution, for
         a range of radiation fields (see text), and three different
         assumptions about the PAH ionization $\fion$ 
         (see Fig.\ \ref{fig:fion}b).
         For radiation fields appropriate to star-forming galaxies,
         $R_{\qpah}^{\rm clip}\approx 0.8_{-0.25}^{+0.3}$, allowing $\qpah$ to
         be estimated reliably from the measured $\Fclip(7.7)/\FTIR$.
         However, for very red starlight spectra (e.q., M31 bulge),
         $R_{\qpah}^{\rm clip}$ increases, 
         because $\Fclip(7.7)/\FTIR$ is suppressed.
         (b) Standard ionization, for a range of radiation fields,
         and three different values of $a_0$ (see Fig.\ \ref{fig:dnda}a).
         }
\end{center}
\end{figure}
\subsection{\label{subsec:sensitivity_to_starlight_spectrum}
            Sensitivity to Starlight Spectrum and Reddening}

The sensitivity of the different emission bands to the spectrum of the
starlight heating the dust and PAHs is explored in 
Figure \ref{fig:bandstrengths_vs_starlight}, which shows feature strengths
(normalized by the total IR power) 
$\Fclip({\rm band})/\FTIR$ for 24 
different examples of starlight
heating.  All cases are for $U=1$, but 12 
different starlight spectra are
considered.
For each starlight spectrum, we compute the IR emission for
diffuse dust, and also for dust in $A_V=2$\,mag clouds with the starlight
incident on the cloud surface.
For each of the single-age stellar populations (3, 10, 100,
300 Myr, and 1 Gyr) we consider both BC03 \citep{Bruzual+Charlot_2003} 
and BPASS \citep{Eldridge+Stanway+Xiao+etal_2017,Stanway+Eldridge_2018}
stellar models.
In Figure \ref{fig:bandstrengths_vs_starlight} 
the cases are ordered by
decreasing fractional power in the $3.3\micron$ feature.

As the (unreddened) radiation spectrum varies from a 3\,Myr old starburst to 
the M31 bulge, $\Fclip(3.3)/\FTIR$ drops by a factor
$\sim$$6$, from 
0.91\% to 0.15\%.
Other bands are less sensitive -- 
$\Fclip(7.7)/\FTIR$ declines by only a factor 
$\sim$$2.5$, from 5.8\% to 2.3\%.
Thus band ratios such as $\Fclip(3.3)/\Fclip(7.7)$ are sensitive to the
illuminating spectrum, as will be further examined below.

For the ``cloud'' cases, reddening of the starlight within the cloud
results in the total emission from the cloud having fractional
band powers $\Fclip/\FTIR$ that are lower than the values for the
unreddened incident spectrum.  For example, compare
the mMMP ISRF and mMMP ISRF, $A_V=2$ cases 
in Figure \ref{fig:bandstrengths_vs_starlight}: 
$\Fclip(3.3\micron)/\FTIR$ is lower 
by a factor of $\sim$2 for the $A_V=2$ cloud.
However, if the incident starlight spectrum is dominated by UV, as for
starbursts with ages $\ltsim 10\Myr$, reddening leads to only a small
decrease in $\Fclip({\rm band})/\FTIR$, because
the starlight power is dominated by far-UV radiation -- the PAHs and
the dust are heated by the same photons.  Even the
UV-sensitive $3.3\micron$
band is minimally affected: $\Fclip(3.3\micron)/\FTIR$ is reduced
by only 10\% in going from unreddened
BC03 $3\Myr$ radiation to the case of an $A_V=2$ cloud 
(see Figure \ref{fig:bandstrengths_vs_starlight}).

\subsection{$\qpah$ Estimation}

The PAH abundance $\qpah$ can be 
estimated from the strength of the observed PAH
emission features.  $\qpah$ estimation is usually done using the
$7.7\micron$ feature, because it is the strongest, and also because it
was well-matched to 
band 4 of the Infrared Array Camera 
\citep[IRAC;][]{Fazio+Hora+Allen+etal_2004}, 
allowing the $7.7\micron$ feature to be
measured efficiently by Spitzer Space Telescope.
$\qpah$ is taken to be proportional to the fraction of the total IR
power appearing in the $7.7\micron$ feature:
\beq
q_{\rm PAH}
=  R_{q_{\rm PAH}}^{\rm clip} \times 
\frac{\Fclip(7.7)}{\FTIR}
~~~,
\eeq
where the factor 
$R_{q_{\rm PAH}}^{\rm clip}$ 
is 
obtained by modeling the PAH emission spectrum from a model with 
known $\qpah$ (e.g., Table \ref{tab:features}).  
In this paper we extract feature fluxes $\Fclip$ using the simple method
described in \S\ref{subsec:feature strengths}.
The factor $R_{q_{\rm PAH}}^{\rm clip}$ 
will in general depend on the spectrum of the starlight responsible for
heating the PAHs and dust, and also on both the
state of ionization and the size distribution
of the PAHs.

Figure \ref{fig:qpah_estimator}a shows 
$R_{q_{\rm PAH}}^{\rm clip}$ 
for a number of different starlight spectra, and for three different assumed
PAH ionization functions $\fion$.
While $R_{\qpah}^{\rm clip}$ varies among the different starlight spectra,
it is gratifying to see that the variations
in $R_{\qpah}^{\rm clip}$ are modest over 
a wide range of starlight spectra that might be appropriate
in star-forming galaxies, ranging from a very young 
3 Myr old starburst
to a 300 Myr old starburst,
with the mMMP radiation field falling in between:
we can generally take 
$R_{\qpah}^{\rm clip}\approx 0.8_{-0.25}^{+0.3}$ 
provided that the stellar population 
is not extremely evolved.  
For the very extreme case of the starlight from the M31 bulge population, 
we have $R_{\qpah}^{\rm clip}\approx 1.6$; if this 
starlight is
reprocessed by dust clouds with $A_V\approx 2$, $R_{\qpah}^{\rm clip}$ 
rises to 
$R_{\qpah}^{\rm clip}\approx2.8$.

Figure \ref{fig:qpah_estimator}a shows that
variations in the PAH ionization also affect $R_{\qpah}^{\rm clip}$, 
because PAH neutrals  
do not radiate as strongly as PAH cations in the $7.7\micron$ C-C band.
Thus the low $\fion$ models have higher 
$R_{\qpah}^{\rm clip}$ values than
the standard model. However, 
$R_{\qpah}^{\rm clip}$ 
is more sensitive to changes in the starlight spectrum than
to variations in $\fion$.

The sensitivity of $R_{\qpah}^{\rm clip}$ to
the assumed size distribution is examined in Figure \ref{fig:qpah_estimator}b.
$R_{\qpah}^{\rm clip}$
varies by only $\sim$$\pm5\%$ for $a_{01}=4\pm1\Angstrom$.
From Figure \ref{fig:qpah_estimator} we see that if we have a good way to
estimate the spectrum of the starlight exciting the PAH emission, we can
estimate $\qpah$ to within $\sim$$\pm10\%$ accuracy from 
$\Fclip(7.7)/\FTIR$.

\subsection{Sensitivity to PAH Size Distribution}

As discussed in Section \ref{sec:pah_size}, 
the band strengths are also sensitive to the properties of the PAH
population, particularly the size distribution and the fractional
ionization $\fion(a)$.
Because the peak temperature reached following absorption of a single
UV photon is determined by the heat capacity of the PAH, 
the emission spectrum depends on the PAH size.
Figure \ref{fig:pah_mmp} shows emission spectra for 
selected PAH sizes when illuminated by the mMMP 
ISRF: the emission shifts systematically 
to longer wavelengths as PAH size increases.
In addition, the PAH properties may themselves change systematically with
size.  \citet{Shannon+Boersma_2019} discussed the effect of PAH size
on the shape of the $7.7\micron$ complex.

\begin{figure}
\begin{center}
\includegraphics[angle=0,width=8.5cm,
                 clip=true,trim= 0.5cm 0.2cm 0.5cm 0.5cm]
{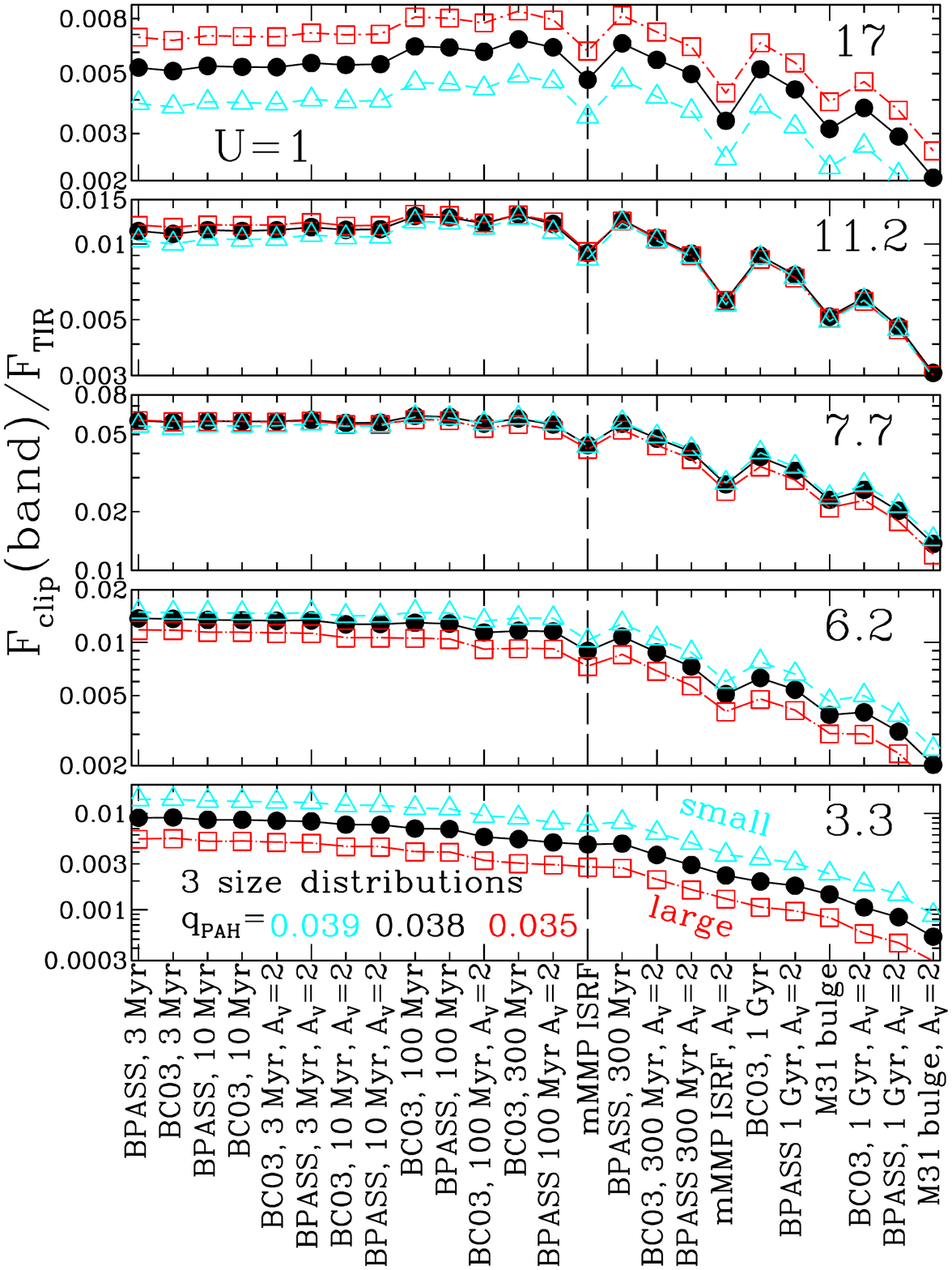}
\includegraphics[angle=0,width=8.5cm,
                 clip=true,trim= 0.5cm 0.2cm 0.5cm 0.5cm]
{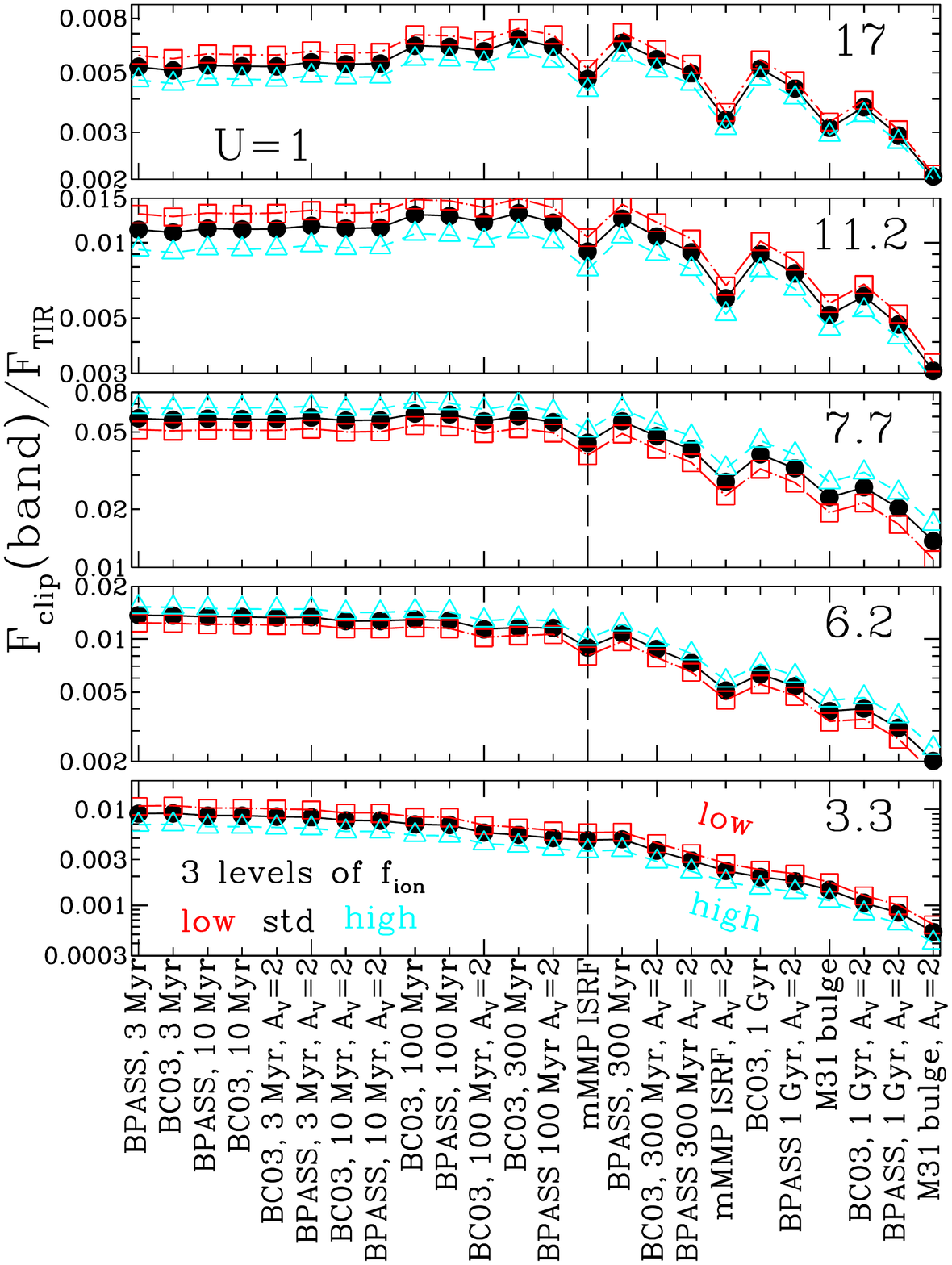}
\caption{\label{fig:PAH_bandstrengths}\footnotesize
         Fractional power $\Fclip/\FTIR$ 
         in selected PAH bands for dust model heated by
         different starlight spectra with $U=1$.
         (a) Standard ionization and the 3 size distributions from
         Fig.\ \ref{fig:dnda}a. 
         (b) Standard size distribution and the 3 ionization
         functions $\fion$ in Fig.\ \ref{fig:fion}b.
         }
\end{center}
\end{figure}

The sensitivity to the PAH size distribution is explored in
Figure \ref{fig:PAH_bandstrengths}a, where 
filled symbols are for
the standard size distribution, 
open squares are for the PAH size 
distribution shifted to peak at larger sizes, 
and open triangles 
are for the PAH size distribution shifted to smaller sizes.
Note that in all cases we hold the lower cutoff fixed at 
$a_{\rm min}=4.0\Angstrom$.

The $3.3\micron$ feature is sensitive to 
variations in the size distribution, because the $3.3\micron$ emission is 
dominated by the smallest PAHs -- 
those with  sufficiently small heat capacities such
that a single $\sim$$10\eV$ photon can heat the PAH to temperatures
$T\gtsim600\K$ where it can radiate effectively at $3.3\micron$.
Size distributions 
shifted to smaller sizes lead to relatively
stronger $3.3\micron$ emission.

In the single-photon heating limit ($U\ltsim 10^3$),
larger PAHs are most efficient for converting absorbed starlight energy
into emission in the 17$\micron$ feature 
(see Figure \ref{fig:conversion_efficiency})
and therefore shifting the PAH size distribution toward larger sizes
raises the $17\micron$ feature strength relative to the other PAH features.

\subsection{\label{subsec:sensitivity_to_ionization}
            Sensitivity to PAH Ionization}

The sensitivity to the PAH ionization balance is explored in Figure
\ref{fig:PAH_bandstrengths}b; 
filled symbols are for the standard
size-dependent ionization fraction $\fion(a)$, while 
open triangles and squares 
are for the ``high'' and ``low'' ionization fractions
illustrated in Figure \ref{fig:fion}.
The model postulates that neutral PAHs have enhanced opacity 
(relative to PAH cations) in the 
$3.3\micron$ C-H stretch,
and reduced opacity (relative to cations) in the $6.2\micron$
and $7.7\micron$ C-C stretching modes.  Models with
``low'' $\fion$ therefore
have stronger emission at $3.3\micron$ and $11.2\micron$,
while models with ``high'' $\fion$ have lower emission in those two
bands.
Models with high $\fion$ have increased emission in the
$6.2\micron$ and $7.7\micron$ bands.
Thus, band ratios such as $F(3.3)/F(7.7)$ or 
$F(11.2)/F(7.7)$ are diagnostic of the
environmental conditions determining the PAH ionization balance.

\begin{figure}[t]
\begin{center}
\includegraphics[angle=0,width=8.5cm,
                 clip=true,trim= 0.5cm 5.0cm 0.5cm 2.5cm]
{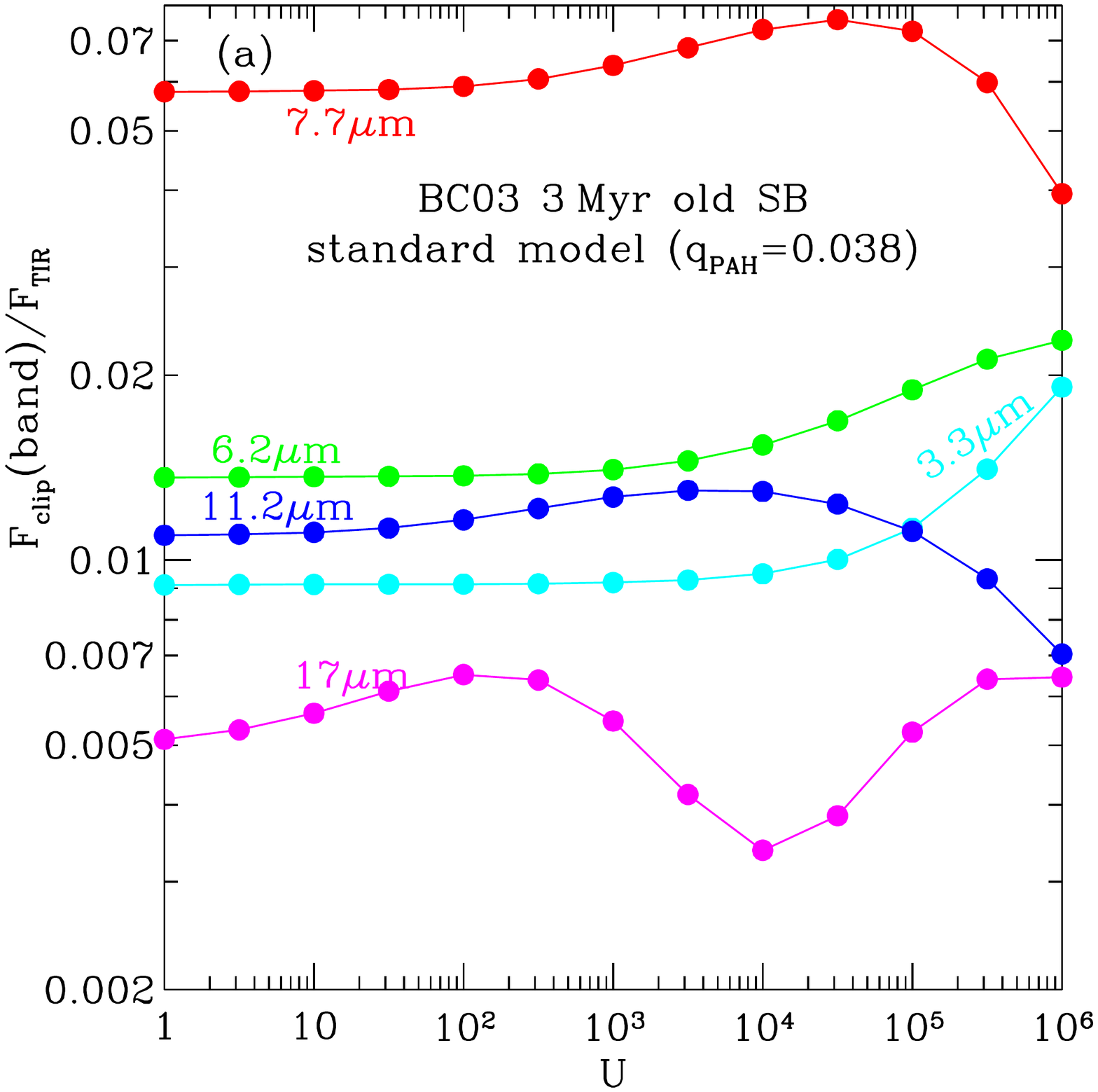}
\includegraphics[angle=0,width=8.5cm,
                 clip=true,trim= 0.5cm 5.0cm 0.5cm 2.5cm]
{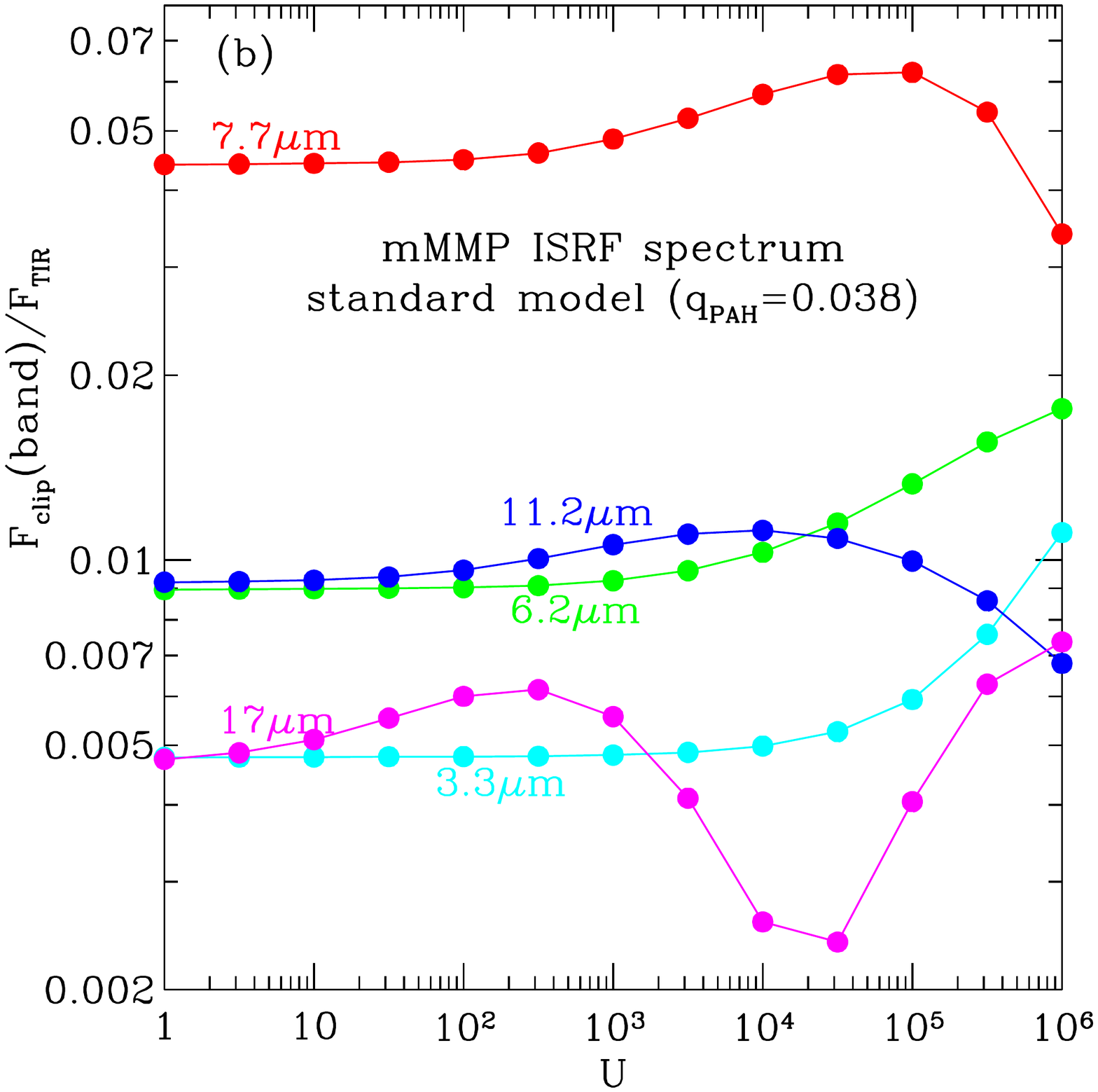}
\caption{\label{fig:PAH_bandstrengths_vs_U}\footnotesize
         ``Clipped'' fractional power in selected 
         PAH bands 
         (see Figure \ref{fig:PAH_features}) 
         for dust model as a function
         of starlight intensity parameter $U$,
         for standard size distribution, standard ionization and 
         starlight from (a) 3 Myr old starburst,
         (b) mMMP interstellar radiation field.
         }
\end{center}
\end{figure}
\begin{figure}
\begin{center}
\includegraphics[angle=0,width=8.5cm,
                 clip=true,trim=0.5cm 5.0cm 0.5cm 2.5cm]
{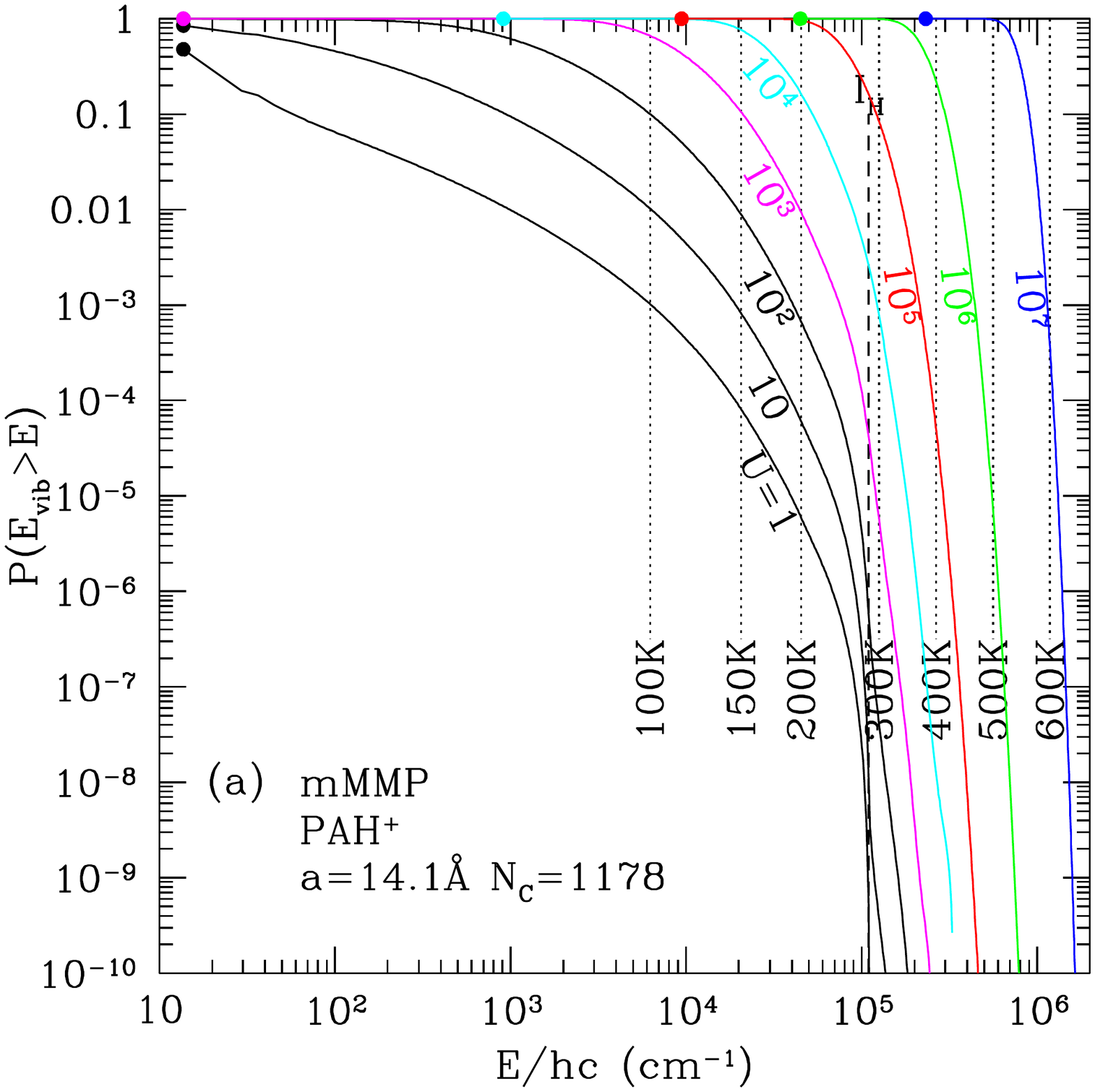}
\includegraphics[angle=0,width=8.5cm,
                 clip=true,trim=0.5cm 5.0cm 0.5cm 2.5cm]
{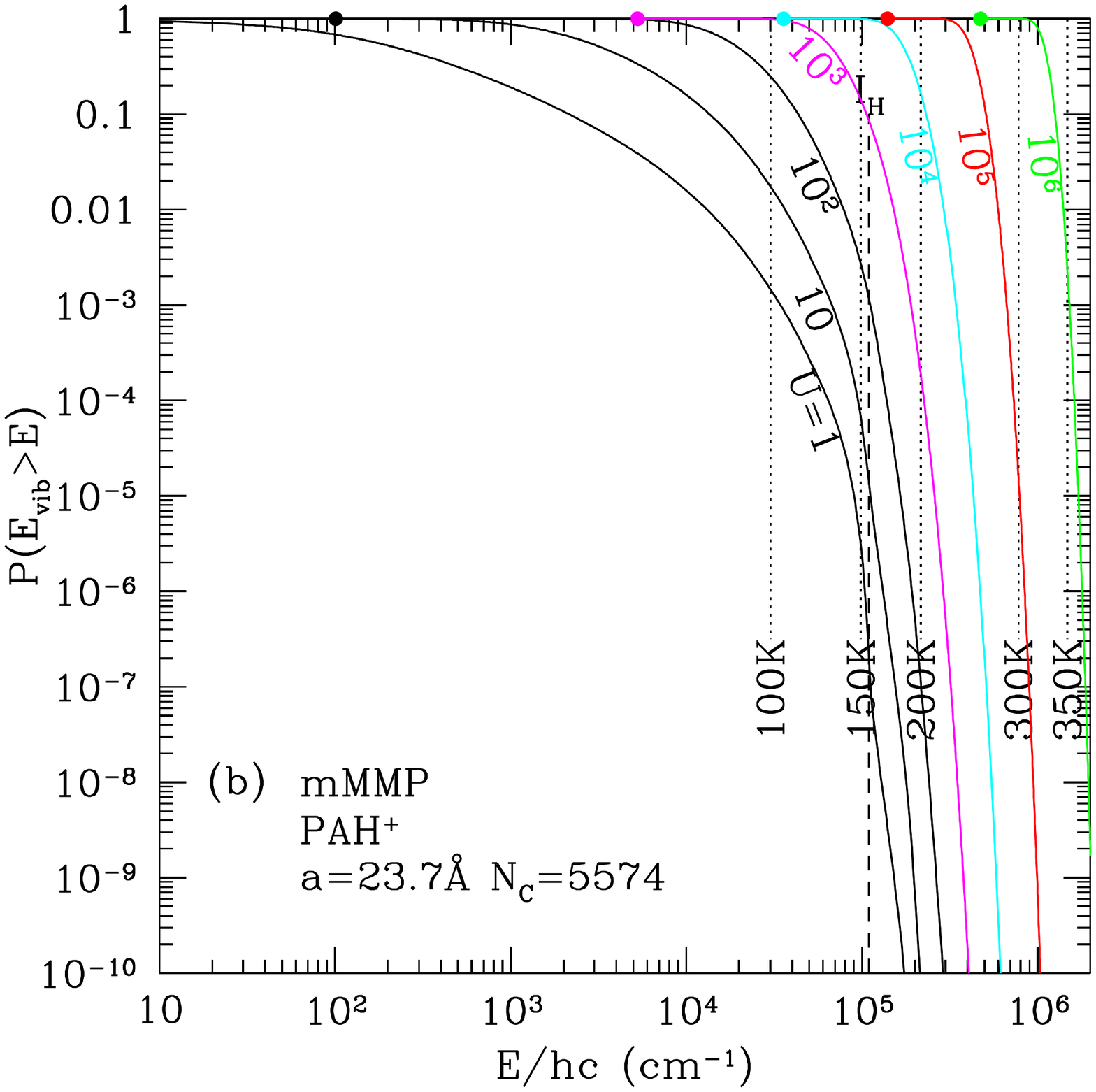}
\caption{\label{fig:large_dpdt}\footnotesize
         Probability $P(E_{\rm vib}>E)$ of having vibrational
         energy greater than $E$,
         for (a) $a=14.1\Angstrom$ ($N_{\rm C}=1180$).
         (b) $a=23.7\Angstrom$ ($N_{\rm C}=5600$).
         Vertical dashed line is $I_{\rm H}=13.6\eV$, the highest photon
         energy present in the illuminating starlight.
         For $U=10^3$ these 2 nanoparticles spend $\sim$$10\%$ of the time
         at $T> 150\K$, radiating effectively in the $17\micron$ complex.
         For $U\geq10^5$, both nanoparticles remain at $T\gtsim 200\K$,
         radiating effectively at $\lambda < 15\micron$.
         }
\end{center}
\end{figure}

\subsection{\label{subsec:sensitivity_to_heating}
            Sensitivity to Starlight Intensity}

At low starlight intensities, the PAH emission is 
excited by single-photon heating, 
and the fractional power emitted in each of the
bands does not depend on the intensity of the starlight -- only on its
spectral shape.
Figure \ref{fig:PAH_bandstrengths_vs_U} 
shows the fractional power $F({\rm band})/\FTIR$
in each of the
five emission features as a function of $U$ for our standard model, with 
Figure \ref{fig:PAH_bandstrengths_vs_U}a calculated for the quite hard
spectrum of a 3 Myr-old starburst, and Figure \ref{fig:PAH_bandstrengths_vs_U}b
calculated for the mMMP spectrum.
For the shortest wavelength bands (e.g., $3.3\micron$),
$\Fclip({\rm band})/\FTIR$ remains relatively constant 
until $U$ reaches very high values.
For longer wavelength bands, $\Fclip({\rm band})/\FTIR$ 
begins to rise when
some part of the PAH population remains warm enough between photon
absorption events to be able to radiate in the band.

As the starlight intensity parameter rises above $U=1$, 
the first band affected
is the $17\micron$ complex, with 
$\Fclip(17)/\FTIR$ 
initially rising, then
dropping as $U$ increases above $\sim$$300$,
and then rising again for $U>10^{4.5}$.  
The $17\micron$ feature is efficiently radiated by
PAH nanoparticles with $N_{\rm C}\approx10^3$ -- $10^4$ C atoms
(see Figure \ref{fig:conversion_efficiency}), and 
for $U\gtsim10^2$ these nanoparticles
do not fully cool between absorption events.  
Thermal emission at $17\micron$ requires temperatures such that
$h\nu/kT \ltsim 5$, or $T\gtsim 170\K$.
Figure \ref{fig:large_dpdt} shows energy distribution functions
for PAHs with $N_{\rm C}\approx 1200$ and $5600$, for selected values of
$U$.  For these two examples, for $U=10^3$ the nanoparticle
spends $\sim$$10\%$ of the time above $T>150\K$, able to radiate in 
the $17\micron$ feature.  This accounts for the initial rise in
$\Fclip(17)/\FTIR$ as $U$ increases to $\sim10^2$.
As $U$ increases beyond $\sim$$10^{2.5}$,
photons absorbed by an already-warm nanoparticle raise it to higher
energies than it would have been able to reach by single-photon heating
at lower $U$.  Therefore,
energy which for $U\ltsim 10^2$ 
would be radiated in
the $17\micron$ complex instead is shifted to shorter wavelengths, e.g.,
the $11.2\micron$ feature.
This explains the drop in $\Fclip(17)/\FTIR$ as
$U$ is increased from $10^2$ to $10^4$.

The decrease in $\Fclip(17)/\FTIR$ to a minimum at $U\approx 10^4$ followed
by a rise to a second peak at $U\approx10^6$ is related to the bimodal
size distribution adopted for the PAHs (see Fig.\ \ref{fig:dnda}a), with 
a second component ($a_{02}=30\Angstrom$) having a 
mass distribution peaking
near $N_C\approx 10^5$.
The larger PAHs in this second component account for the second peak in
$\Fclip(17)/\FTIR$ at $U\approx10^6$.
For very high $U$ the larger PAHs are
heated to $T\gtsim 150\K$ (see Figure \ref{fig:large_dpdt})
and contribute to the $17\micron$ feature,
accounting for the rise in $\Fclip(17)/\FTIR$ 
for $U\gtsim10^{4.5}$ in
Figures \ref{fig:PAH_bandstrengths_vs_U}.
For smoother size distributions, the variation of
$\Fclip(17)/\FTIR$ would have been reduced.
Another complicating factor is that as $U$ reaches $\sim$$10^3$, 
the silicate material in the astrodust grains 
begins to radiate in the $18\micron$ silicate feature
(see Figure \ref{fig:dustmix}a).
The $18\micron$ silicate emission profile interferes with 
the simple method 
used here for extraction of the flux in the $17\micron$ feature, which 
assumes a simple ``baseline'' between $15.5$ and $18.5\micron$.

Similar behavior is seen for other bands.
$\Fclip(11.2)/\FTIR$ initially rises, and then declines for 
$U\gtsim 10^4$ as the PAHs become hot enough to shift power to shorter
wavelengths.
$\Fclip(7.7)/\FTIR$ rises as $U$ increases to $\sim$$10^{4.5}$,
followed by a decline as power is shifted to shorter wavelengths.
$\Fclip(6.2)/\FTIR$ and $\Fclip(3.3)/\FTIR$ have not yet
peaked for the highest intensities $U=10^6$ considered here.

\subsection{\label{subsec:band_ratios}
            Band Ratios}

Above we have investigated how PAH band intensities, 
relative to total infrared (TIR),
are affected by the 
spectrum and intensity of the starlight, and by the PAH size
distribution and ionized fraction.  Because the band intensities are
proportional to the PAH abundance, which can vary, it is useful to
see how PAH band {\it ratios} are affected by the starlight properties,
and by the PAH size distribution and ionization.
We emphasize that, although trends in ratios among features are
conserved, the method employed to recover the feature strengths
will affect the median ratio values, sometimes significantly.
See Smith et al., in preparation, for a full suite of comparisons.

\begin{figure}
\begin{center}
\includegraphics[angle=0,width=8.5cm,
                 clip=true,trim=0.5cm 5.0cm 0.5cm 2.5cm]
{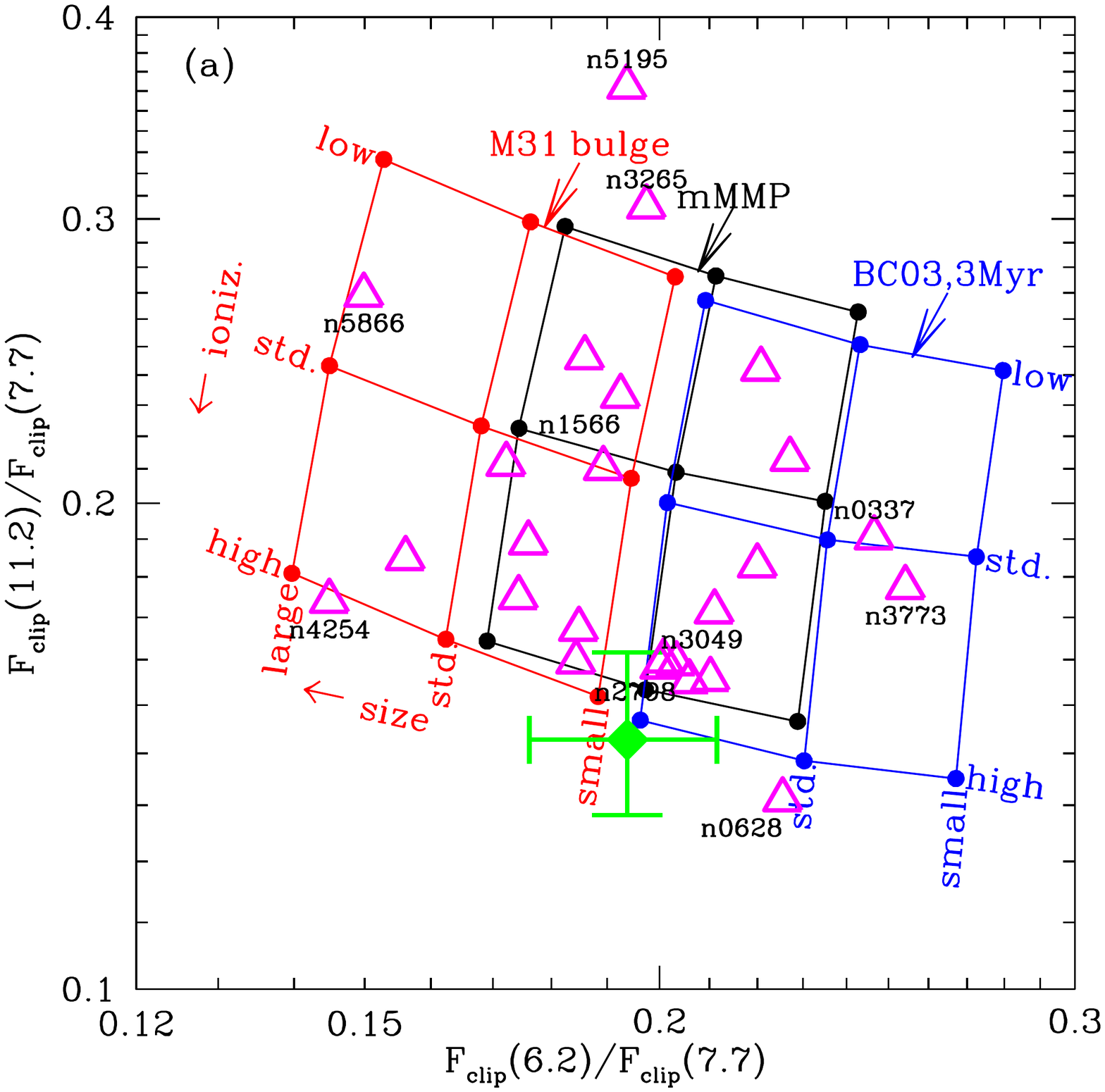}
\includegraphics[angle=0,width=8.5cm,
                 clip=true,trim=0.5cm 5.0cm 0.5cm 2.5cm]
{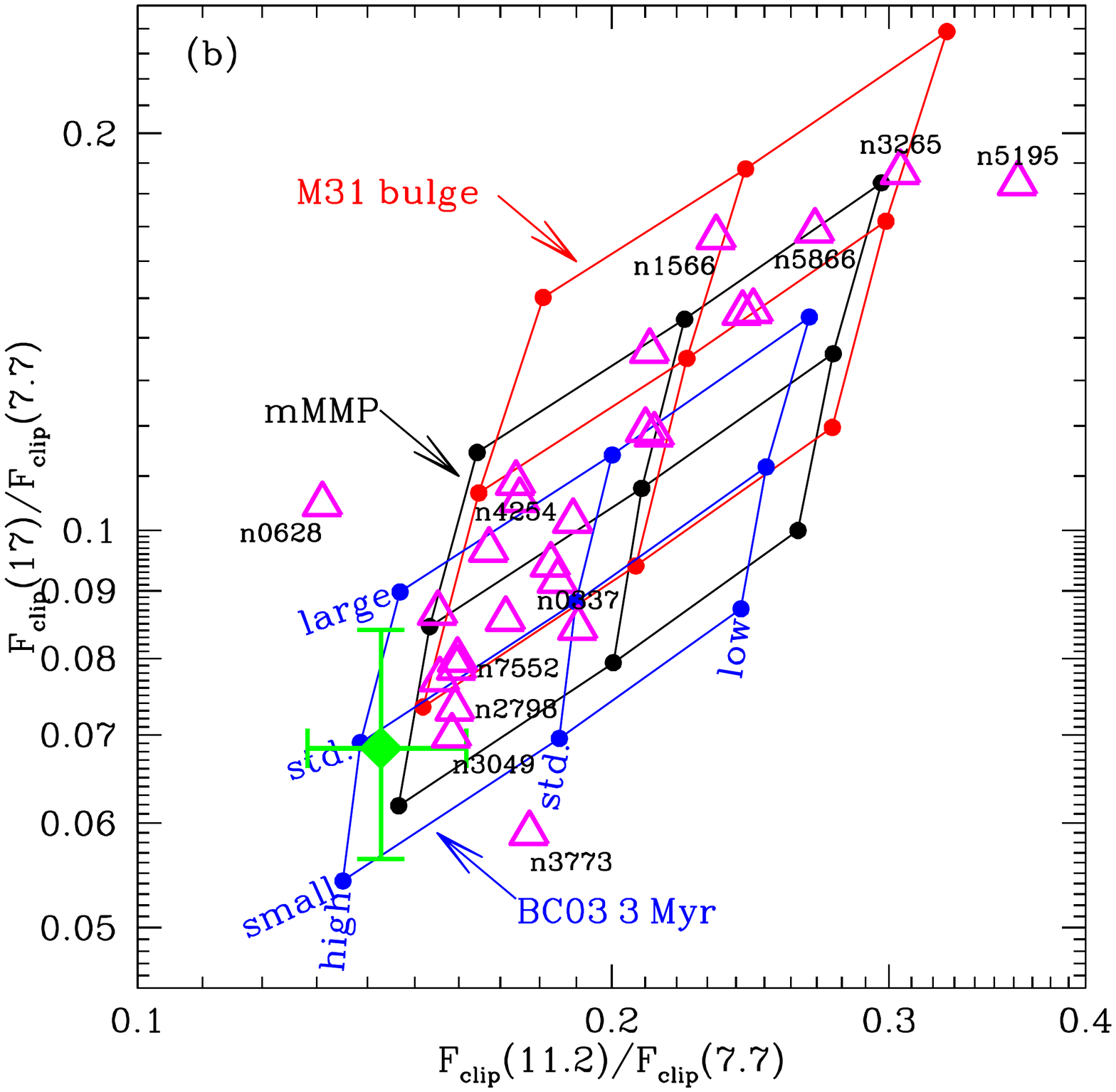}
\includegraphics[angle=0,width=8.5cm,
                 clip=true,trim=0.5cm 5.0cm 0.5cm 2.5cm]
{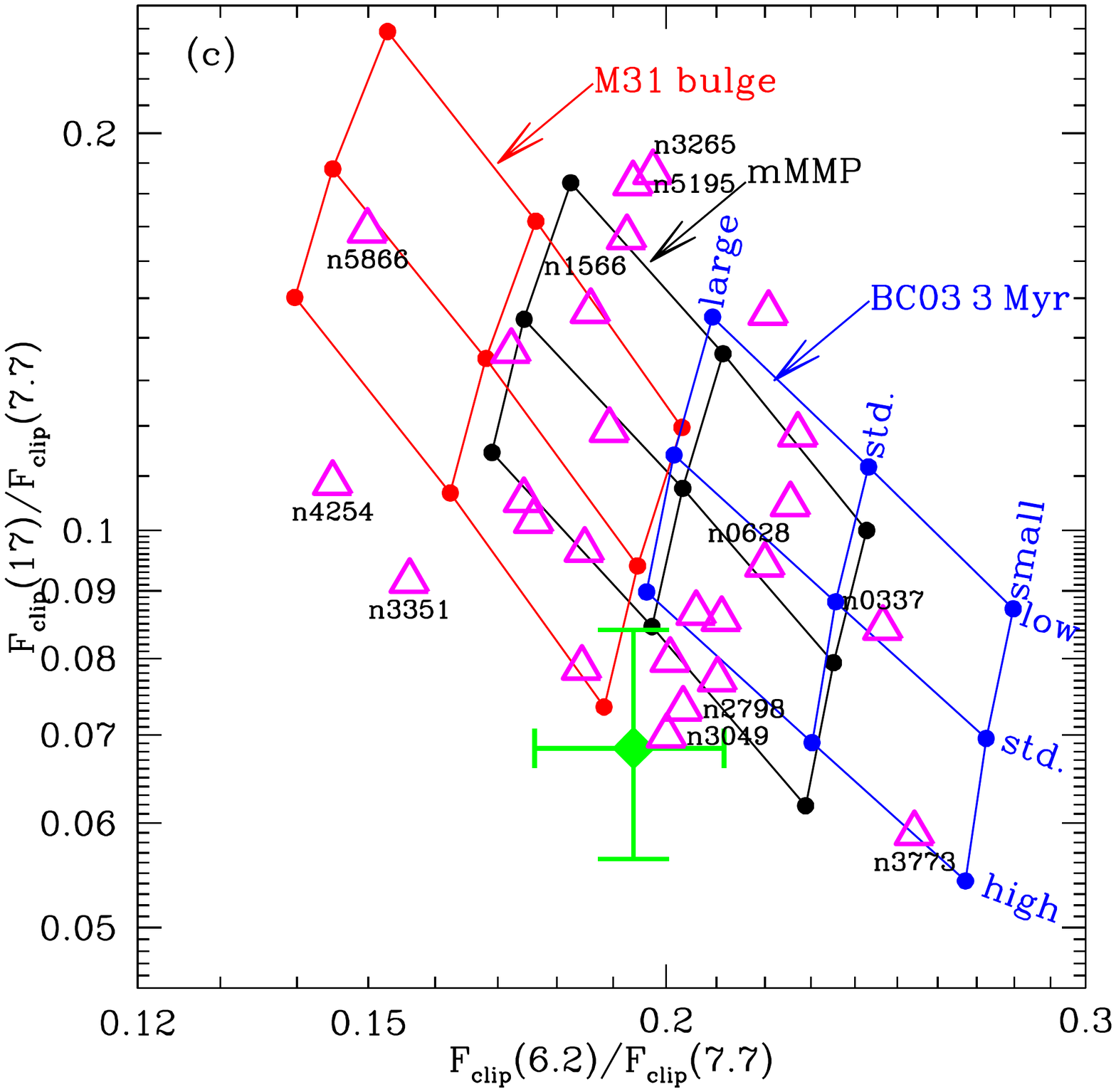}
\includegraphics[angle=0,width=8.5cm,
                 clip=true,trim=0.5cm 5.0cm 0.5cm 2.5cm]
{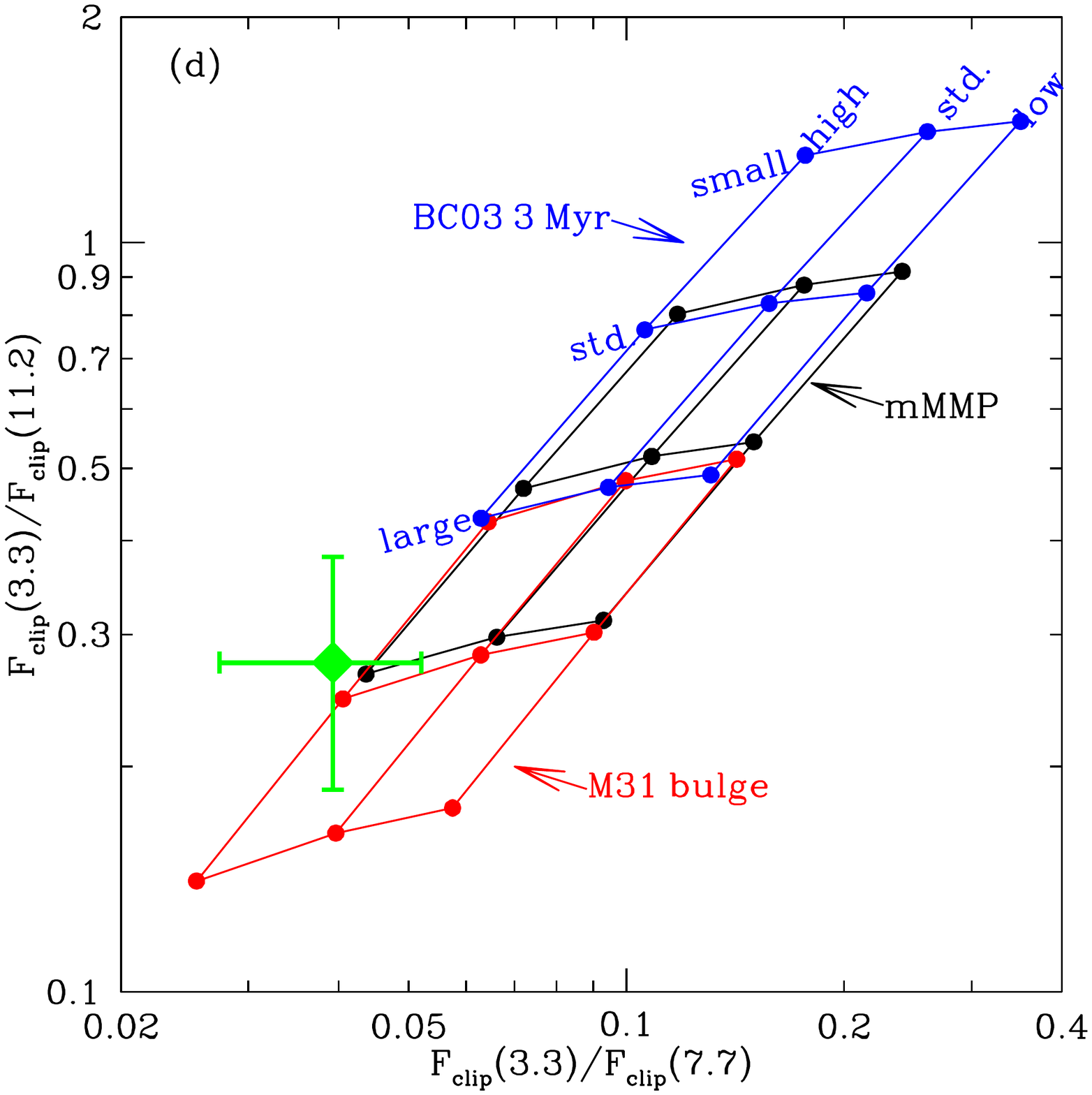}
\caption{\label{fig:band_ratios}\footnotesize
         Selected model PAH band ratios calculated 
         for three different starlight spectra
         (3\,Myr-old starburst, mMMP radiation field, and M31 bulge spectrum).
         Band ratios shown are for $U=1$, but apply for
         $U\ltsim 10^3$.  Results (dots) 
         are shown for three size distributions 
         (``small,'' ``std,'' and ``large''),
         and three ionization models (``low,'' ``std,'' and ``high'').
         Triangles show measured ratios for 25 SINGS galaxies (see text).
         The green
         diamond shows band ratios for the 1C galaxy sample
         of \citet{Lai+Smith+Baba+etal_2020}, with 
         error bars showing the 1st-9th decile range within the 1C sample.
         }
\end{center}
\end{figure}

\citet{Lai+Smith+Baba+etal_2020} present $2.7$--$28\micron$ spectra of
galaxies based on Spitzer and AKARI spectroscopy.
Their ``1C'' sample 
consists of 60 galaxies drawn
from the 113 galaxies in their ``PAH bright'' sample.
The 1C sample galaxies were selected to have
strong PAH emission but weak silicate features 
(either in absorption or emission) 
in order to minimize the effects of reddening.
We have applied our simple feature extraction procedure to the 
\citet{Lai+Smith+Baba+etal_2020} 1C ``template'' spectrum 
after subtraction of emission lines from ions and H$_2$ 
(Thomas Lai 2020, private communication).
The green diamonds in Figures \ref{fig:band_ratios}a-d 
show the observed 
band ratios for the 
\citet{Lai+Smith+Baba+etal_2020} 1C galaxy sample.
The error bars shown 
correspond to the 1st and 9th deciles 
for the 1C galaxy sample (Thomas Lai 2020,
private communication).

We have also applied the feature extraction method described
in Section \ref{subsec:feature strengths} to 25
galaxies
from the Spitzer Infrared Nearby Galaxy Survey (SINGS) 
\citep{Kennicutt+Armus+Bendo+etal_2003,Smith+Draine+Dale+etal_2007}, 
after removing emission lines 
(see spectra in Appendix \ref{app:SINGS}).
Attenuation by dust (though modest) was also corrected for.
The band ratios for each galaxy are plotted as 
triangles in Figures \ref{fig:band_ratios}a-c.
Some of the galaxy points are identified.

Each plot shows a grid for each of three
starlight spectra: 
a 3\,Myr-old starburst, the mMMP local ISRF, and the M31 bulge stars.
The 3 Myr-old starburst and M31 bulge spectra 
span the range from very UV-bright to very red, with the mMMP spectrum
falling in-between
(see Figure \ref{fig:isrf}).  The mMMP spectrum is a good
estimate for a typical star-forming galaxy.

The model
band ratios change when the illuminating stellar spectra are varied,
with $\Fclip(6.2)/\Fclip(7.7)$ 
dropping by a factor $\sim$$1.5$,
$\Fclip(3.3)/\Fclip(7.7)$ decreasing by a factor $\sim$$2.5$, and 
$\Fclip(17)/\Fclip(7.7)$ increasing by a factor $\sim$$2$, as the
starlight is varied from the $3\Myr$ starburst to the M31 bulge.
Varying the PAH size distributions from ``small'' to ``large''
affects the band ratios similarly to changing the spectrum from
the 3\,Myr starburst to the M31 bulge.

The band ratios for the SINGS galaxies 
in Figures \ref{fig:band_ratios}a-c 
are generally within the region spanned by the
considered variations in PAH size, ionization, and starlight spectra.
The ``1C'' galaxy sample 
has $\Fclip(6.2)/\Fclip(7.7)$ in the middle of the model
range, but $\Fclip(11.2)/\Fclip(7.7)$ 
is lower than the models, and
lower than 24 of the 25 SINGS galaxies shown.
Similarly, the 1C galaxy sample has $\Fclip(17)/\Fclip(7.7)$ lower
than 24 of the 25 SINGS galaxies shown.

Because the $7.7\micron$ emission is 
primarily from PAH cations, $F(11.2)/F(7.7)$ is
expected to be sensitive to the PAH ionization
and this is seen in Figures \ref{fig:band_ratios}a and b.
As expected,
varying $\fion$ in the model
has little effect on $\Fclip(6.2)/\Fclip(7.7)$
(the $6.2\micron$ feature and the $7.7\micron$ complex 
are both attributed primarily to cations), but
$\Fclip(11.2)/\Fclip(7.7)$ 
decreases by a factor $\sim$$1.5$ as
$\fion$ varies from the ``low'' to ``high'' 
examples in Figure \ref{fig:fion}b.
We see in Figures \ref{fig:band_ratios}a,b that our
standard model heated by the mMMP starlight with
$U\ltsim 10^4$ gives
$\Fclip(11.2)/\Fclip(7.7)$ 
and $\Fclip(6.2)/\Fclip(7.7)$ close to
observed values for the SINGS galaxies, and the 
considered variations in
size distribution and $\fion$ appear able to accomodate the observed
spread in $\Fclip(6.2)/\Fclip(7.7)$ and $\Fclip(11.2)/\Fclip(7.7)$.

Figure \ref{fig:band_ratios}c shows $\Fclip(17)/\Fclip(7.7)$ vs.\
$\Fclip(6.2)/\Fclip(7.7)$.  $\Fclip(6.2)/\Fclip(7.7)$ is relatively
insensitive to ionization, because both features are thought to be dominated
by cations.  However, changing the size, or changing the starlight spectrum,
does affect $\Fclip(6.2)/\Fclip(7.7)$, 
as already seen in Figure \ref{fig:band_ratios}a.

Figure \ref{fig:band_ratios}d shows how
$\Fclip(3.3)/\Fclip(7.7)$ and $\Fclip(3.3)/\Fclip(11.2)$
respond to changes in illuminating spectrum, size distribution,
and ionization.
Harder spectra (e.g., the 3\,Myr SB) lead to higher values of 
$\Fclip(3.3)/\Fclip(7.7)$ and $\Fclip(3.3)/\Fclip(11.2)$
because the larger photon energies lead to higher peak temperatures, enhancing
the $3.3\micron$ emission.
Similarly, smaller grains lead to higher $\Fclip(3.3)/\Fclip(7.7)$
and $\Fclip(3.3)/\Fclip(11.2)$.
Thus, $\Fclip(3.3)/\Fclip(7.7)$ provides information on size and spectrum,
while $\Fclip(11.2)/\Fclip(7.7)$ 
(see Figures \ref{fig:band_ratios}a and b) helps constrain $\fion$.
We note, however, that our model calculations tend to predict values of
$\Fclip(3.3)/\Fclip(7.7)$ 
that are significantly larger than observed
for the 1C sample --
this is further discussed in \S\ref{subsec:modeling 3.3} below.
We also see in Figure \ref{fig:band_ratios}d that the $3.3/11.2$ band ratio
is sensitive to the PAH size distribution, because only the smallest
PAHs become hot enough to radiate at $3.3\micron$ 
(see Figure \ref{fig:conversion_efficiency}).

\begin{figure}[b]
\begin{center}
\includegraphics[angle=0,width=8.5cm,
                 clip=true,trim=0.5cm 5.0cm 0.5cm 2.5cm]
{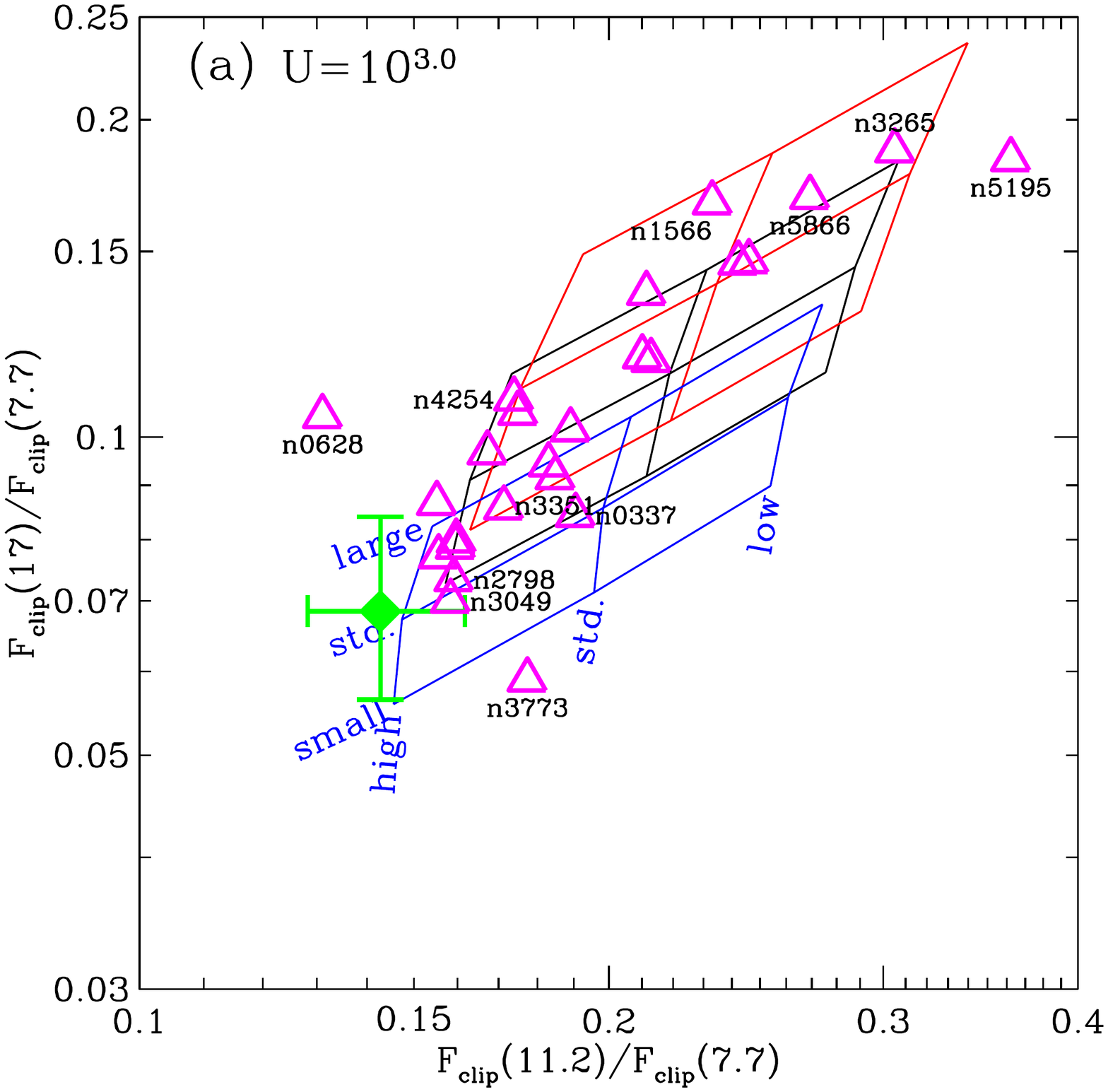}
\includegraphics[angle=0,width=8.5cm,
                 clip=true,trim=0.5cm 5.0cm 0.5cm 2.5cm]
{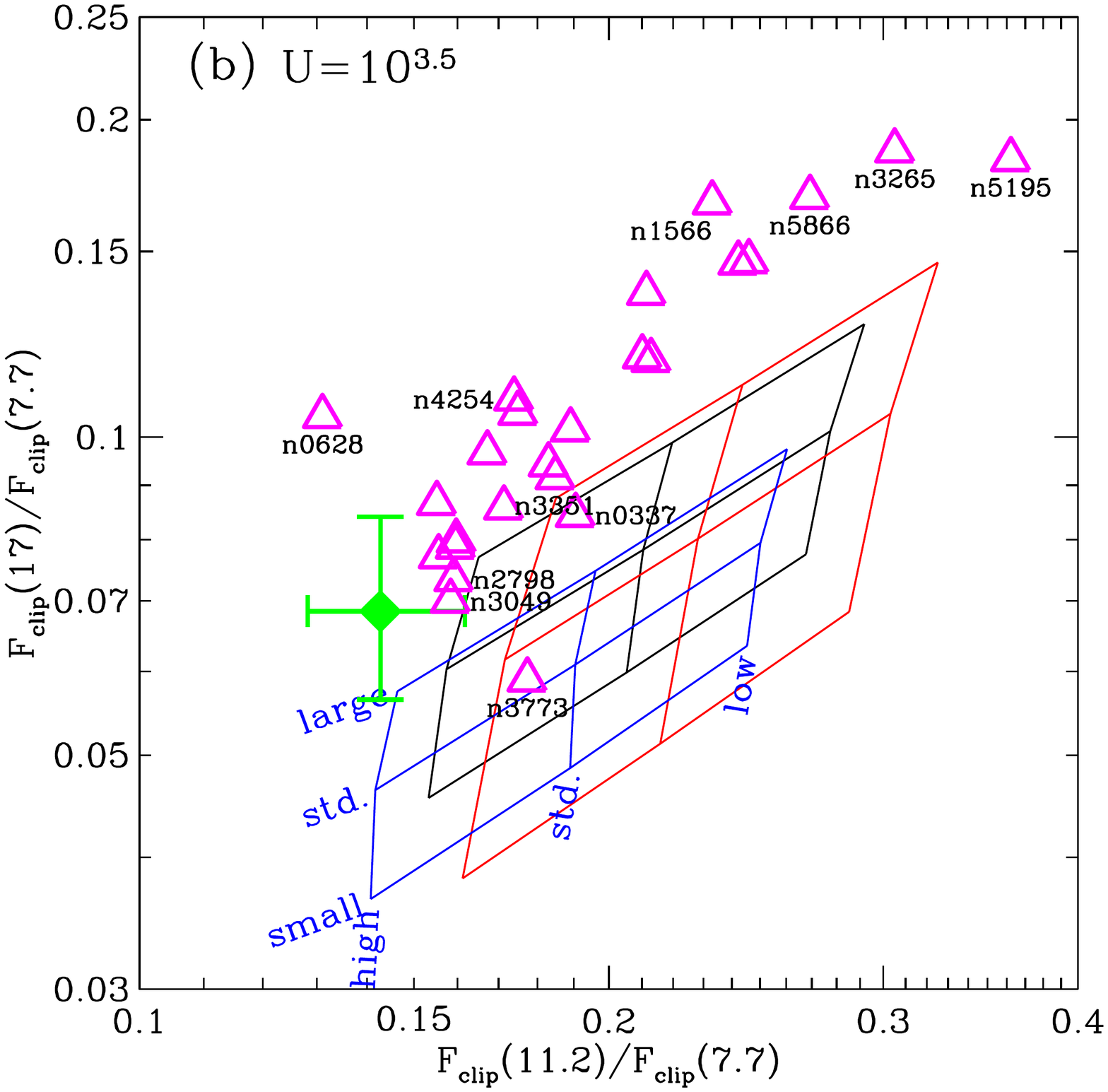}
\caption{\label{fig:band_ratios_2}\footnotesize
         As in Figure \ref{fig:band_ratios}b, 
         but for $U=10^3$ and $U=10^{3.5}$.
         Note the decrease in the relative strength of the 17$\micron$
         feature as $U$ increases beyond $10^3$ 
         (see also Figure \ref{fig:PAH_bandstrengths_vs_U}).
         }
\end{center}
\end{figure}

\subsection{\label{subsec:modeling 17}
         Modeling the 17$\micron$ Emission Feature}

Emission features in the $16-19\micron$ range were reported 
in a number of Galactic objects by 
\citet{vanKerckhoven+Hony+Peeters+etal_2000}.
A  characteristic
$17\micron$ complex of features, 
first identified in NGC\,7331
\citep{Smith+Dale+Armus+etal_2004},
is prominent in the emission from star-forming galaxies
\citep{Smith+Draine+Dale+etal_2007}.
While not yet definitively
identified with specific PAH vibrational modes, its correlation
with other PAH bands makes it appear likely 
that the $17\micron$ feature is also emission from PAHs.
\citet{vanKerckhoven+Hony+Peeters+etal_2000} and
\citet{Boersma+Bauschlicher+Allamandola+etal_2010} discuss the
types of C-C-C bending modes that might be responsible for the 
$16$--$19\micron$ features.
The DL07
PAH model used here includes opacity at $17\micron$ 
consistent with the observed $17\micron$
emission from galaxies.

Figure \ref{fig:band_ratios}b shows
$\Fclip(17)/\Fclip(7.7)$ vs.\ 
$\Fclip(11.2)/\Fclip(7.7)$.
The 1C sample average has
$\Fclip(17)/\Fclip(7.7)$ below the model grids, and
lower than all but one
of the SINGS galaxies shown.
While most of the SINGS galaxies are consistent with our model grids,
some have $\Fclip(17)/\Fclip(7.7)$ below the model grids
in Figure \ref{fig:band_ratios}c.
The 3 galaxies with the lowest values of $\Fclip(17)/\Fclip(7.7)$
are NGC2798, NGC3049, and NGC3773.  Figure \ref{fig:sings} shows that
each of these galaxies has a continuum that is strongly
rising from $20\micron$ to $30\micron$, indicative of heating by
radiation fields with $U\gtsim 10^3$ (see Figure \ref{fig:dustmix} for
model SEDs calculated for $U=10^3$ and $U=10^4$).  
Figure \ref{fig:PAH_bandstrengths_vs_U} showed that $\Fclip(17)/\FTIR$
is expected to decrease as $U$ is increased from $10^3$ to $10^4$, 
suggesting that high $U$ values may
explain the low $\Fclip(17)/\Fclip(7.7)$ seen for some galaxies.

In Figures \ref{fig:band_ratios_2}a,b we
show $\Fclip(17)/\Fclip(7.7)$ and $\Fclip(11.2)/\Fclip(7.7)$
calculated for $U=10^3$ and $10^{3.5}$.
If the heating rate parameter $U$ is increased to $10^{3.0}$ and 
$10^{3.5}$,
$\Fclip(17)/\Fclip(7.7)$ is reduced, but $\Fclip(11.2)/\Fclip(7.7)$ is
hardly affected.  This would be one way to lower $\Fclip(17)/\Fclip(7.7)$
without changing the PAH properties or size distribution.

Thus, the observed low values of $\Fclip(17)/\Fclip(7.7)$ 
for NGC2798, NGC3049,
and NGC3779 might be explained by values of $U$ 
consistent with the observed $20$--$30\micron$ continuum
in these galaxies.

\subsection{\label{subsec:modeling 3.3}
         Modeling the 3.3$\mu$m Emission Feature}
As already seen in Figure \ref{fig:band_ratios}a, the 
values of $\Fclip(11.2)/\Fclip(7.7)$ 
and $\Fclip(6.2)/\Fclip(7.7)$ 
observed for the SINGS galaxies are
in approximate agreement with the model calculations.
However, for our standard size distribution ($a_0=4\Angstrom$),
standard $\fion$, and the mMMP starlight specrum,
the model calculations predict $\Fclip(3.3\micron)/\Fclip(7.7\micron)$
ratios (see Figure \ref{fig:band_ratios}d),
or $\Fclip(3.3\micron)/\Fclip(11.2\micron)$ ratios 
(see Figure \ref{fig:band_ratios}c),
that are significantly larger (by about a factor $\sim$$2$)
than the observed values for the \citet{Lai+Smith+Baba+etal_2020} 1C
spectra.

It is possible that the band strengths adopted
for the 3.3$\micron$ band are too large.
However, the adopted band strengths (from DL07) 
$\int C_{\rm abs}d\lambda^{-1}=(3.94,0.89)\times10^{-18}\cm$ per C-H bond
in (neutral, ionized) PAHs 
appear to be broadly consistent with results from 
theoretical calculations \citep[see, e.g.,][]{Malloci+Joblin+Mulas_2007,
Bauschlicher+Ricca+Boersma+Allamandola_2018,Yang+Li+Glaser_2020b}.
With our adopted band strength, the 
PAH nanoparticles in the model
would account for only $\sim$$1/3$ of the observed interstellar
3.3$\micron$ absorption feature.\footnote{
If ${\rm C/H} = 37 {\rm ppm}$ is in PAHs with H:C::1:3, then
the $3.3\micron$ band strengths adopted here 
for PAH neutrals and cations
(see Table \ref{tab:Drude} in Appendix \ref{app:Cabs for PAHs}) imply 
\beq \nonumber
\frac{\int \Delta \tau\, d\lambda^{-1}}{\NH} 
= 4.9 \left(0.23+0.77f_n\right)\times 10^{-23}\cm\,\Ha^{-1}
\eeq
where $f_n<1$ is the fraction of the C-H bonds that are in neutral PAHs.
The observed absorption toward Cyg OB2-12 in the $3.289\micron$
aromatic C-H stretch 
is $7.1\times10^{-23}\cm\,\Ha^{-1}$
\citep{Hensley+Draine_2020}.}
If the band strength adopted here is correct, 
and $f_n\approx 0.5$, most of the observed 3.3$\micron$
absorption must come from aromatic material in larger grains.

For the mMMP spectrum (which we suggest is appropriate for 
typical star-forming galaxies),
the $\Fclip(3.3)/\Fclip(7.7)$ ratio could be brought into agreement
by shifting to the ``large'' size distribution ($a_0=5\Angstrom$).
However, we suspect that the overprediction of 
$\Fclip(3.3\micron)/\Fclip(7.7\micron)$
may be mainly attributable to neglect of other energy loss channels:
photoelectric emission, photodesorption, and fluorescence
\citep{Allamandola+Tielens+Barker_1989}.
Our calculation of the vibrational excitation of PAHs assumed that absorption
of a photon converts the full photon energy $h\nu$ into
vibrational excitation, with the vibrational energy then
being removed only by infrared emission.
However:
\begin{enumerate}
\item High energy photons can photoionize PAHs.  When a
photoionization takes place, only a fraction of the photon energy appears
as ``heat''.  
\item The electronically-excited state resulting from absorption of a UV
photon may sometimes deexcite by luminescence: 
emission of an optical photon before
``internal conversion'' is able to transfer all of the electronic energy to
the vibrational modes
\citep[see reviews by][]
{Witt+Vijh_2004,Witt+Lai_2020}.
\item A PAH with a high vibrational temperature
may sometimes be able to radiate an
optical photon via ``Poincare fluorescence'' (also known as 
``recurrent fluorescence'')
\citep{Leger+Boissel+dHendecourt_1988,Lai+Witt+Crawford_2017}.
\item A PAH with a large amount of vibrational energy per degree
of freedom will sometimes break a C-H bond, ejecting a hydrogen atom
\citep[see, e.g.,][]{Marciniak+Joblin+Mulas+etal_2021}.
Neglect of the energy lost to bond-breaking
will lead to overestimation of the
$3.3\micron$ emission, which depends on the high-$T$ tail of the
temperature distribution function.
\end{enumerate}

PAHs with our ``standard'' size distribution $dn/da$
and ionization fraction
$\fion(a)$, heated by
the mMMP starlight spectrum thought to be appropriate for normal
star-forming galaxies,
may in fact be consistent with observations
when the above energy loss channels are included when calculating the
PAH temperature distribution functions and emission.
This will be the subject of future work.

\section{\label{sec:summary}
         Summary}

The principal results of this study are as follows:
\begin{enumerate}
\item Using a model that includes a PAH population and a population of
larger astrodust grains \citep{Draine+Hensley_2021a,Hensley+Draine_2021b}
containing amorphous silicates, 
carbonaceous material, and other compounds, 
we calculate the infrared emission spectrum for
a range of illuminating radiation field spectra, 
a wide range of starlight intensities, 
and for the full size distribution of particles in the dust model.
We consider starlight spectra ranging from the far-UV-bright spectrum of
a 3\,Myr-old starburst, to the very red
spectrum of the highly evolved population of
stars in the M31 bulge.

\item We also consider heating of dust in clouds, with the dust inside
the cloud irradiated by a reddened (and weakened) radiation field.
For the mMMP starlight spectrum, which may be representative of the
diffuse radiation in a star-forming galaxy, we find that for fixed
PAH abundance, the cloud spectrum has
the fractional power in the $3.3\micron$ feature reduced by a
factor $\sim$$1.7$, 
but the fractional power in the longer wavelength PAH features, in particular
the $7.7\micron$ complex,
is only reduced by a factor $\sim$$1.3$.

\item The PAH abundance parameter $\qpah$ can be 
estimated using
the fractional power in the ``clipped'' $7.7\micron$ feature:
$\qpah = R_{\qpah}^{\rm clip} \times [\Fclip(7.7\micron)/\FTIR]$.
The coefficient $R_{\qpah}^{\rm clip}$ is sensitive to the spectrum of the
starlight heating the dust.
We estimate $R_{\qpah}^{\rm clip}$ 
for a variety of starlight spectra and assumptions
about the PAH size distribution and ionization.

\item The sensitivity of the PAH emission features to variations in 
the PAH size distribution
is studied.  The $3.3\micron$ feature shows the greatest sensitivity to
variations in the PAH size distribution, as these affect the abundance of the
smaller PAH nanoparticles that account for most of the $3.3\micron$ emission. 
The $3.3\micron$ and $11.2\micron$ features are sensitive to ionization state.
The $\Fclip(6.2\micron)/\Fclip(7.7\micron)$ 
feature ratio shows factor of $\sim$$1.5$
variations as we vary the PAH size distribution 
(see Figure \ref{fig:band_ratios}a).

\item The sensitivity of the PAH emission features to the PAH ionization
fraction are studied.  For changes that correspond to factor of $\sim$2
changes in the effective ionization parameter, we find variations by
factors $\sim$$1.6$ in 
$\Fclip(11.2\micron)/\Fclip(7.7\micron)$ (see Figure
\ref{fig:band_ratios}a and b),

\item The $17\micron$ feature strength is sensitive to the intensity
of the starlight heating the dust.  The relatively low values of
$\Fclip(17)/\Fclip(7.7)$ observed in some SINGS galaxies
and in the 1C galaxy sample of \citet{Lai+Smith+Baba+etal_2020}
may be explained
by high values of the heating rate parameter $U\approx 10^3$.

\item The treatment of PAH excitation by starlight used here has neglected
several energy loss channels, 
resulting in overestimation of the $3.3\micron$
emission for a given PAH size distribution, as discussed in
\S\ref{subsec:modeling 3.3}.

\item A library of dust and PAH emission spectra for a wide range of
starlight spectra and starlight intensities is available on-line 
(\url{https://doi.org/10.7910/DVN/LPUHIQ}).
A companion paper
(Smith et al.\ 2021, in preparation) will use
this library to interpret observed spectra using the PAHFIT 
spectral-fitting code.
\end{enumerate}

\acknowledgements
This research was supported in part by
NSF grants AST-1408723 and AST-1908123,
and NASA grant NSSC19K0572.
We especially 
thank Thomas Lai for making available the observed galaxy spectra
in advance of publication, and for providing a line-subtracted average
1C spectrum.
Finally, we thank the anonymous referee for a detailed and expert
report that helped us improve the manuscript.

\appendix

\section{\label{app:Cabs for PAHs}
         PAH Absorption Cross Sections}

In this work we follow the recommendations of DL07 to estimate
photon absorption cross sections for PAHs.
For a PAH with $N_{\rm C}$ carbon atoms, we take a weighted average of
$N_{\rm C}\Gamma(\lambda)$, 
where $\Gamma(\lambda)$ is an estimate for ``pure PAH''
absorption per C atom, and
$C_{\rm abs}^{\rm (gra)}(\lambda)$, 
the absorption cross section for a graphite sphere:
\beq
C_{\rm abs}(\lambda) = \xi_{\rm PAH}N_C \Gamma(\lambda) + 
(1-\xi_{\rm PAH})C_{\rm abs}^{\rm (gra)}(\lambda)
~~.
\eeq
The ad-hoc weighting factor $\xi_{\rm PAH}$ varies between 0.99 and 0
with increasing $N_C$:
\beq
\xi_{\rm PAH} = 0.99\times {\rm min}\left[1,\frac{5\times10^4}{N_C}\right]
~~.
\eeq
We use 
$C_{\rm abs}^{\rm (gra)}$ estimated for turbostratic graphite by
\citet{Draine_2016}, using Maxwell-Garnet effective medium theory with
$\epsilon(\bE\parallel c)$ for the matrix and $\epsilon(\bE\perp c)$ for
inclusions.
The ``pure PAH'' contribution to the cross section per C is given by
\beqa
\Gamma(\lambda) &~=~& S_1(\lambda) + (1.35x-3)\times10^{-18}\cm^2 
\hspace*{10em} {\rm for~} 10 < x < 15
\\
&=& (66.302 -24.367x + 2.950x^2 - 0.1057x^3)\times10^{-18}\cm^2
\hspace*{1.8em} {\rm for~} 7.7 < x < 10
\\
&=& S_2(\lambda) + c_0+c_1x+c_2(x-5.9)^2 + c_3(x-5.9)^3
\hspace*{4em} {\rm for~} 5.9 < x < 7.7
\\
&=& S_2(\lambda) + c_0 + c_1x
\hspace*{16em} {\rm for~} 3.3 < x < 5.9
\\
&=& 34.58\times10^{-18-3.431/x}\cm^2\, C(\lambda/\lambda_c) + \sum_{j=3}^{30} S_j(\lambda)
\hspace*{6.5em} {\rm for~} x < 3.3
\eeqa
where 
\beqa
x&\equiv& \micron/\lambda
\\
c_0 &~=~& 1.8687\times10^{-18}\cm^2
\\
c_1 &=& 1.905\times10^{-19}\cm^2
\\
c_3 &=& 4.175\times10^{-19}\cm^2
\\
c_4 &=& 4.37\times10^{-20}\cm^2
\\
S_j(\lambda) &\equiv&
\frac{2}{\pi}\sum \frac{\gamma_j\lambda_j\sigma_{{\rm int},j}}
{(\lambda/\lambda_j-\lambda_j/\lambda)^2 + \gamma_j^2}
\eeqa
and
\beq
C(y) \equiv \frac{1}{\pi}\arctan\left[\frac{10^3(y-1)^3}{y}\right]+\frac{1}{2}
\eeq
is the cutoff function proposed by \citet{Desert+Boulanger+Puget_1990}.
The wavelength $\lambda_c$ for the onset of PAH electronic absorption is
taken to be
\citep[see the discussion in][]{Salama+Bakes+Allamandola+Tielens_1996}
\beqa
\lambda_c &=& \frac{0.951\micron}{1+3.616/\sqrt{M}}
\hspace*{2em}{\rm for~ neutral~ PAHs} ~~,
\\
&=&
\frac{1.125\micron}{1+2.567/\sqrt{M}}
\hspace*{2em}{\rm for~PAH~cations}~~,
\eeqa
where the number of rings $M$ is taken to be
\beqa
M &~=~& 0.3 N_{\rm C} \hspace*{1em}{\rm for~}N_{\rm C}\leq 40
\\
&&0.4 N_{\rm C} \hspace*{1em}{\rm for~}N_{\rm C} > 40 ~~.
\eeqa
The central wavelength $\lambda_j$, fractional width $\gamma_j$,
and integrated absorption $\sigma_j$ for the Drude profiles are given 
in Table \ref{tab:Drude}.  
Most of the adopted parameters are unchanged from DL07.
However, the strength of the feature at $14.19\micron$ ($j=23$) and
the strength of the $17\micron$ complex ($j=26,27,28$) have been increased by 33\% to improve
agreement with observed spectra.

\begin{figure}
\begin{center}
\includegraphics[angle=0,width=8.5cm,
                 clip=true,trim=0.5cm 5.0cm 0.5cm 2.5cm]
{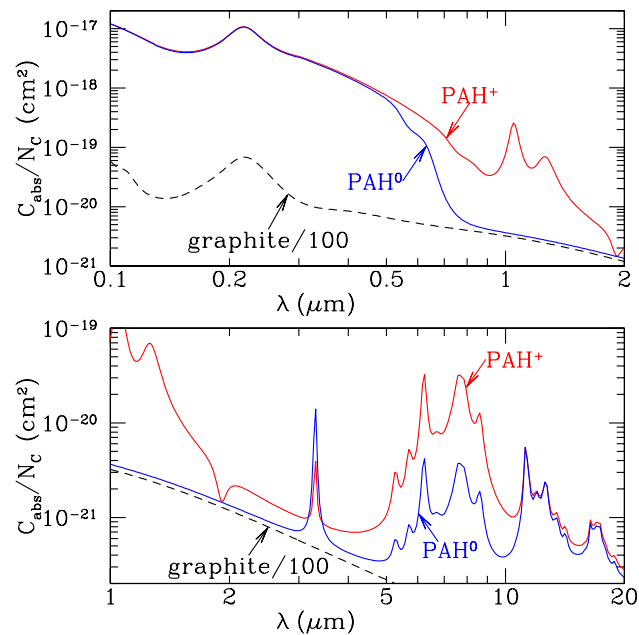}
\caption{\label{fig:Cabs}\footnotesize
         Model absorption cross section per C atom for $a=0.0006\micron$
         PAH$^0$ and PAH$^+$ with $N_{\rm C}=90$ C atoms.
         Also shown is $1\%$ of the cross section per C for turbostratic
         graphite.}
\end{center}
\end{figure}

Figure \ref{fig:Cabs} shows the absorption cross sections (per C atom)
used to model PAH neutrals and cations, for the specific example of
PAHs with $N_{\rm C}=90$ C atoms.

\begin{table}[b]
\newcommand \notea {a}
\newcommand \noteb {b}
\newcommand \notec {c}
\newcommand \noted {d}
\newcommand \notee {e}
\begin{center}
\footnotesize
\caption{\label{tab:Drude}PAH Resonance Parameters$^{\rm \notea}$}
\vspace*{-1.0em}
\begin{tabular}{l l l c c l l}
\hline\hline
&&&\multicolumn{2}{c}{\underline{$\sigma_{{\rm abs}j}\equiv\int \sigma_{{\rm abs},j} d\lambda^{-1}$}}\\
&$\lambda_j$&&Neutral&Ionized\\
$j$&($\mu$m)&$\gamma_j$&($10^{-20}\cm/{\rm C}$) &($10^{-20}\cm/{\rm C}$)&Tentative Identification & Note  \\
\hline
1 & 0.0722 & 0.195 & $7.97\times10^7$ & $7.97\times10^7$ & $\sigma\rightarrow\sigma^*$ transition in aromatic C & \noteb\\
2 & 0.2175 & 0.217 & $1.23\times10^7$ & $1.23\times10^7$ & $\pi\rightarrow\pi^*$ transition in aromatic C & \noteb \\
3 & 1.050  & 0.055 & 0                & $2.0\times10^4$  & Weak electronic transition in PAH cation & c \\
4 & 1.260  & 0.11  & 0                & $7.8\times10^3$  & Weak electronic transition in PAH cation & c \\
5 & 1.905  & 0.09  & 0                & $-146.5$           & ?                                        & c \\
6 & 3.300  & 0.012 & 394(H/C)         & $89.4$(H/C)        & Aromatic C-H stretch                     & d \\
7 & 5.270  & 0.034 & 2.5              & $20$               & C-H bend + C-H stretch combination mode  & \\
8 & 5.700  & 0.035 & 4                & $32$               & C-H and + C-H stretch combination mode   & \\
9 & 6.220  & 0.030 & 29.4             & $235$              & Aromatic C-C stretch (in-plane)          & \\
10& 6.690  & 0.070 & 7.35             & $59$               & ?                                        & \\
11& 7.417  & 0.126 & 20.8             & 181              & Aromatic C-C stretch                     & \\
12& 7.598  & 0.044 & 18.1             & 163              & Aromatic C-C stretch                     & \\
13& 7.850  & 0.053 & 21.9             & 197              & C-C stretch + C-H bending                & \\
14& 8.330  & 0.052 & 6.94(H/C)        & 48.4(H/C)          & C-C stretch + C-H bending?               & \\
15& 8.610  & 0.039 & 27.8(H/C)        & 194(H/C)         & C-H in-plane bending                     & \\
16& 10.68  & 0.020 & 0.3(H/C)         & 0.3(H/C)         & C-H out-of-plane bending, solo?          & \\
17& 11.23  & 0.012 & 18.9(H/C)        & 17.7(H/C)        & C-H out-of-plane bending, solo           & \\
18& 11.33  & 0.032 & 52(H/C)          & 49(H/C)          & C-H out-of-plane bending, solo           & \\
19& 11.99  & 0.045 & 24.2(H/C)        & 20.5(H/C)        & C-H out-of-plane bending, duo            & \\
20& 12.62  & 0.042 & 34.8(H/C)        & 31.0(H/C)          & C-H out-of-plane bending, trio           & \\
21& 12.69  & 0.013 & 1.3(H/C)         & 1.3(H/C)         & C-H out-of-plane bending, trio           & \\
22& 13.48  & 0.040 & 8.0(H/C)         & 8.0(H/C)         & C-H out-of-plane bending, quartet?       & \\
23& 14.19  & 0.025 & 0.60             & 0.60             & C-H out-of-plane bending, quartet?       & \notee \\
24& 15.90  & 0.020 & 0.04             & 0.04              & ?                                       & \\
25& 16.447 & 0.014 & 0.5             & 0.5              & C-C-C bending?                          & \\
26& 17.04  & 0.065 & 2.99             & 2.99             & C-C-C bending?                           & \notee \\
27& 17.375 & 0.012 & 0.15             & 0.15             & C-C-C bending?                           & \notee \\
28& 17.87  & 0.016 & 0.090            & 0.090            & C-C-C bending?                           & \notee \\
29& 18.92  & 0.10  & 0.10             & 0.17             & C-C-C bending?                           & \\
30& 15.    & 0.8   & 50.              & 50.              & large-scale bending modes                & \\
\hline
\multicolumn{6}{l}{\notea~~values taken from DL07 except as noted.}\\
\multicolumn{6}{l}{\noteb~~\citet{Li+Draine_2001b}}\\
\multicolumn{6}{l}{\notec~~\citet{Draine+Li_2007}}\\
\multicolumn{6}{l}{\noted~~\citet{Mattioda+Allamandola+Hudgins_2005}, \citet{Mattioda+Hudgins+Allamandola_2005}}\\
\multicolumn{6}{l}{\notee~~increased by 33\% from DL07}\\
\end{tabular}
\end{center}
\end{table}

\section{\label{app:SINGS}
         Sample Spectra}

\begin{figure}
\begin{center}
\newcommand{\figwidtha}{3.65cm}
\newcommand{\figwidthb}{3.4cm}

\includegraphics[angle=0,width=\figwidtha,
                 clip=true,trim=0.5cm 5.0cm 0.5cm 2.5cm]
{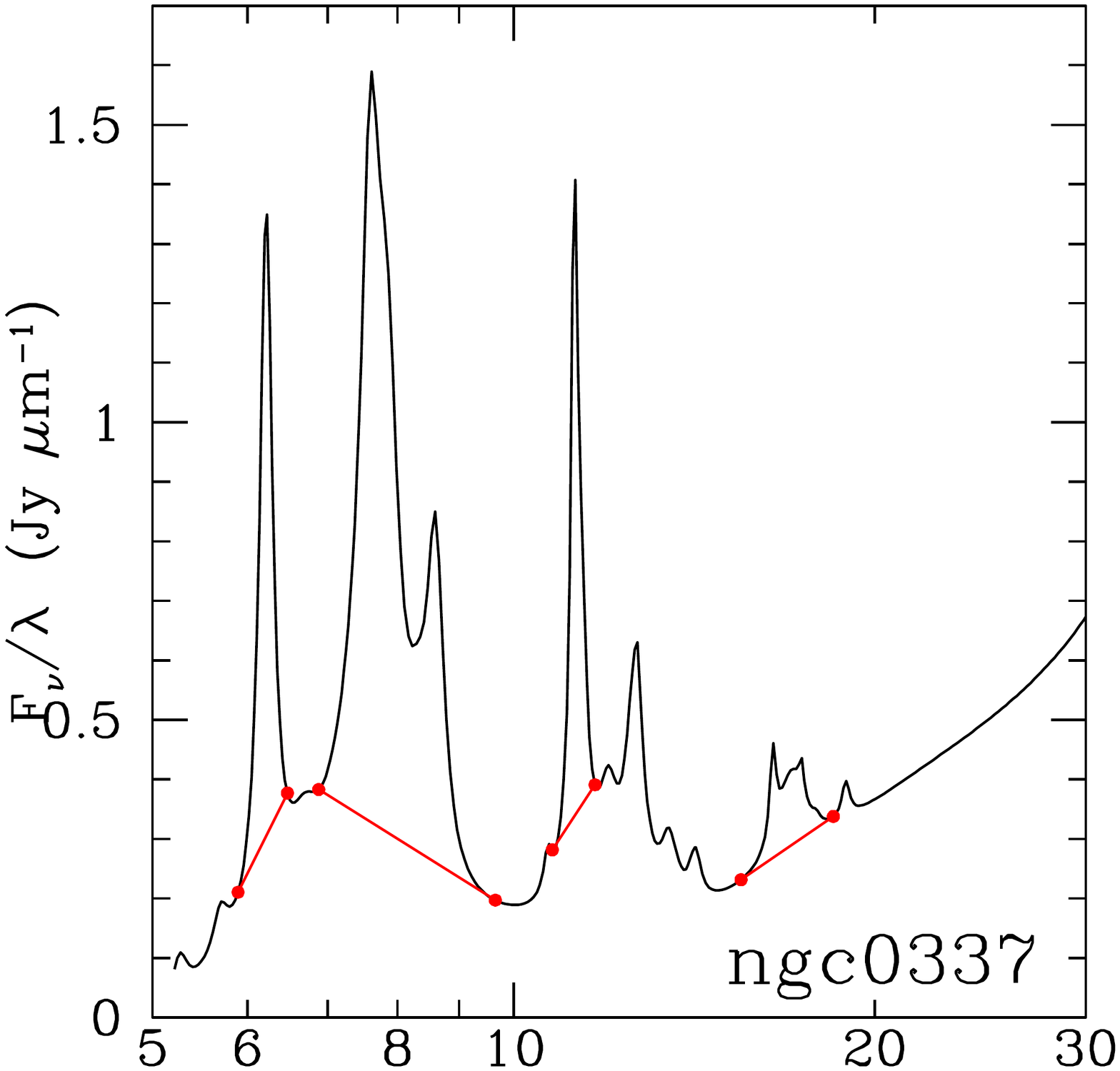}
\includegraphics[angle=0,width=\figwidthb,
                 clip=true,trim=0.5cm 5.0cm 1.75cm 2.5cm]
{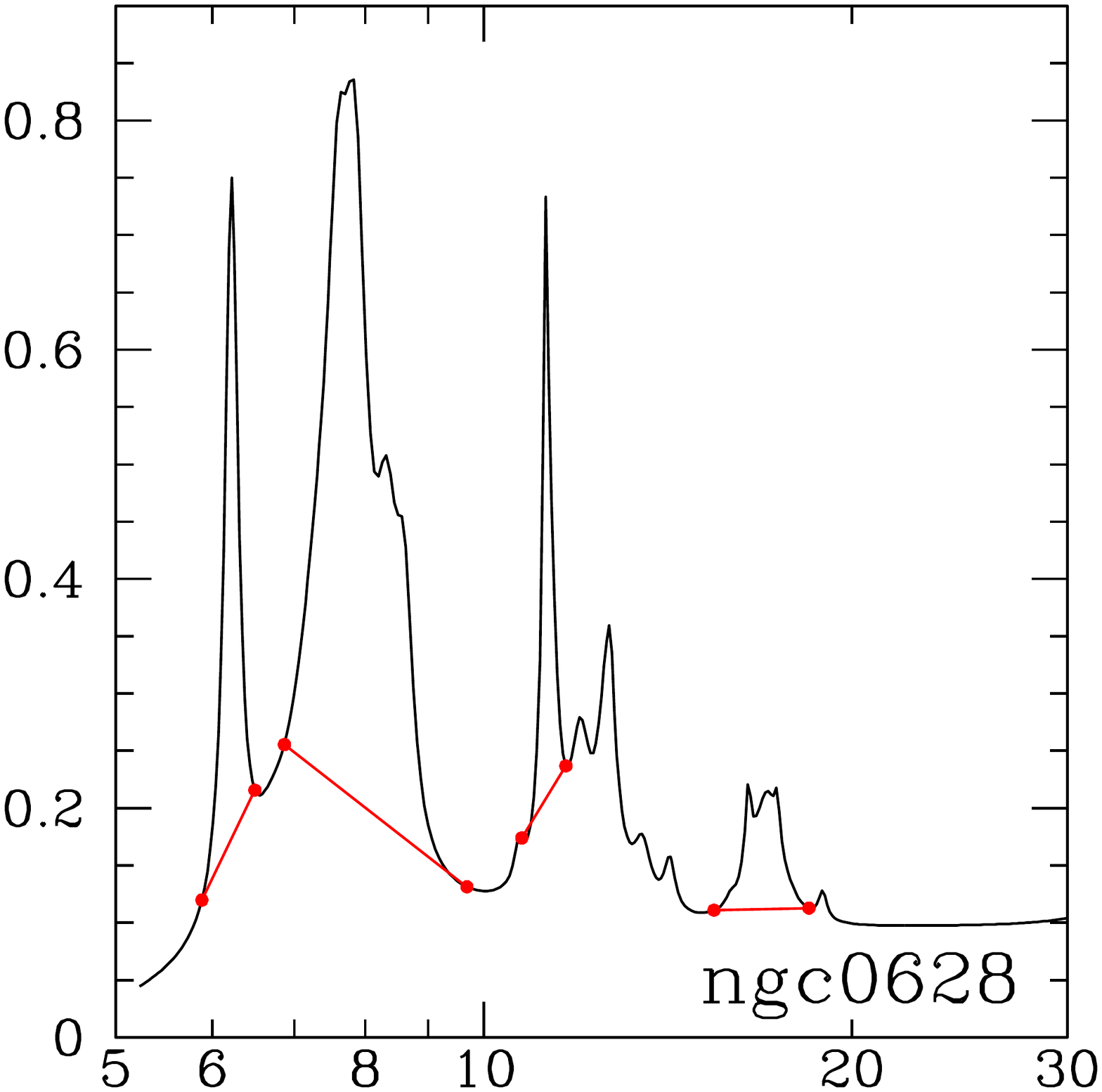}
\includegraphics[angle=0,width=\figwidthb,
                 clip=true,trim=0.5cm 5.0cm 1.75cm 2.5cm]
{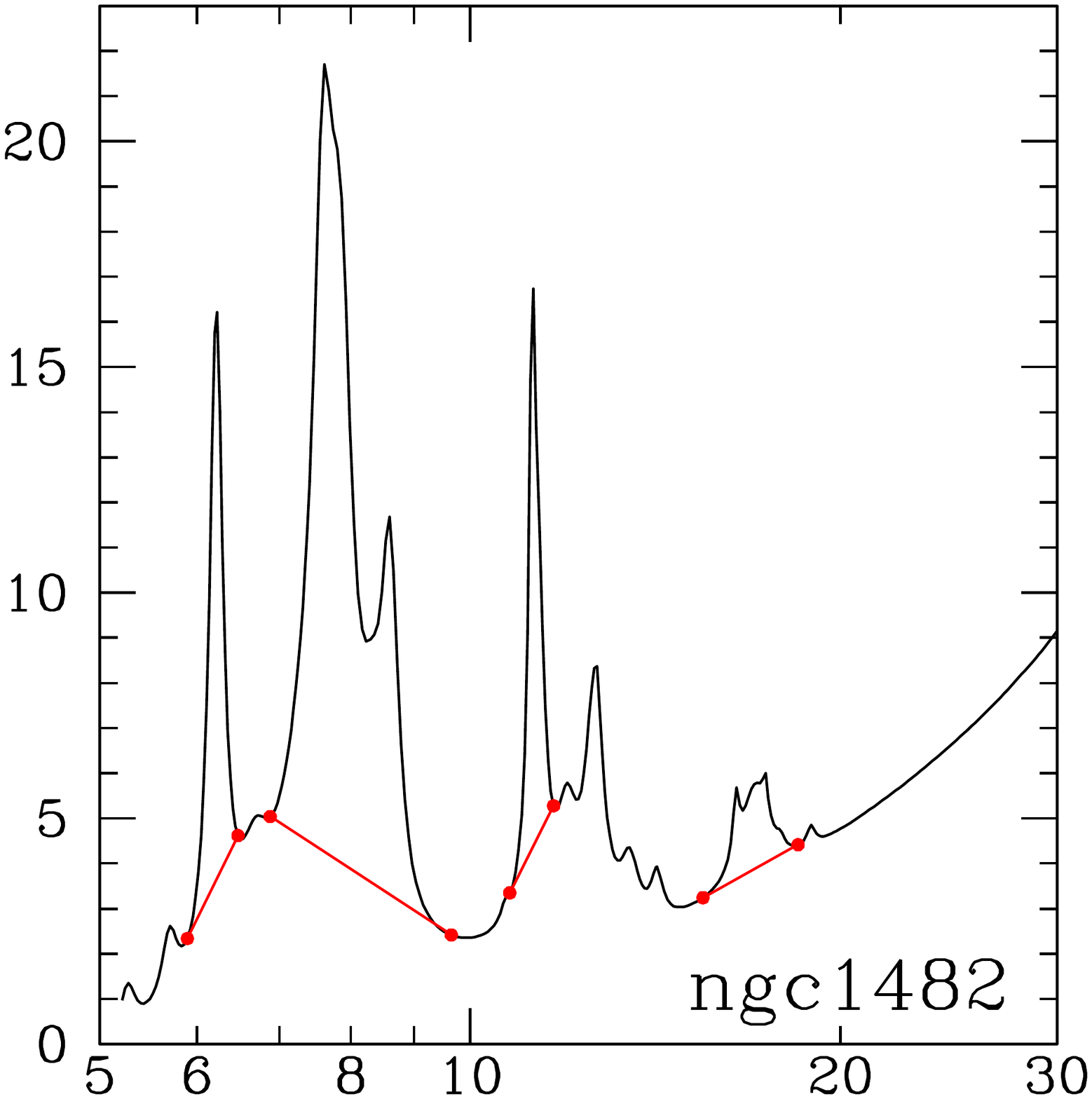}
\includegraphics[angle=0,width=\figwidthb,
                 clip=true,trim=0.5cm 5.0cm 1.75cm 2.5cm]
{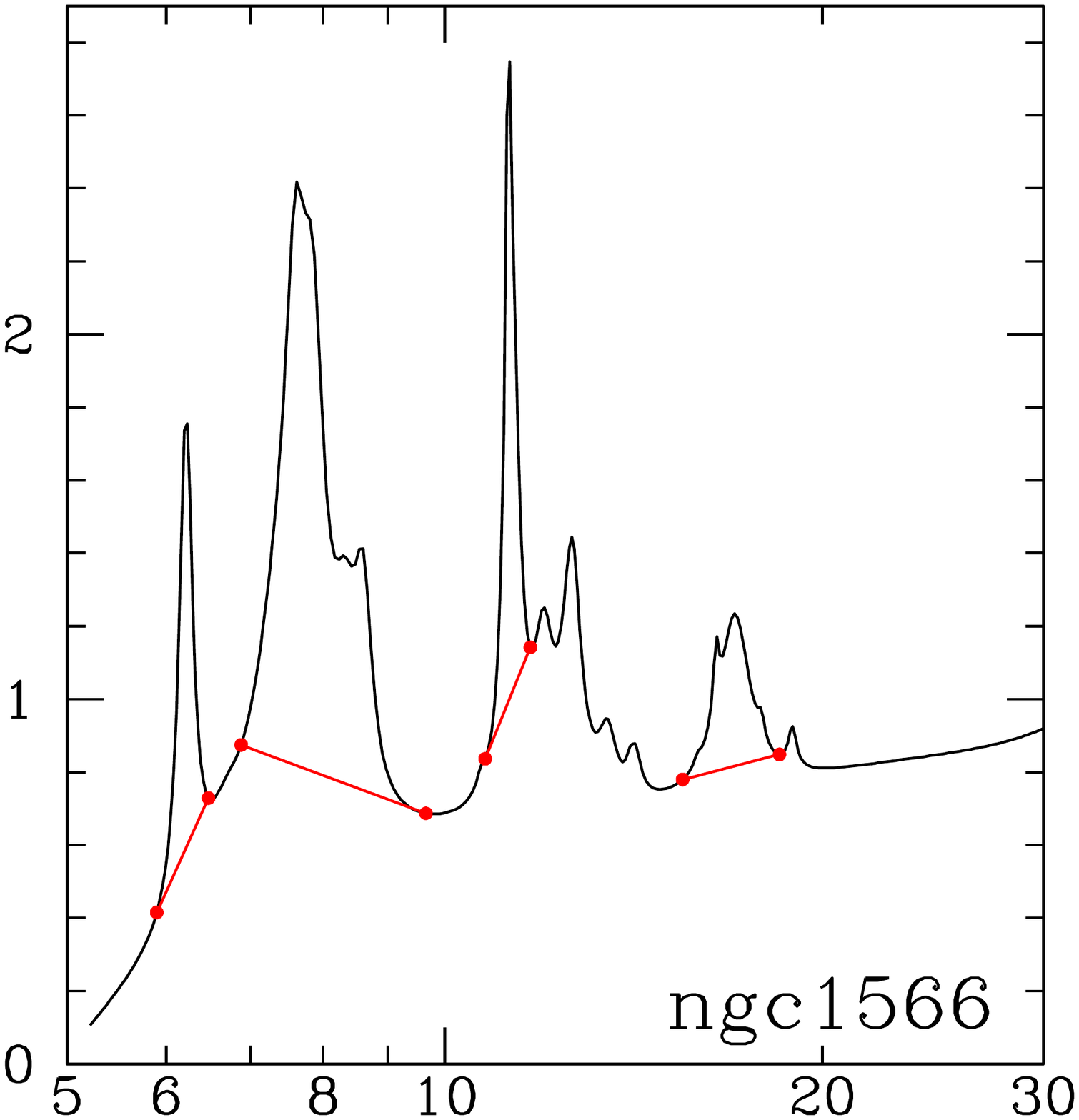}
\includegraphics[angle=0,width=\figwidthb,
                 clip=true,trim=0.5cm 5.0cm 1.75cm 2.5cm]
{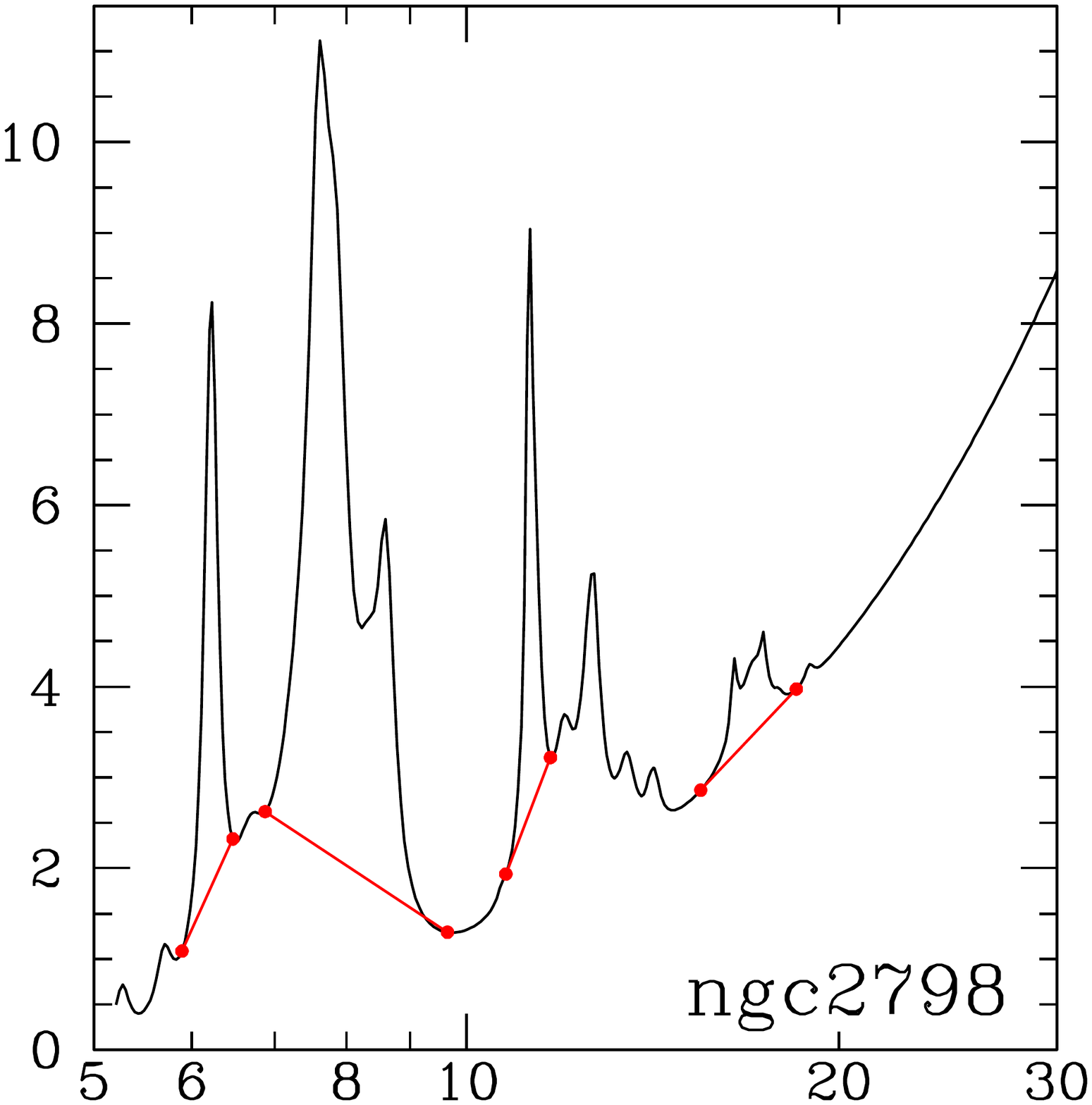}
\includegraphics[angle=0,width=\figwidtha,
                 clip=true,trim=0.5cm 5.0cm 0.5cm 2.5cm]
{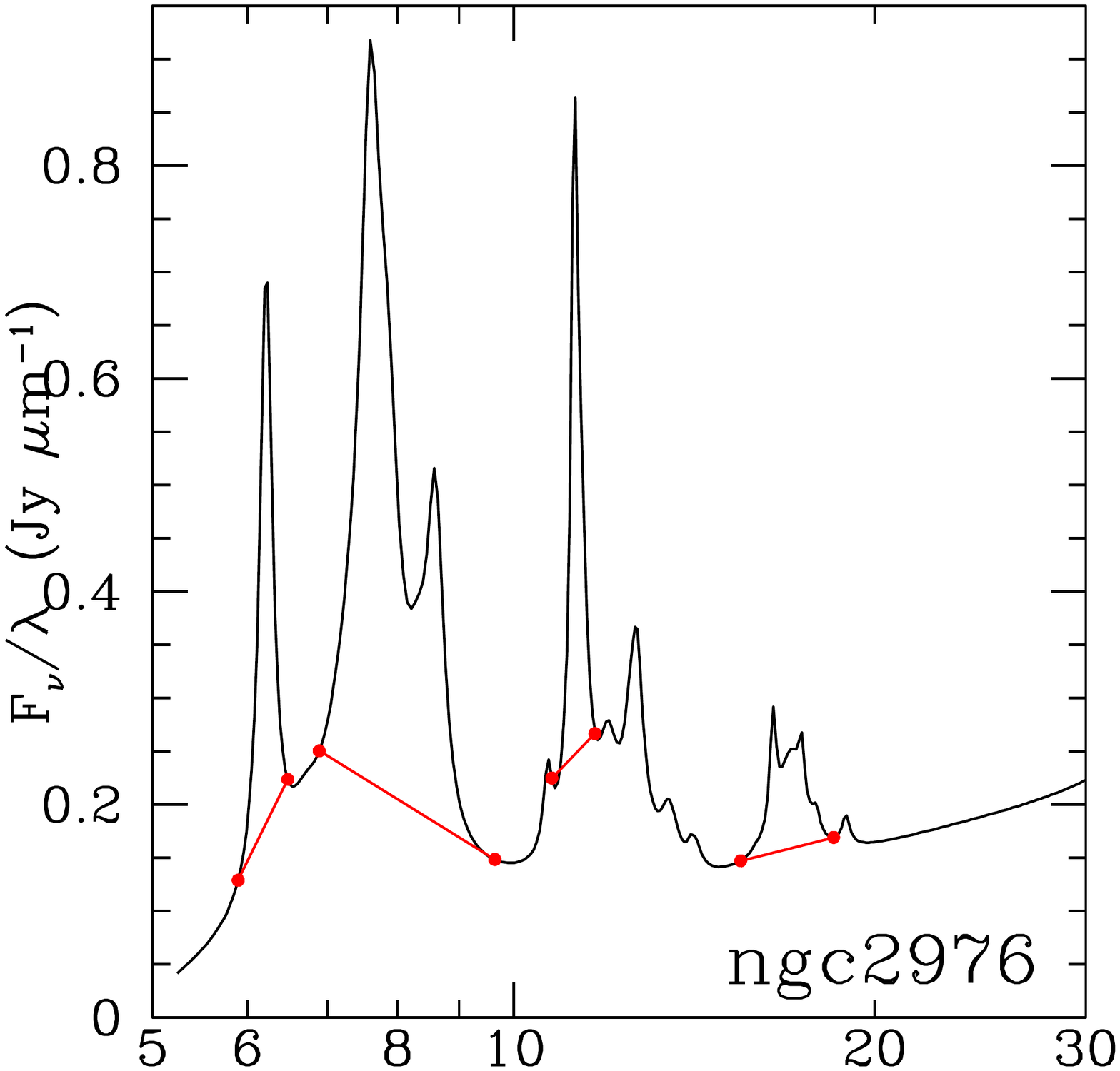}
\includegraphics[angle=0,width=\figwidthb,
                 clip=true,trim=0.5cm 5.0cm 1.75cm 2.5cm]
{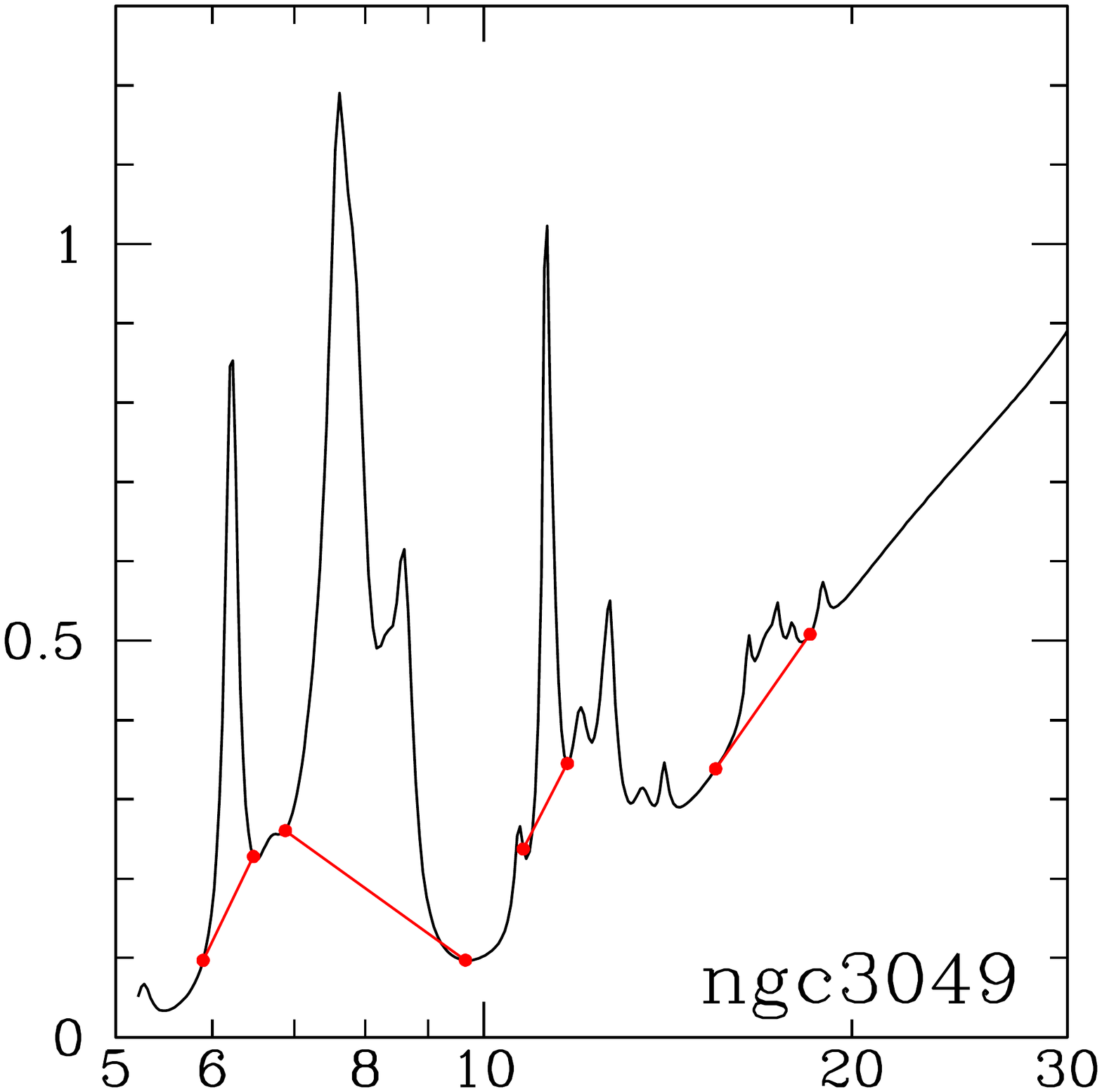}
\includegraphics[angle=0,width=\figwidthb,
                 clip=true,trim=0.5cm 5.0cm 1.75cm 2.5cm]
{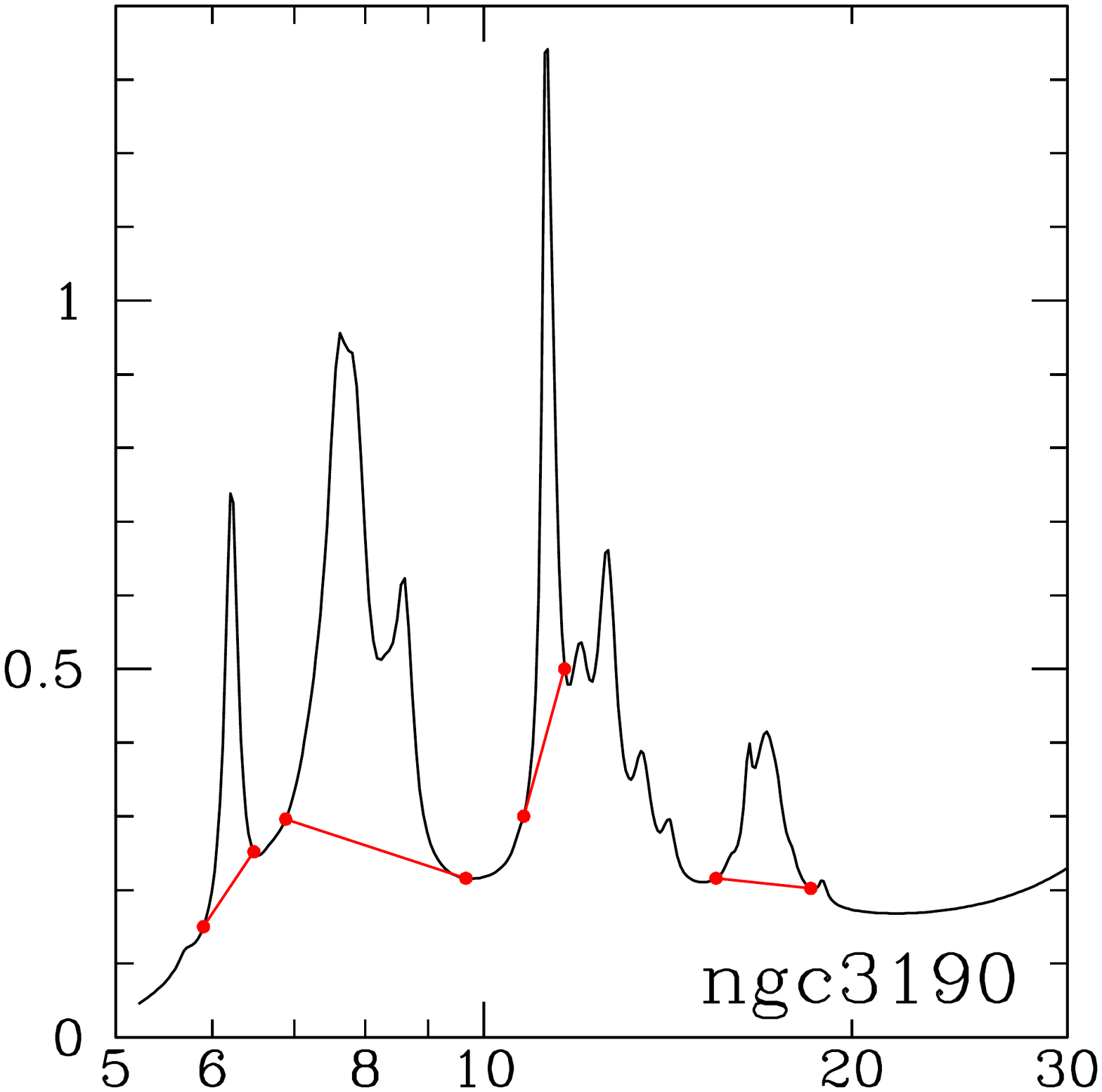}
\includegraphics[angle=0,width=\figwidthb,
                 clip=true,trim=0.5cm 5.0cm 1.75cm 2.5cm]
{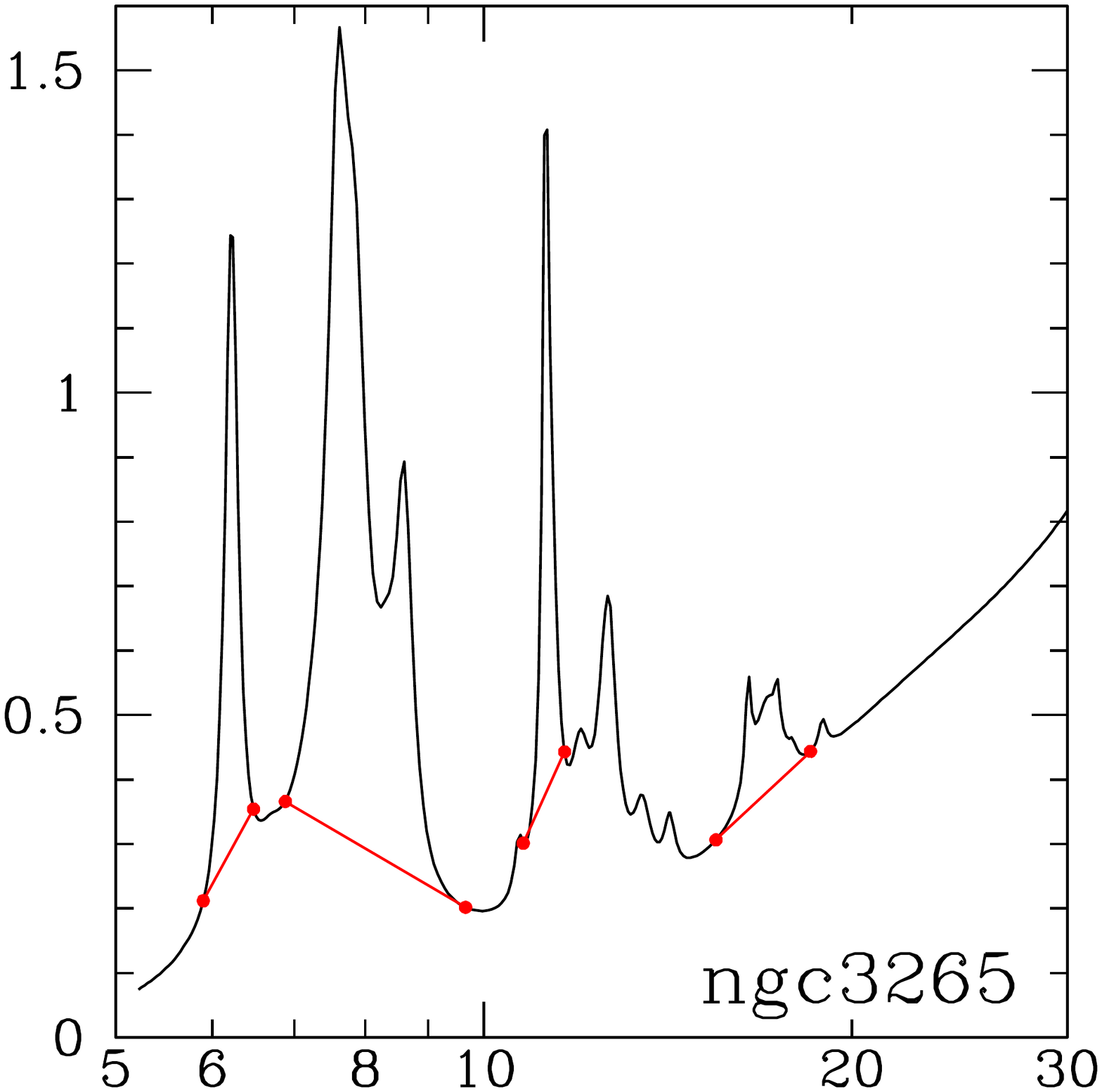}
\includegraphics[angle=0,width=\figwidthb,
                 clip=true,trim=0.5cm 5.0cm 1.75cm 2.5cm]
{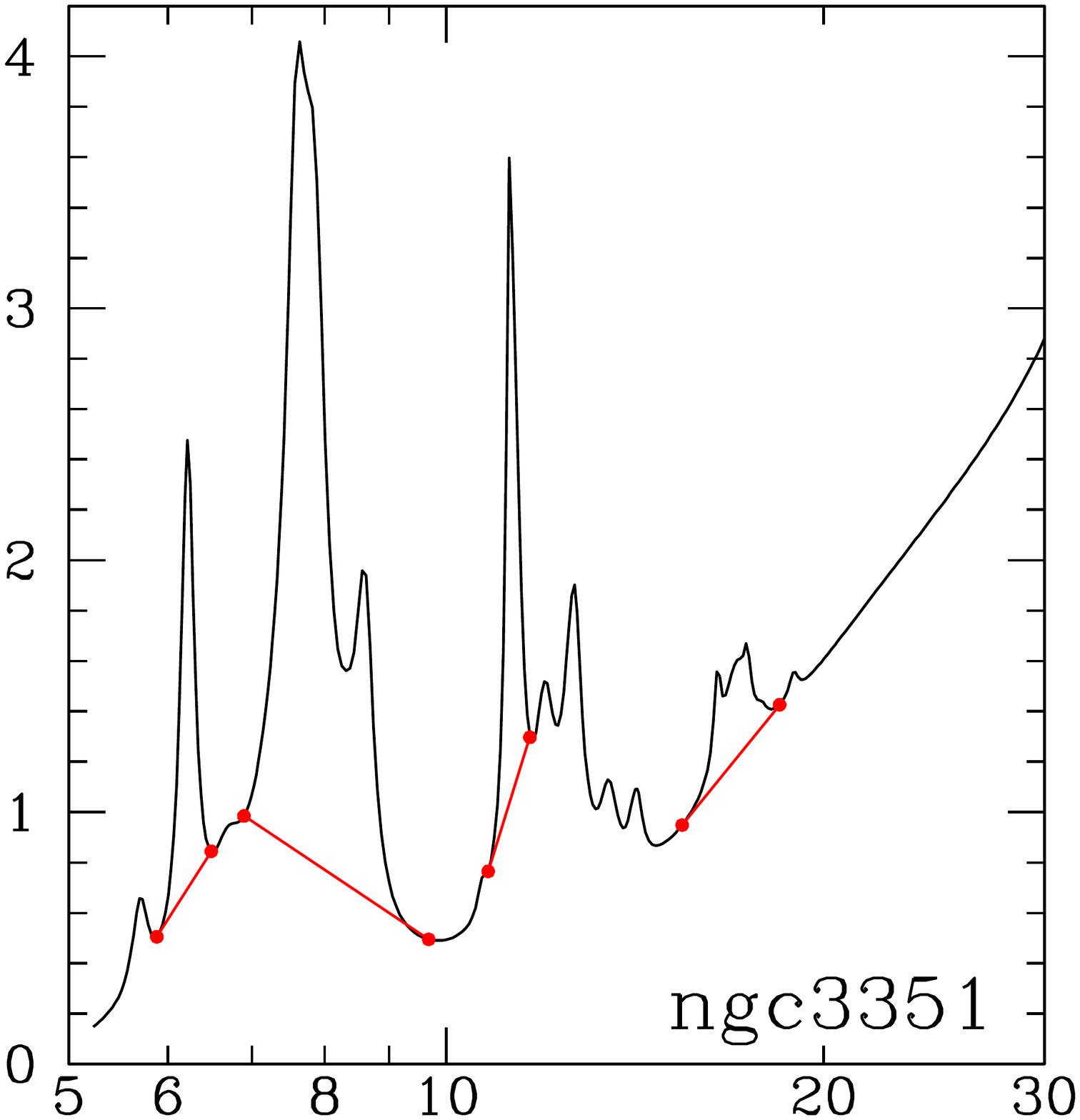}
\includegraphics[angle=0,width=\figwidtha,
                 clip=true,trim=0.5cm 5.0cm 0.5cm 2.5cm]
{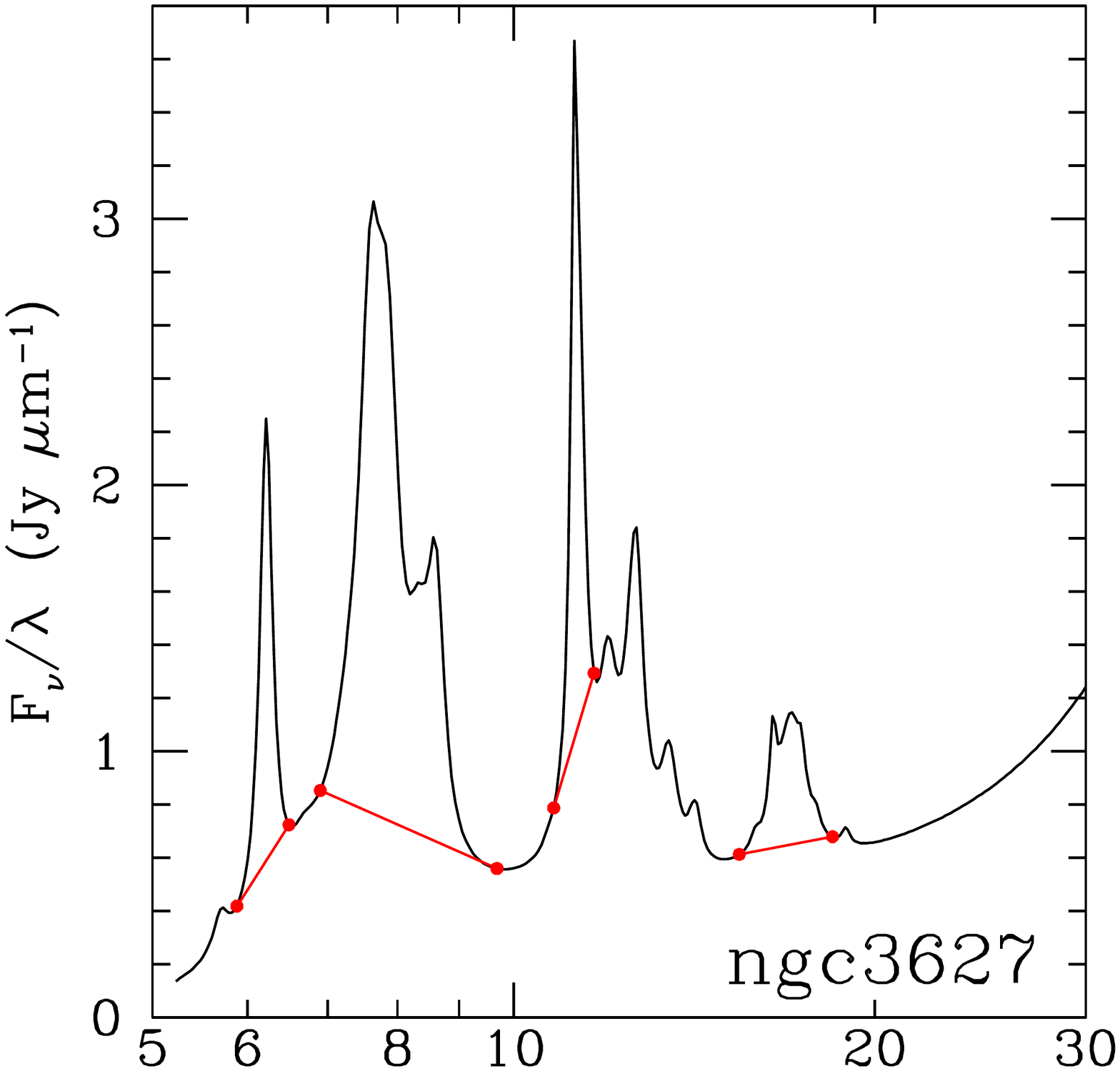}
\includegraphics[angle=0,width=\figwidthb,
                 clip=true,trim=0.5cm 5.0cm 1.75cm 2.5cm]
{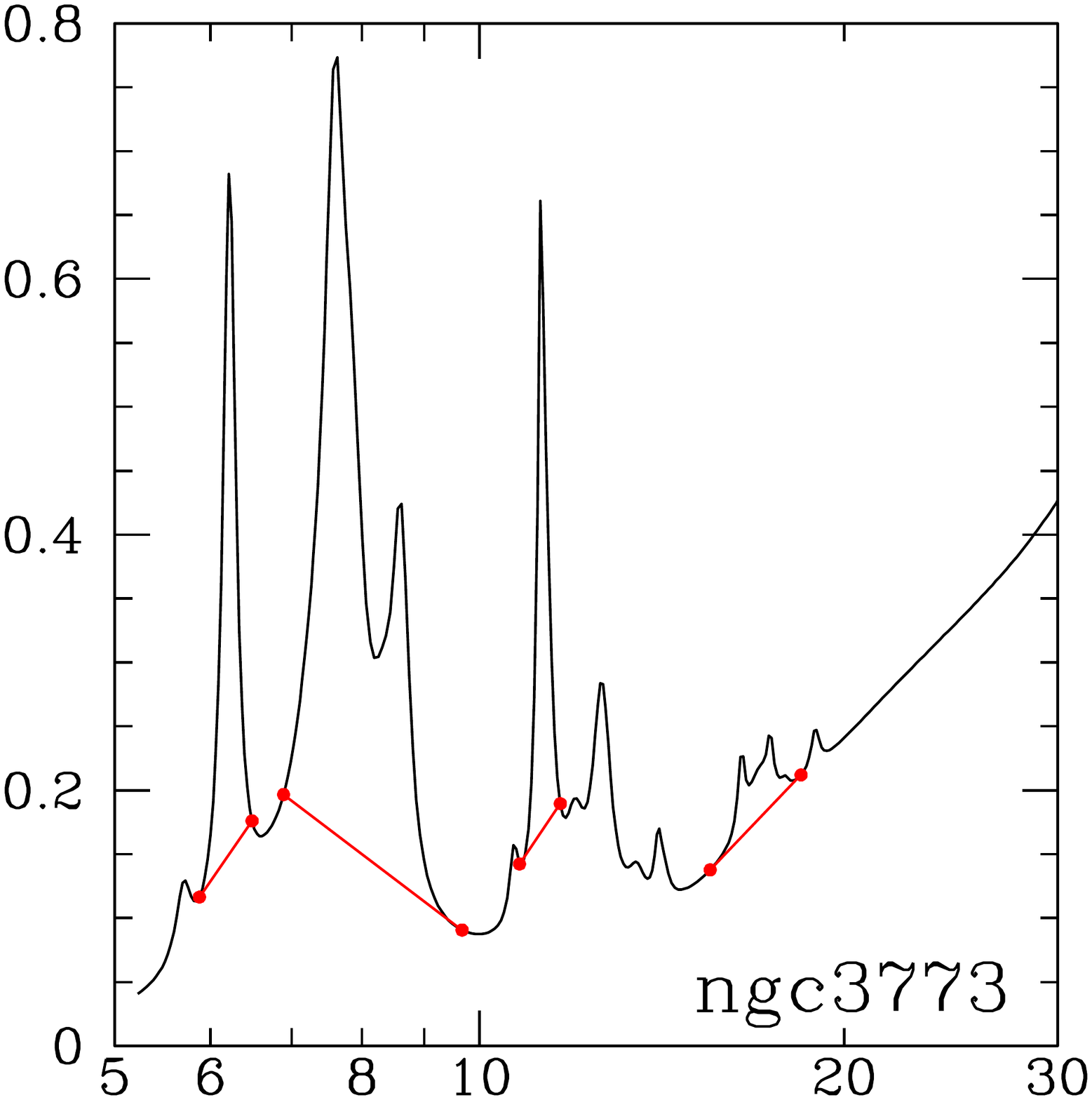}
\includegraphics[angle=0,width=\figwidthb,
                 clip=true,trim=0.5cm 5.0cm 1.75cm 2.5cm]
{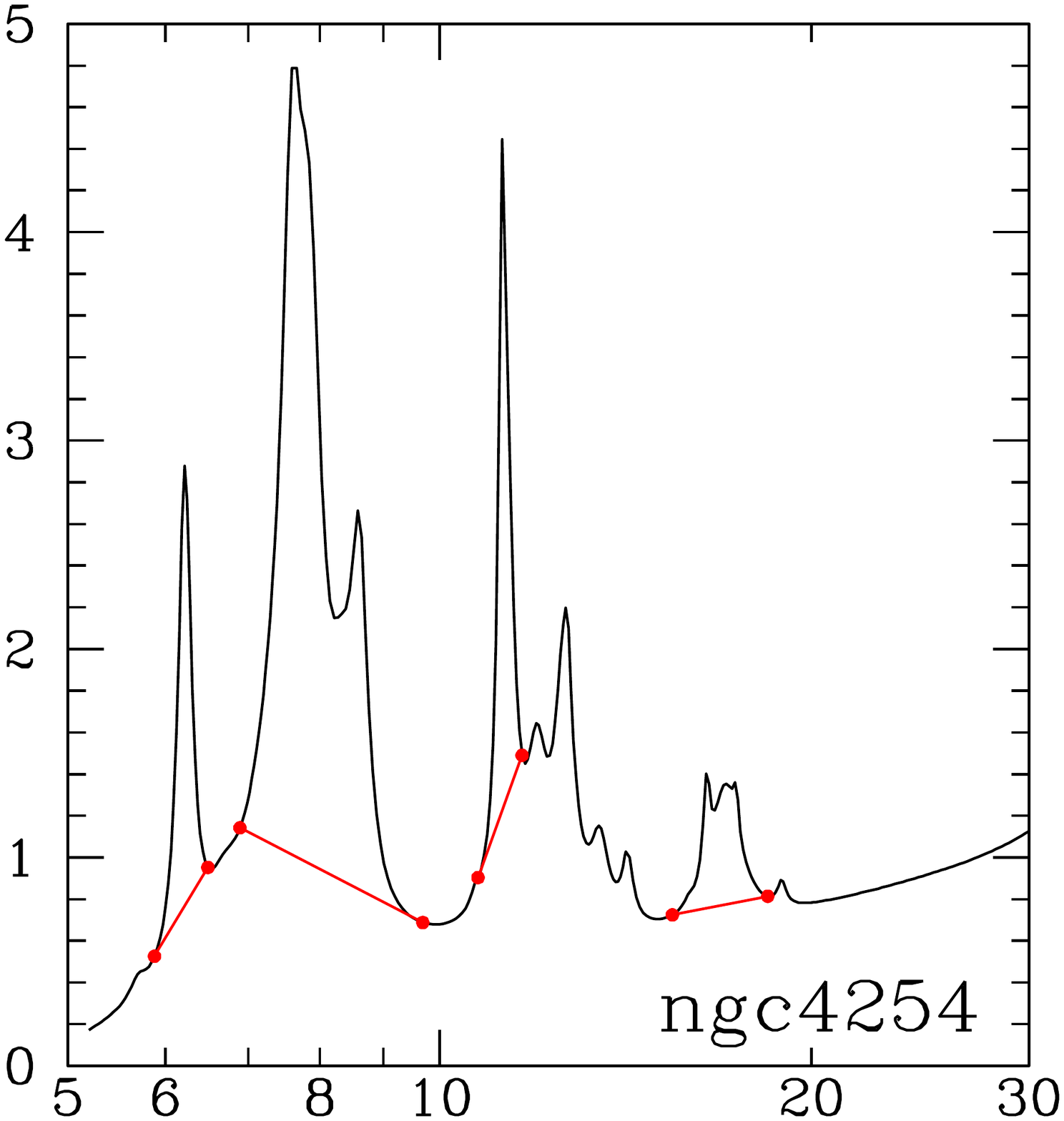}
\includegraphics[angle=0,width=\figwidthb,
                 clip=true,trim=0.5cm 5.0cm 1.75cm 2.5cm]
{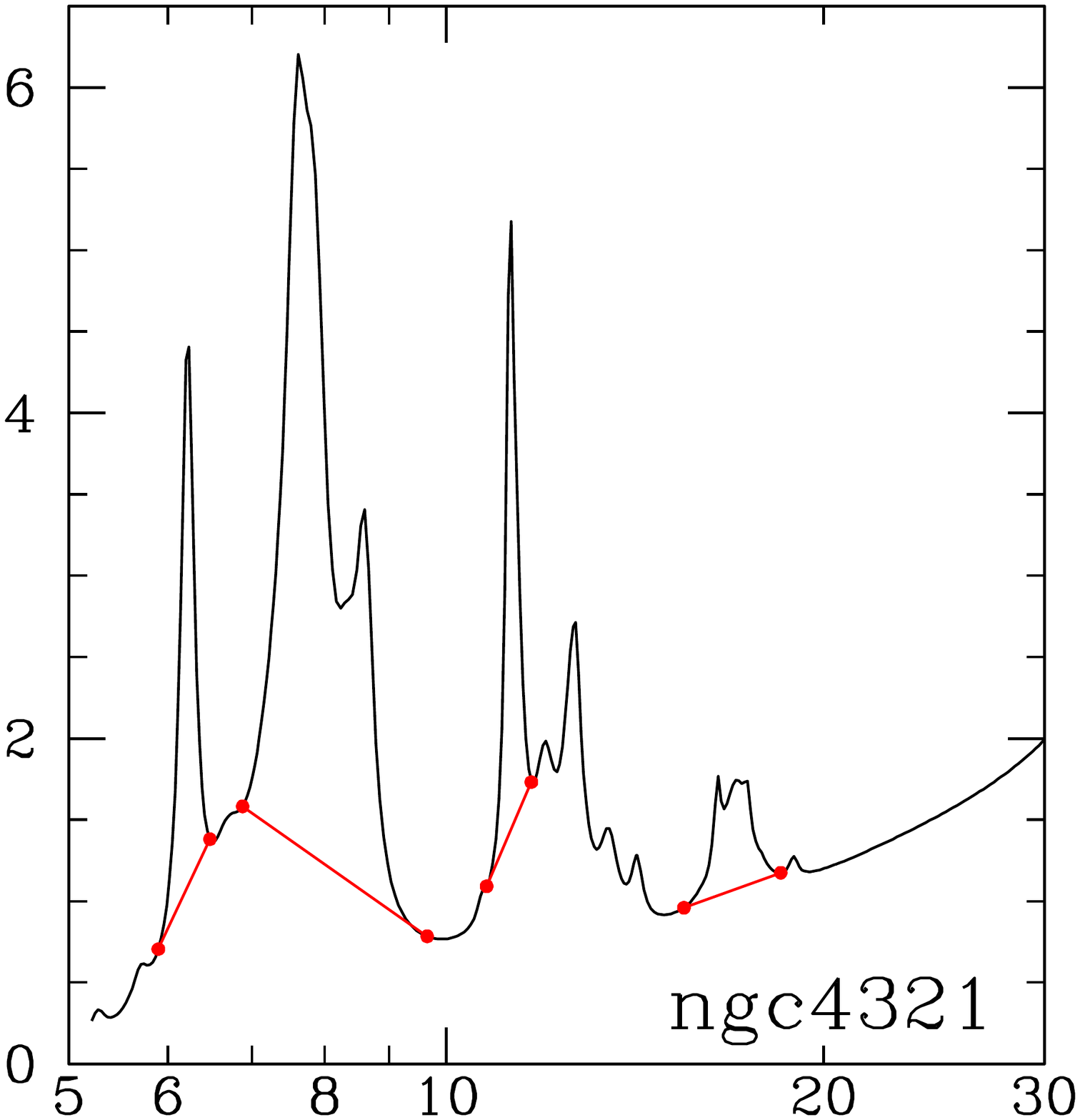}
\includegraphics[angle=0,width=\figwidthb,
                 clip=true,trim=0.5cm 5.0cm 1.75cm 2.5cm]
{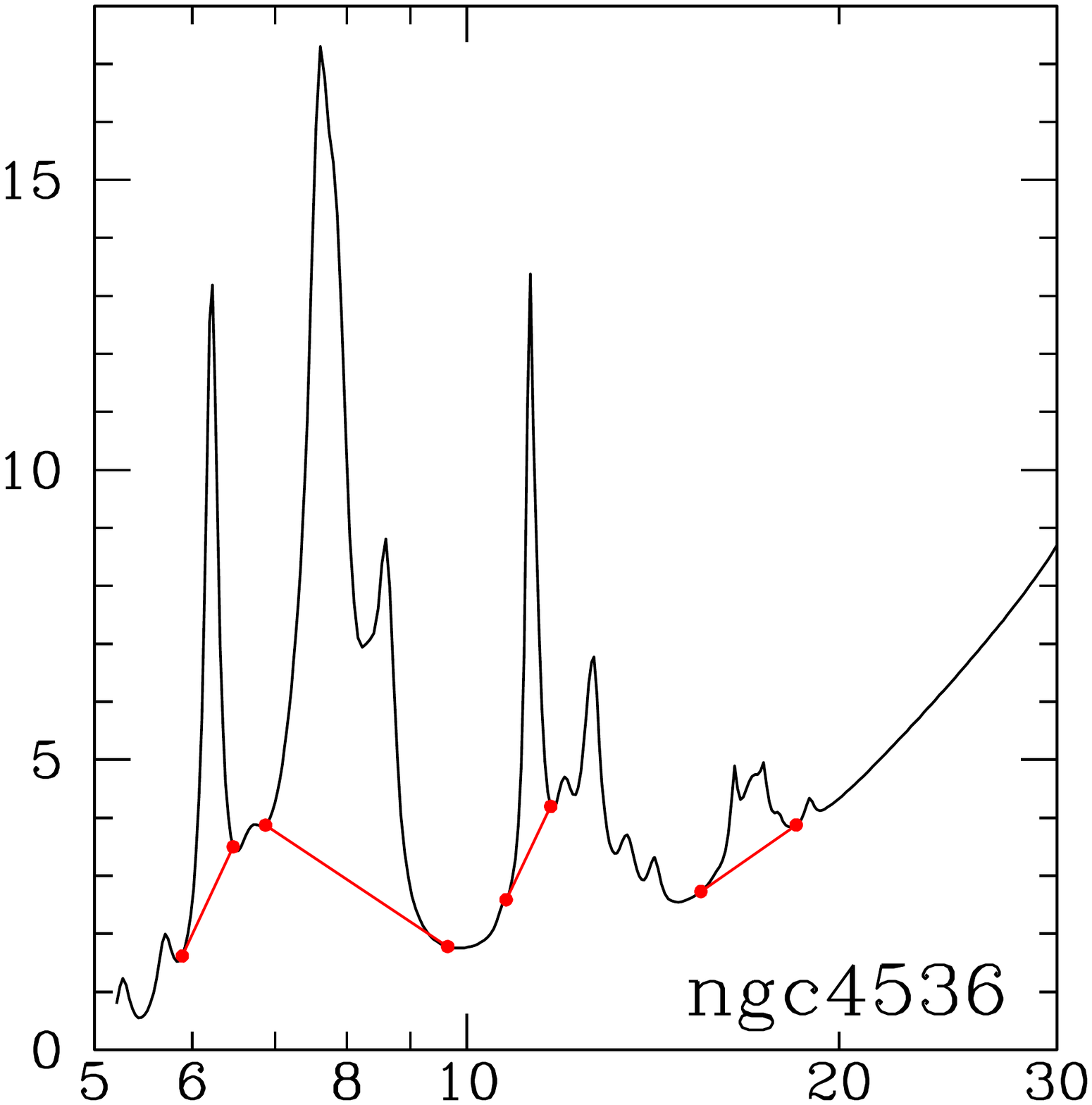}
\includegraphics[angle=0,width=\figwidtha,
                 clip=true,trim=0.5cm 5.0cm 0.5cm 2.5cm]
{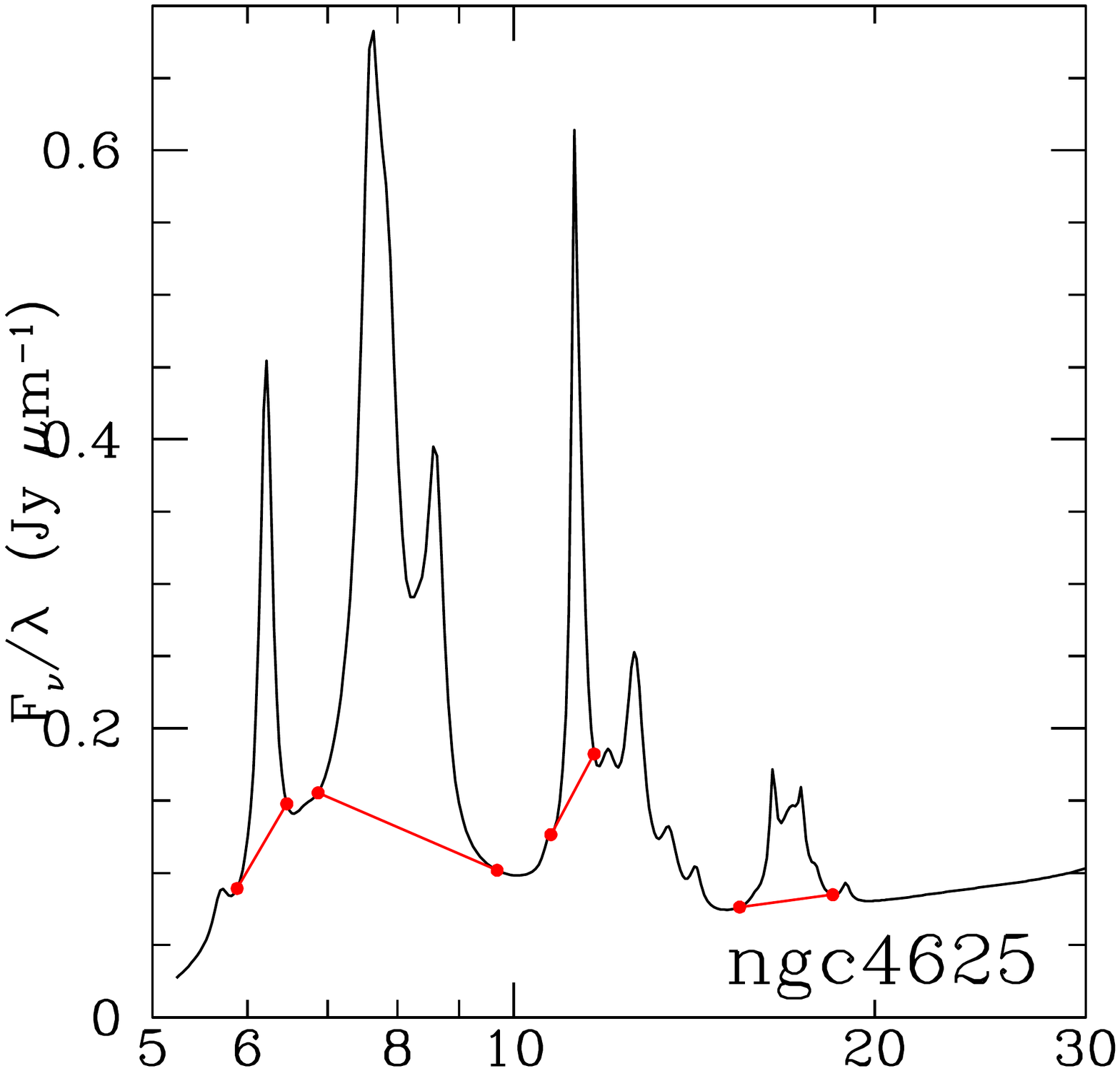}
\includegraphics[angle=0,width=\figwidthb,
                 clip=true,trim=0.5cm 5.0cm 1.75cm 2.5cm]
{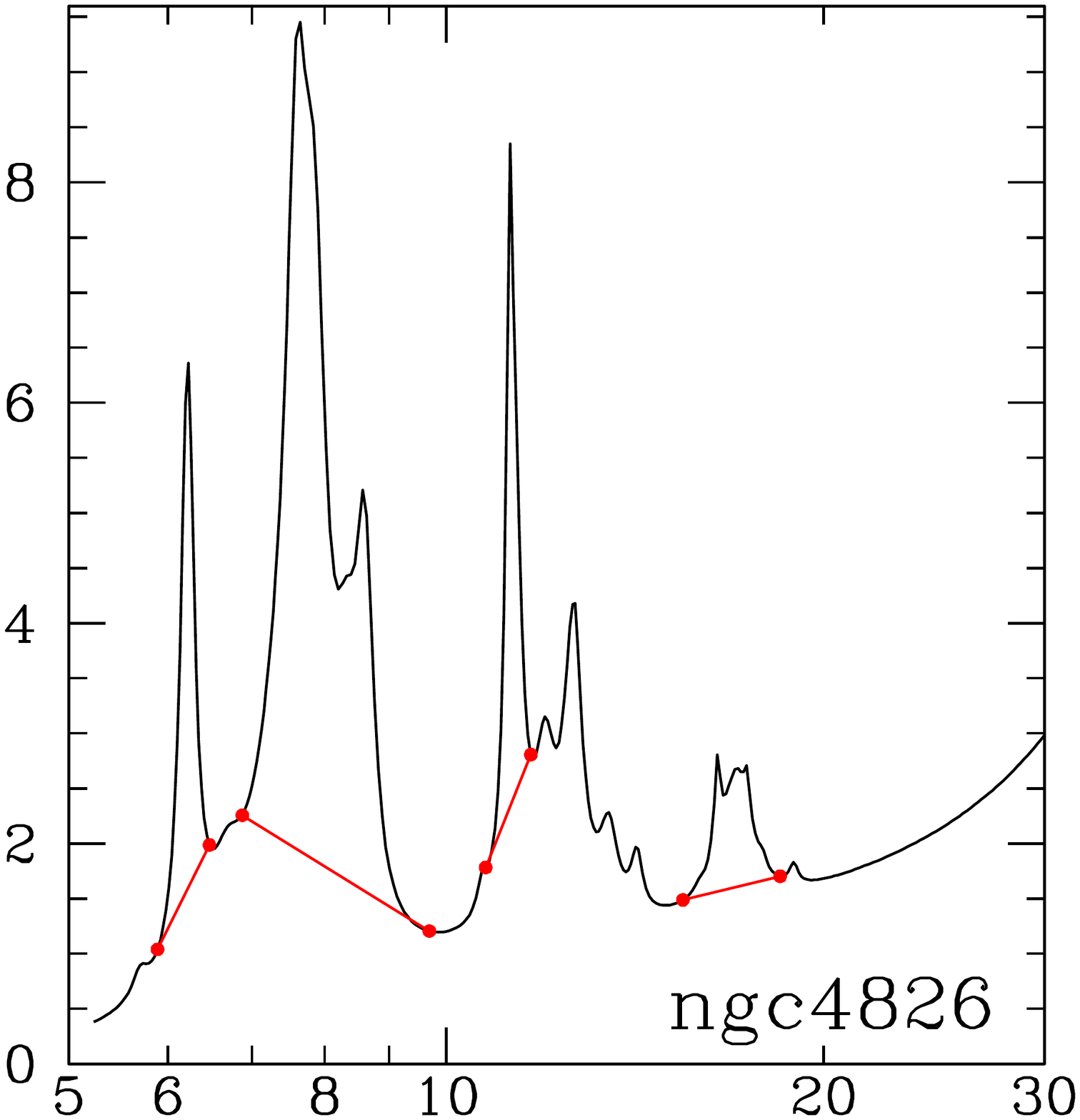}
\includegraphics[angle=0,width=\figwidthb,
                 clip=true,trim=0.5cm 5.0cm 1.75cm 2.5cm]
{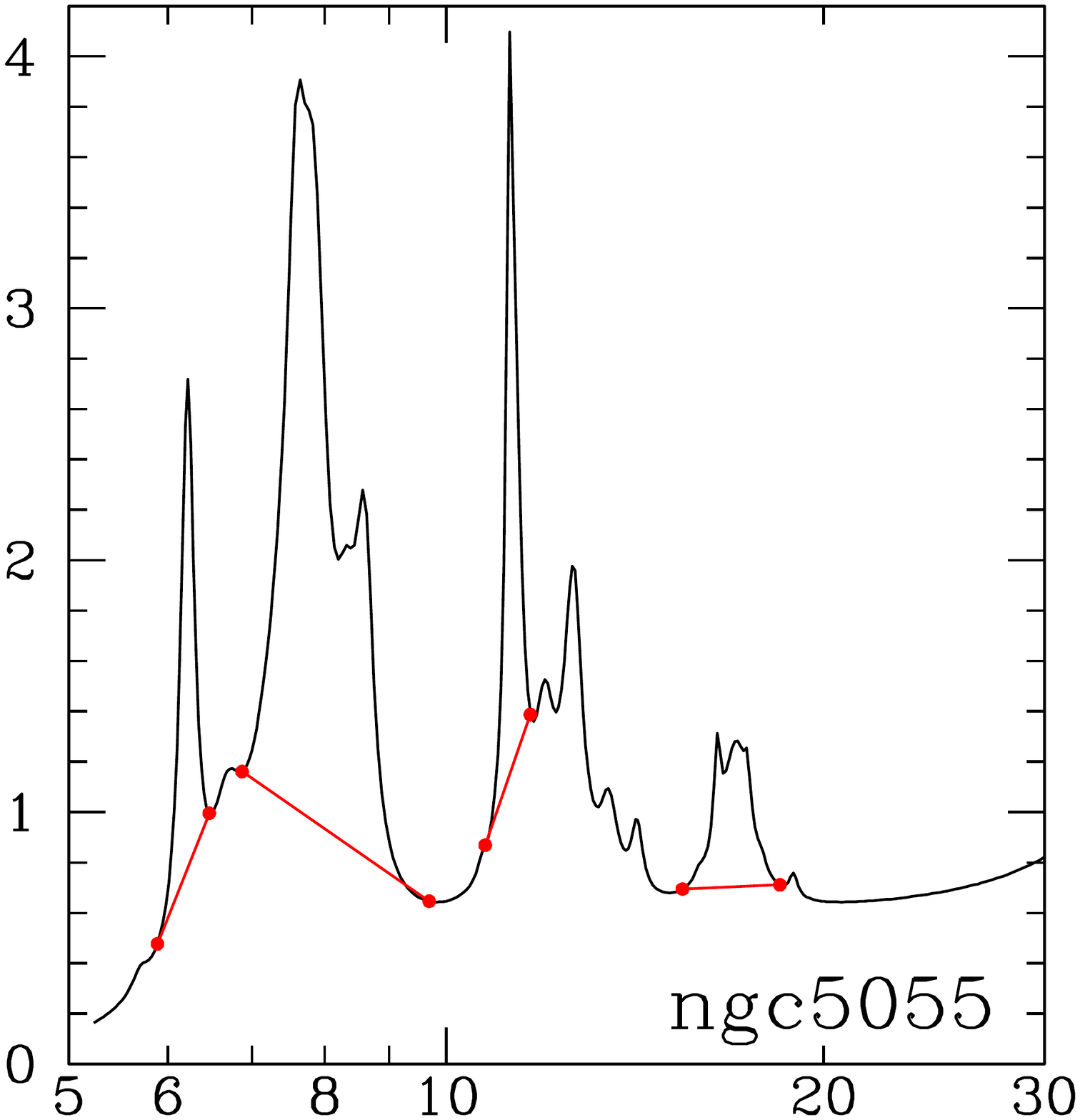}
\includegraphics[angle=0,width=\figwidthb,
                 clip=true,trim=0.5cm 5.0cm 1.75cm 2.5cm]
{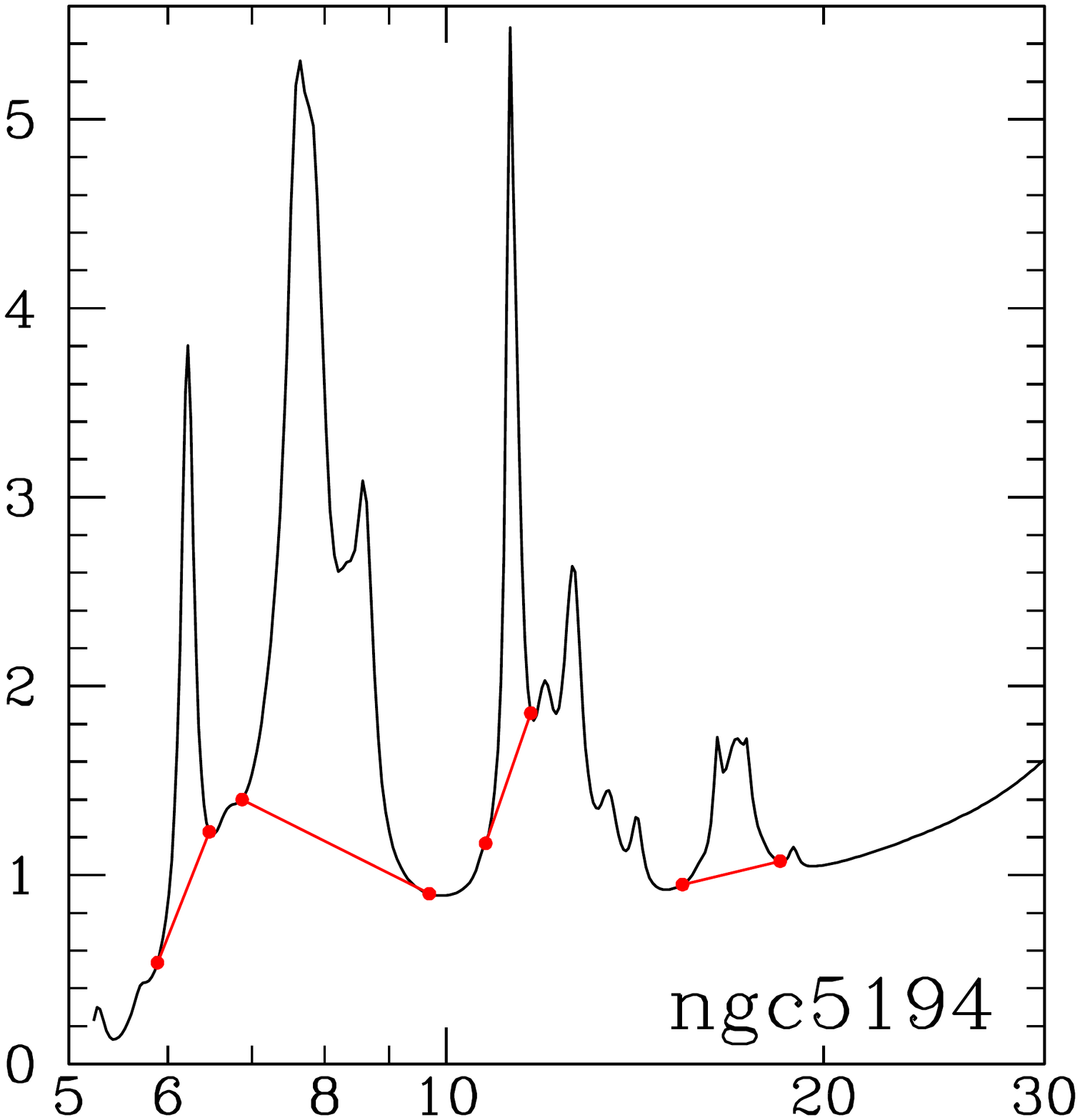}
\includegraphics[angle=0,width=\figwidthb,
                 clip=true,trim=0.5cm 5.0cm 1.75cm 2.5cm]
{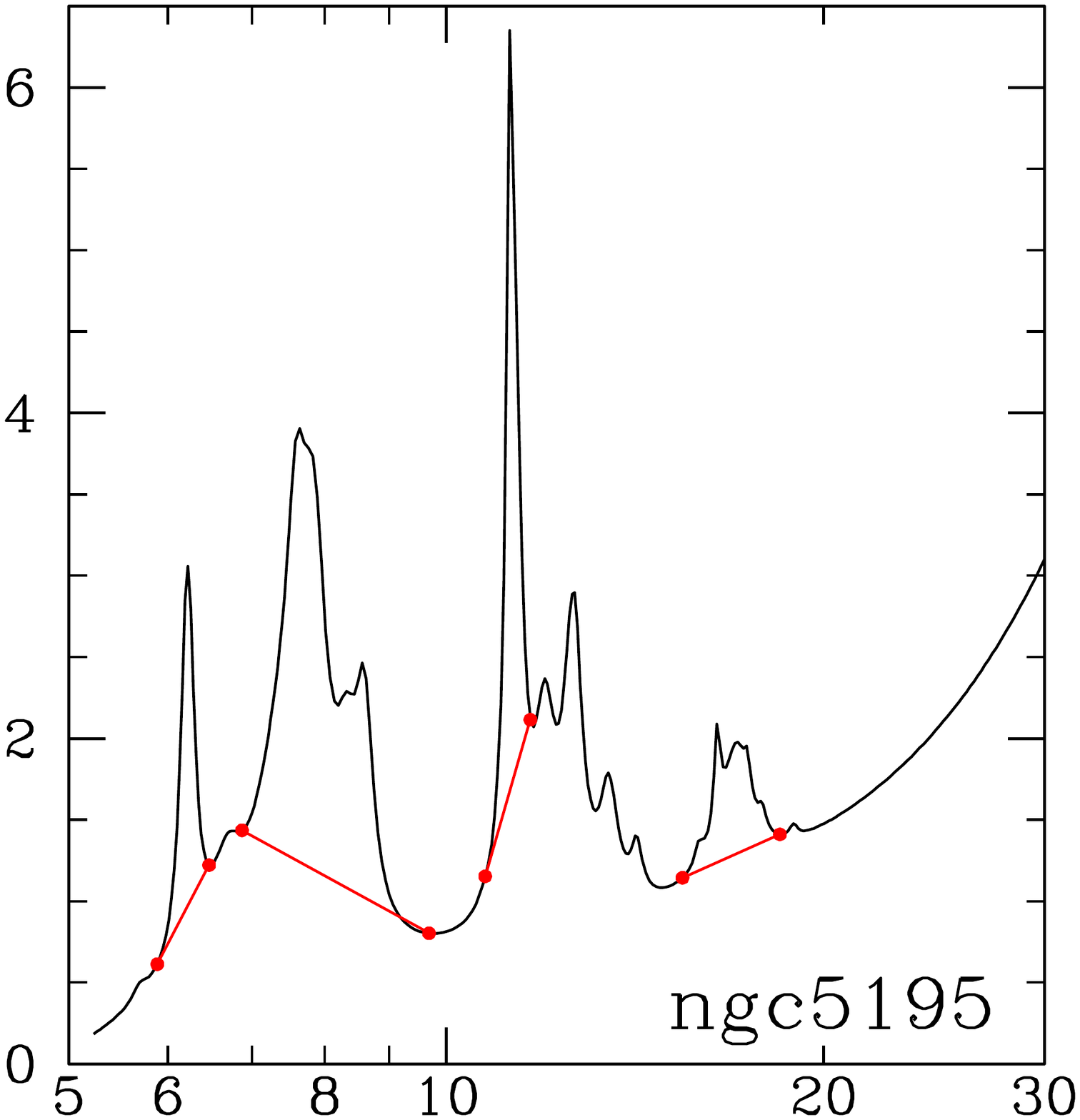}
\includegraphics[angle=0,width=\figwidtha,
                 clip=true,trim=0.5cm 5.0cm 0.5cm 2.5cm]
{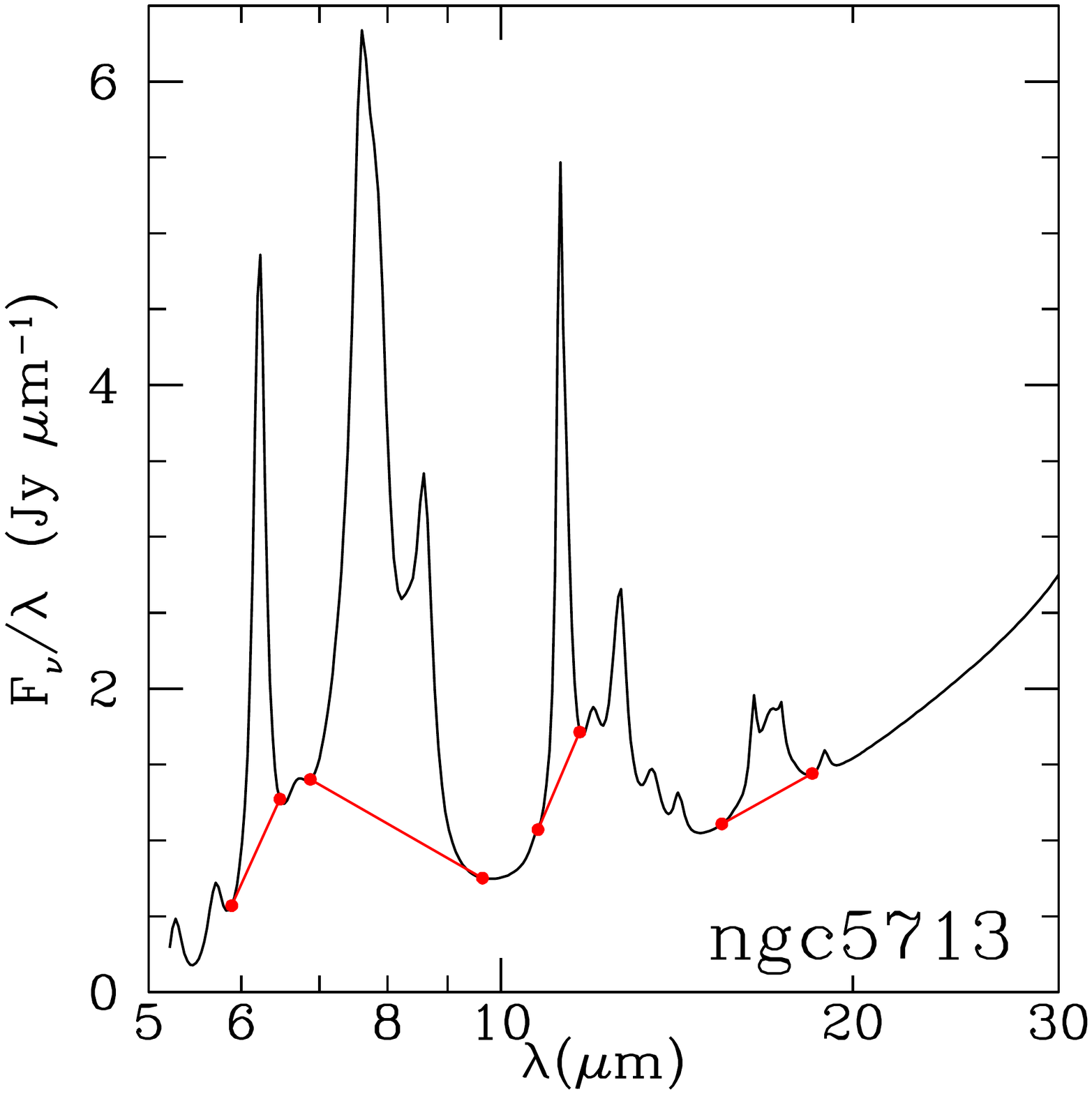}
\includegraphics[angle=0,width=\figwidthb,
                 clip=true,trim=0.5cm 5.0cm 1.75cm 2.5cm]
{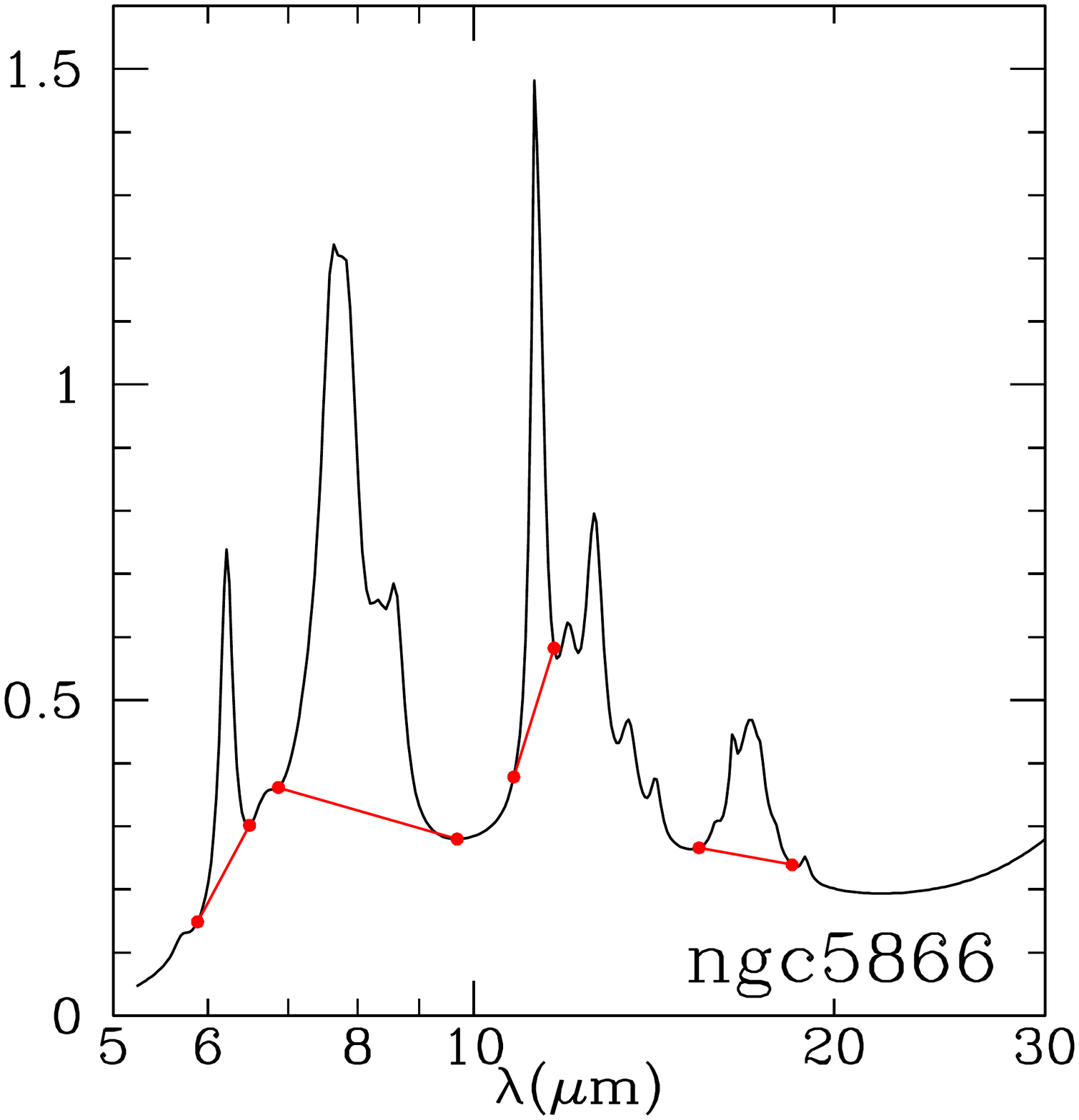}
\includegraphics[angle=0,width=\figwidthb,
                 clip=true,trim=0.5cm 5.0cm 1.75cm 2.5cm]
{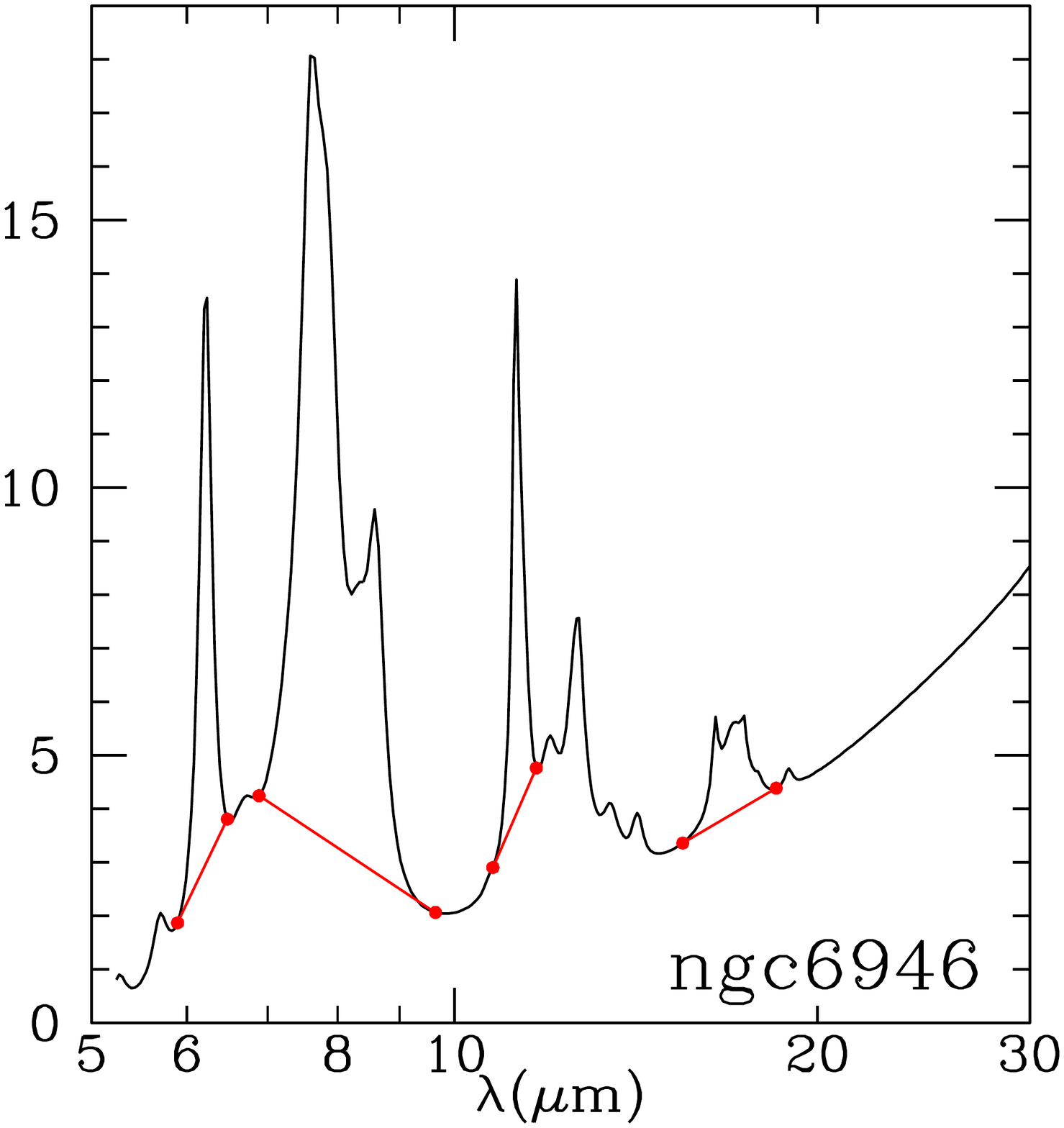}
\includegraphics[angle=0,width=\figwidthb,
                 clip=true,trim=0.5cm 5.0cm 1.75cm 2.5cm]
{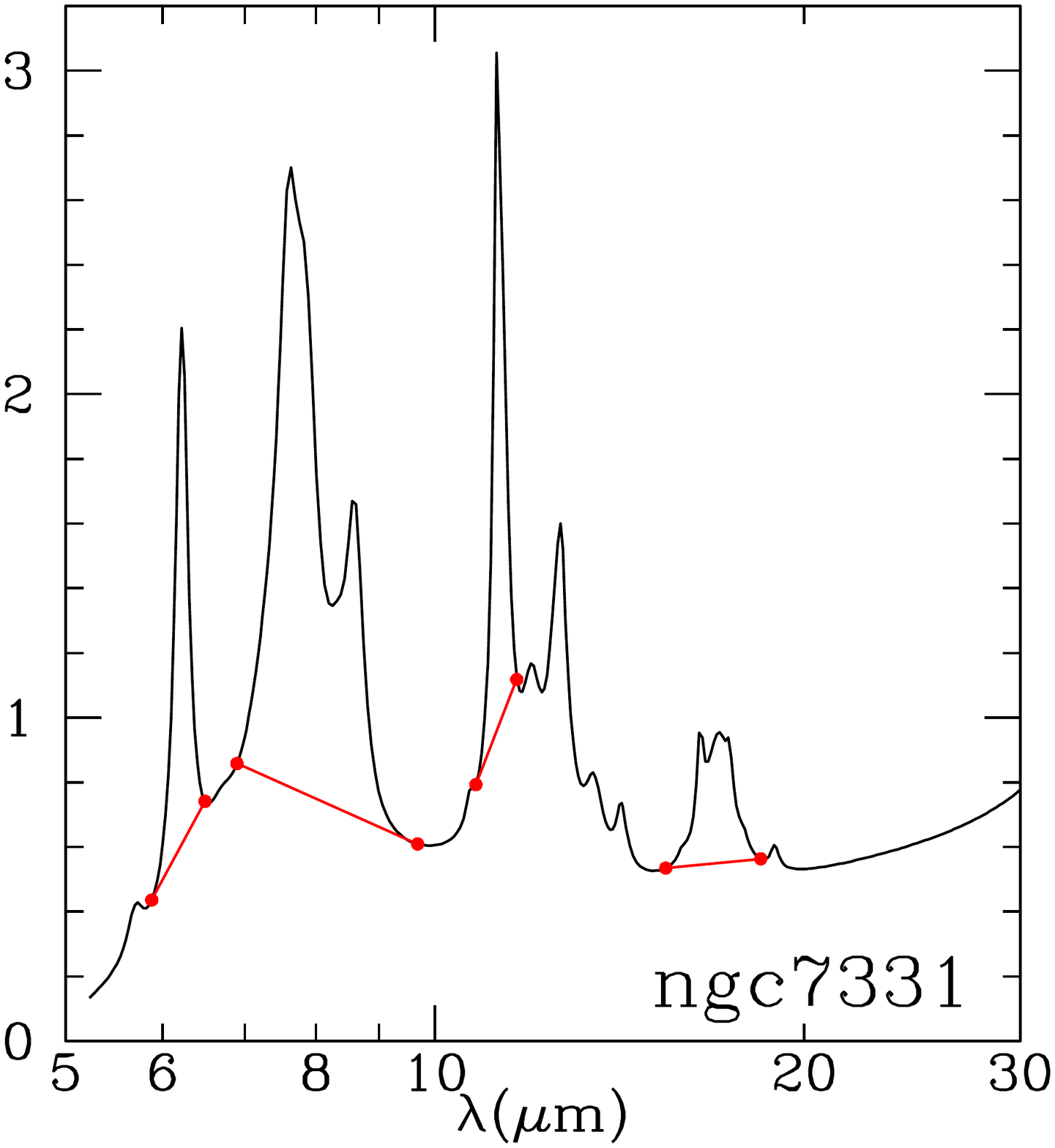}
\includegraphics[angle=0,width=\figwidthb,
                 clip=true,trim=0.5cm 5.0cm 1.75cm 2.5cm]
{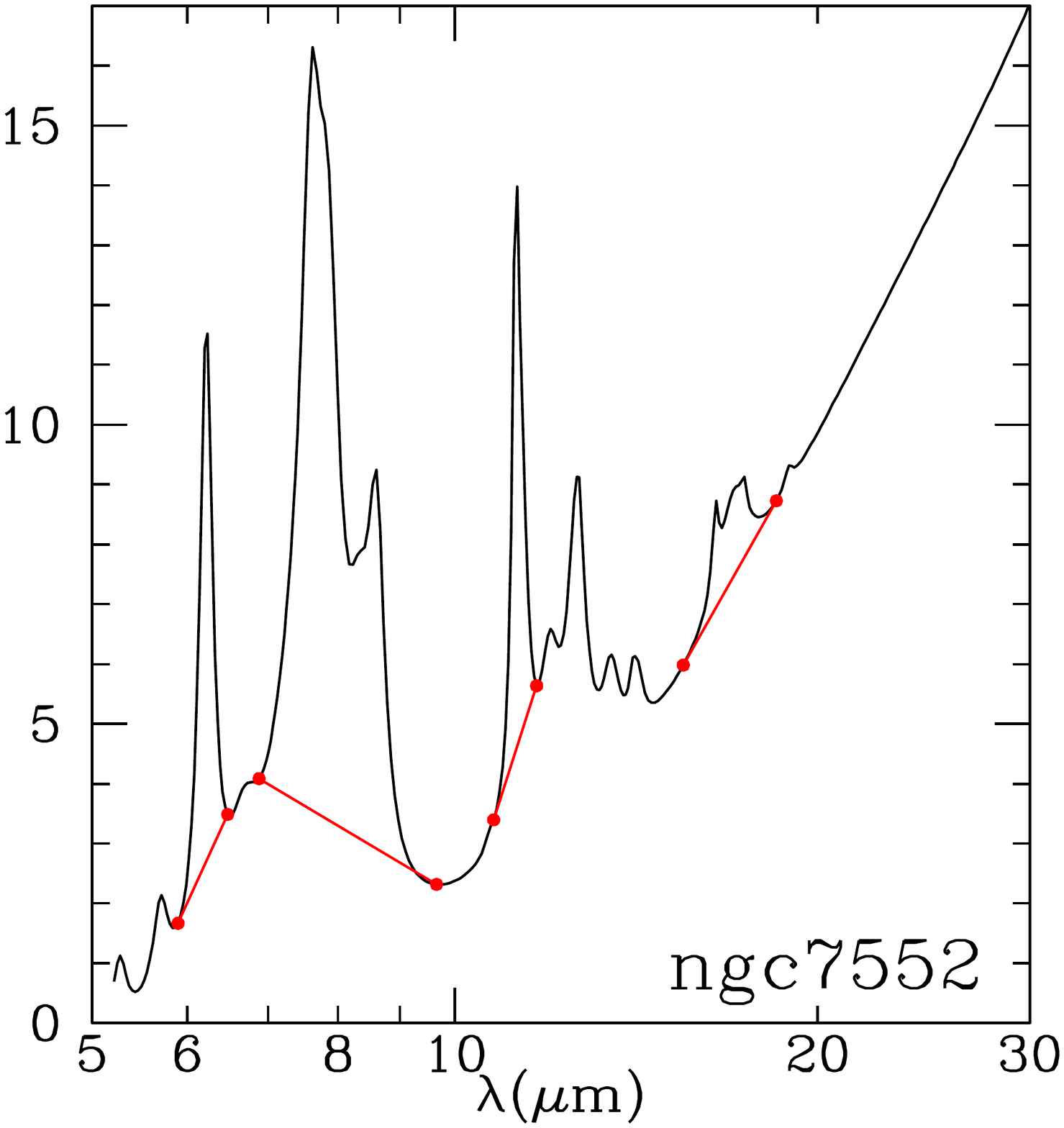}
\caption{\footnotesize \label{fig:sings}
         Line-subtracted and attenuation-corrected spectra of 
         25 SINGS galaxies \citep{Smith+Draine+Dale+etal_2007}, 
         showing clip lines used for extracting $\Fclip$.
         }
\end{center}
\end{figure}

Spectra for 25 of the SINGS galaxies \citep{Smith+Draine+Dale+etal_2007}, 
after removal of emission lines
and approximate subtraction of starlight, are shown in Figure \ref{fig:sings}.
For four galaxies\footnote{NGC1482, NGC 4536, NGC5866, NGC6946}
the observed
spectra showed evidence of extinction in the silicate feature; the spectra
used here have been corrected for extinction.
For each spectrum the ``clip lines'' are shown in red.

\bibliography{/u/draine/work/libe/btdrefs}

\end{document}